\documentclass[reprint,showkeys,amsmath,amssymb,prb,]{revtex4-2}
\usepackage{amsmath}
\usepackage{graphicx}
\usepackage{dcolumn}
\usepackage{bm}
\usepackage{array}
\usepackage{multirow}
\usepackage{mathrsfs}
\usepackage{placeins}
\usepackage{hyperref}
\usepackage{xcolor}
\usepackage{booktabs}

\begin{document}

\preprint{APS/123-QED}

\title{Persistent Homology for Structural Characterization in Disordered Systems}

\author{An Wang$^{1}$}
\email[Corresponding author: ]{amturing@outlook.com}

\author{Li Zou$^{2}$}
\affiliation{$^{1}$Department of Chemistry, University of Warwick, Coventry CV4 7AL, United Kingdom\\
$^{2}$Department of Cognitive Robotics, Delft University of Technology, 2628 CD Delft, The Netherlands}

\begin{abstract}

We propose a unified framework based on persistent homology (PH) to characterize both local and global structures in disordered systems. It can simultaneously generate local and global descriptors using the same algorithm and data structure, and has shown to be highly effective and interpretable in predicting particle rearrangements and classifying global phases. We also demonstrated that using a single variable enables a linear SVM to achieve nearly perfect three-phase classification. Inspired by this discovery, we define a non-parametric metric, the Separation Index (SI), which not only achieves this classification without sacrificing significant performance but also establishes a connection between particle environments and the global phase structure. Our methods provide an effective framework for understanding and analyzing the properties of disordered materials, with broad potential applications in materials science and even wider studies of complex systems.

\end{abstract}


\maketitle

\section{Introduction}

Local and global structural characterizations emphasize different aspects of materials, with the former focusing on microscopic features like coordination environment~\cite{zhang2019tuning,li2020modulating}, short-range order (SRO)~\cite{cowley1965short,capella1978unitarity,ferrari2023simulating}, bond angles and lengths~\cite{fuller1959hydrogen,geisinger1985molecular,laskowski1993main,shirley1995approach}, and the latter on macroscopic features like long-range order (LRO)~\cite{cowley1965short,hohenberg1967existence,white2013long,yang1962concept}, phase structure~\cite{pershan1988structure,izyumov2012phase,hirschfelder1937theory,takayama1976amorphous}, lattice constants~\cite{jette1935precision,alexander1998amorphous,drabold2009topics}, and overall symmetry~\cite{glazer2012space,aroyo2013international,de1974pseudo,dressel2014chiral,agarwala2020higher}. In conclusion, local characterization focuses on the environment and structure of individual particles or regions, while global characterization refers to the overall topology or geometry of the material.

In most cases, local characterization is straightforward because the geometric and interaction environment of particles has clear physical significance. However, most global characterization methods at present rely on simple operations, such as averaging, concatenation, or transformations of local features, to derive global representations from local characterizations. As a result, local and global characterizations often originate from different mathematical frameworks, algorithms, or data structures.

For instance, the radial distribution function (RDF) characterizes the global structure of a system by analyzing its average density distribution, which is obtained by averaging the local density of individual particles across the system~\cite{bernal1964bakerian}. Besides, in the context of feature engineering for machine learning (ML), Atom-centered symmetry functions (ACSF)~\cite{behler2007generalized,behler2011atom} and smooth overlap of atomic positions (SOAP)~\cite{bartok2013representing} characterize the local environment of individual particles using coordinate-independent functions and Gaussian smoothing combined with spherical harmonics expansion. For global characterization, the ACSF and SOAP vectors of individual particles are typically concatenated into a single vector or subjected to straightforward transformations, such as compression into vectors of uniform length. Moreover, order parameters~\cite{landau1936theory,landau1937theory} are commonly used to quantify the degree of order and symmetry. For local characterization, the Steinhardt's bond-orientational order parameter measures local structural symmetry using spherical harmonics~\cite{steinhardt1983bond}, while the ten Wolde's approach extends this by defining bonding criteria between particles and constructing bonding networks to better distinguish ordered and disordered environments~\cite{ten1995numerical}. For global characterization, the bond-orientational order parameters for all particles are typically averaged to quantify the global degree of order, providing an estimate of LRO formation~\cite{steinhardt1983bond,ten1995numerical}. Recently, we have noted that the introduction of the perspective of inversion symmetry and its breaking~\cite{zaccone2011approximate,milkus2016local,liu2022local,zaccone2023theory} has brought new insights to the relevant field. Additionally, the static structure factor (SSF) investigates multiscale order by mapping the average density distribution of particles into the frequency domain using Fourier transformations, with the ability to capture SRO and LRO by adjusting the wave vector~\cite{patterson1934fourier}. Medium-range order (MRO) also plays a significant role in understanding the relationship between microscopic and macroscopic properties of materials, particularly in phase behaviors~\cite{sheng2006atomic,bates1999block}. However, the extraction of MRO, such as local connectivity~\cite{thorpe1983continuous} or motifs~\cite{bernal1964bakerian,frank1952supercooling,finney1970random}, indeed still relies on averaging or statistical processing of local information of particles.

The above methods are already well-established with few shortcomings in effectiveness and performance, but their primary limitation lies in interpretability. Aggregating local features into a global representation by simple averaging, concatenation, or transformations often fails to capture or explain the complex mechanisms through which microscopic structures interact and transition into macroscopic properties. Complexity science emphasizes that macroscopic phenomena, such as self-organized criticality~\cite{bak1987self,olami1992self} or phase transitions~\cite{christensen2005complexity}, emerge from nonlinear, cross-scale interactions rather than simple additive contributions~\cite{anderson1972more,bak1987self,christensen2005complexity}. Existing approaches lack a unified mathematical framework to explain these interactions, making it difficult to bridge the microscopic and macroscopic scales. Therefore, an ideal global characterization method should not only accurately describe the overall structure of the system but also provide a mechanistic model that explains how microscopic structures influence macroscopic properties.

In recent years, deep learning has found broad use in physics~\cite{carleo2019machine,mehta2019high,radovic2018machine,carleo2017solving}. For instance, Graph Neural Networks (GNNs)~\cite{scarselli2008graph} aim to combine local and global features in graphs by iteratively aggregating and updating node representations through message passing, but they lack interpretability and rely on predefined graph structures and initial mappings at the node, edge, and graph levels, which limits their flexibility. Physics-Informed Neural Networks (PINNs)~\cite{raissi2019physics} integrate physical laws into models but rely on explicit equations, such as PDEs, making them less suitable for problems that cannot be precisely described by analytical expressions. Both of them are task-driven, making them better suited for ML tasks rather than as tools for traditional physics based on mathematical derivation.

Our goal is not to compare with or surpass the existing descriptors in terms of performance, nor to replace the automated feature engineering of deep learning, which is already highly effective~\cite{bapst2020unveiling}. Instead, we aim to develop a unified mathematical framework that seamlessly transitions between local and global representations. In our framework, local characterizations describe the structure of a neighborhood centered around a particle, and as the neighborhood radius increases, it gradually incorporates structural information over a larger area, eventually encompassing the entire system. This approach naturally enables a transition from local to global characterization, as both the neighborhood around a particle and the global system can fundamentally be represented as point clouds that differ only in scope.

Topological Data Analysis (TDA) provides a robust and effective tool for representing and analyzing point cloud data, and through persistent homology (PH), a method from algebraic topology~\cite{munkres2018elements}, we can encode the topological information of point clouds into a vectorizable representation~\cite{edelsbrunner2002topological,zomorodian2004computing,ghrist2008barcodes}. It is notable that PH has already seen successful applications, such as in studying the global topological characteristics of osmolyte molecular aggregation and hydrogen-bonding networks~\cite{xia2019persistent}, integrating machine learning to predict the structure-energy relationships of molecular clusters~\cite{chen2020topology}, and employing weighted approaches to uncover local topological features and propose new quantitative tools~\cite{anand2020weighted}. PH has also been employed to reveal SRO and MRO in silica glass and Cu-Zr metallic glass~\cite{hiraoka2016hierarchical}. Additionally, the Persistence Weighted Gaussian Kernel (PWGK), developed based on PH, has demonstrated stability and efficiency in feature engineering for machine learning (ML) tasks~\cite{kusano2016persistence}. Furthermore, PH-based ML potential descriptors enable efficient prediction of the physical properties of amorphous carbon without the need for hyperparameter tuning~\cite{minamitani2023persistent}. These works have greatly inspired us.

Based on this approach, we have developed a framework that enables both local and global characterization of disordered systems. This framework can serve as a supplement to traditional physical methods, providing a unified representation that seamlessly interprets how microscopic structures influence macroscopic properties. Besides, it can also generate the high-performance descriptors, enriching the feature engineering or initial structural mapping in downstream tasks of interpretable machine learning.

In our applications, we explore two main pathways within the PH-based framework we developed: a ML approach and a traditional physics (non-ML) approach. In the first pathway, we applied ML to two key tasks: classifying particle rearrangements by labeling particles as “soft” or “hard” and identifying the global phase structure, distinguishing between liquid, amorphous, and crystalline phases. The two tasks above are closely related. Based on physical intuition, solids mostly contain hard particles, while liquids are primarily soft. Additionally, highly ordered and symmetric structures are also largely composed of hard particles. Interestingly, we find that high classification accuracy across the three phases in multi-particle systems can be achieved using just a single variable. By leveraging the optimal hyperplane from a linear SVM model, we define a scalar field called “Global Softness”, which represents the average distance of all samples to the hyperplane and effectively captures the overall fluidity trend in the system.

The second pathway uses a traditional physics approach without ML. We analyzed the topology of both local particle environments and the global systems using our proposed new metric, the Separation Index (SI). The SI maintains a comparable level of performance to that achieved by the machine learning model, aligning with the finding that high accuracy of phase classification can be achieved with a single variable. By adjusting the neighborhood radius, SI also functions as a mechanistic model, illustrating how long-range order and global symmetry in crystalline materials emerge from the local environment of hard particles. The calculated results of SI align with those of Global Softness, offering a unified interpretation of fluidity and structural order through both ML and traditional physics perspectives.

We further explored how SI and Global Softness detect phase transitions. The results show that SI can capture the onset point of structural transitions between any two phases, especially a shift from a high-energy phase to a low-energy phase. Besides, Global Softness effectively identifies changes in system fluidity trends. Ultimately, these two pathways, i.e, the ML approach and the traditional physics (non-ML) approach have achieved a harmonious integration.

This article is organized as follows:

\begin{enumerate}
    \item Sec.~\ref{sec:methods} introduces the theories and methods involved in this article, including the persistent homology (PH), unified structural characterization, and both ML-based and non-ML approaches.

    \item Sec.~\ref{sec:data_set_generation} introduces the methods of data generation, including the protocols of molecular dynamics (MD) simulations, the approach of group division, the method of sample labeling, and the process of dataset construction.
    
    \item Sec.~\ref{sec:results} introduces the computational experiments, presents the results, and provides an analysis, including the outcomes and interpretations of both ML-based and non-ML approaches. Additionally, it explores phase transitions and their onset by tracking changes in various metrics along the simulated trajectories.

    \item Sec.~\ref{sec:conclusion} concludes the article and discusses the directions of future research.
\end{enumerate}

\section{Methods}
\label{sec:methods}

\subsection{Topological Methods}

\subsubsection{Vietoris-Rips (VR) Complex}
\label{subsec:vr_complex}

A multi-particle system can be represented as a point cloud $P = \{ p_1, p_2, \ldots, p_N \}$, where each point $p_i = (x_i, y_i, z_i)$ lies in a subspace of $\mathbb{R}^3$. The distance matrix $D$ is defined by $D_{ij} = | p_i - p_j |_2$, representing the Euclidean distance between points $p_i$ and $p_j$.

Given a distance threshold $\epsilon$, the Vietoris-Rips (VR) complex, a type of abstract simplicial complex (simply as ``simplicial complex'' or ``complex''), $\mathrm{VR}(P, \epsilon)$ can be constructed. Some points forms a simplex if the pairwise distances between them are less than or equal to $\epsilon$. Formally, a simplex $\sigma = [p_{i_0}, p_{i_1}, \ldots, p_{i_k}] \subseteq P$ belongs to $\mathrm{VR}(P, \epsilon)$ if $D_{i_j i_l} \leq \epsilon$ for all $j, l \in \{0, 1, \ldots, k\}$.

The process of building $\mathrm{VR}(P, \epsilon)$ begins at treating each point in $P$ as a 0-simplex (vertex). If two points $p_i$ and $p_j$ satisfy $D_{ij} \leq \epsilon$, they will be connected by a 1-simplex (edge). The 1-dimensional part of $\mathrm{VR}(P, \epsilon)$ is thus a graph composed of vertices and edges. Similarity, if the pairwise distance between three points $p_i, p_j, p_k$ satisfies $D_{ij}, D_{ik}, D_{jk} \leq \epsilon$, they form a 2-simplex (triangle), representing the 2-dimensional part of the complex.

The simplices in a complex are not isolated but rather interconnected by sharing vertices and edges, forming a higher-dimensional structure. For instance, If two triangles $[p_i, p_j, p_k]$ and $[p_j, p_k, p_l]$ share an edge $[p_j, p_k]$, they are connected to each other by this shared boundary, forming part of the complex. This process generalizes to higher dimensions, yielding a simplicial complex that includes points, edges, triangles, and higher-order simplices. The construction is controlled by $\epsilon$, and different $\epsilon$ corresponds to different geometric realizations of the point cloud, enabling the extraction of its geometric and topological information at varying scales.

\subsubsection{Homology Groups}
The VR complex reveals topological features such as connected components, loops, and voids, which describe the topological relationships between points. These features are characterized by homology groups~\cite{poincare1899complementa,veblen1912application}, denoted as $H_k$, where $k$ is the dimension of the homology group. Specifically, $H_0$ represents connected components, $H_1$ represents cycles (1-dimensional loops), and $H_2$ represents cavities (2-dimensional voids).

The \textit{homology group} is a fundamental tool from algebraic topology~\cite{munkres2018elements} used to classify topological spaces~\cite{hausdorff1914grundzuge,kuratowski1920notion,alexandroff1929memoire} by analyzing the structure of simplicial complexes. The boundary of a simplex is the most significant to understand the homology. Specifically, a $k$-simplex is a $k$-dimensional analogy of a triangle. For instance, a point is a 0-simplex, an edge is a 1-simplex, a triangle is a 2-simplex, a tetrahedron is a 3-simplex, etc. The boundary of a $k$-simplex is composed of its $(k-1)$-dimensional faces. Each $(k-1)$-dimensional face is defined by a subset of the vertices of the $k$-simplex. For example, the boundary of a solid triangle is all three edges of this triangle, and the face of this solid triangle means the hollow triangle with its interior removed.

The \textit{boundary operator} $\partial_k$ maps each $k$-simplex to its boundary, consisting of $(k-1)$-simplices: $\partial_k: C_k \to C_{k-1}$. Here, $C_k$ represents the $k$-chain group of a chain complex, composed of formal linear combinations of $k$-simplices over a chosen coefficient group, i.e., the cyclic group $\mathbb{Z}_{\mathfrak{p}}$. Specifically, an element $\sigma \in C_k$ can be written as a finite sum: $\sigma = c_1\sigma_1 + c_2\sigma_2 + \dots + c_n\sigma_n$, where $\sigma_i$ are $k$-simplices, and $c_i \in \mathbb{Z}_{\mathfrak{p}}$ are coefficients subject to modular arithmetic, meaning all addition and scalar multiplication are performed modulo $\mathfrak{p}$. A key property of the boundary operator is that the boundary of a boundary is always zero, i.e., $\partial_{k-1}(\partial_k(\sigma)) = 0$ for any $k$-simplex $\sigma$. This ensures that the boundary of a closed structure, such as a cycle, has no further boundary, which is essential for identifying topological features like cycles.

The \textit{kernel} of $\partial_k$, $\ker(\partial_k)=\{\sigma\in C_k\mid\partial_k(\sigma)=0\}$, consists of $k$-chains whose boundaries are zero—these are the cycles (closed loops) in the space. All operations in $C_k$ are performed modulo $\mathfrak{p}$, meaning the coefficients of $k$-chains belong to the cyclic group $\mathbb{Z}_{\mathfrak{p}}$. When $\mathfrak{p}=2$, the coefficients are reduced to binary values (0 or 1), so $k$-chains and their boundaries are reduced to binary relationships, simplifying analysis by focusing on the presence (1) or absence (0) of topological features.

The \textit{image} of $\partial_{k+1}$, $\operatorname{im}(\partial_{k+1})=\{\partial_{k+1}(\sigma)\mid\sigma\in C_{k+1}\}$, contains the boundaries of $(k+1)$-chains, which are the boundaries of higher-dimensional simplices. The homology group $H_k=\ker(\partial_k)/\operatorname{im}(\partial_{k+1})$ identifies $k$-dimensional cycles (or loops) that are not boundaries of higher-dimensional objects.

The \textit{Betti number} $\beta_k$, which is the rank of $H_k$, indicates the number of independent $k$-dimensional features. Specifically, $\beta_0=\text{rank}(H_0)$ describes the number of connected components, with $\beta_0=\dim(\ker(\partial_0))-\dim(\operatorname{im}(\partial_1))$; $\beta_1=\text{rank}(H_1)$ counts independent cycles, with $\beta_1=\dim(\ker(\partial_1))-\dim(\operatorname{im}(\partial_2))$; and $\beta_2=\text{rank}(H_2)$ measures independent cavities, with $\beta_2=\dim(\ker(\partial_2))-\dim(\operatorname{im}(\partial_3))$.

\subsubsection{Persistent Homology}
\label{subsec:persistent_homology}

\begin{figure}[htbp]
\includegraphics[width=0.5\textwidth]{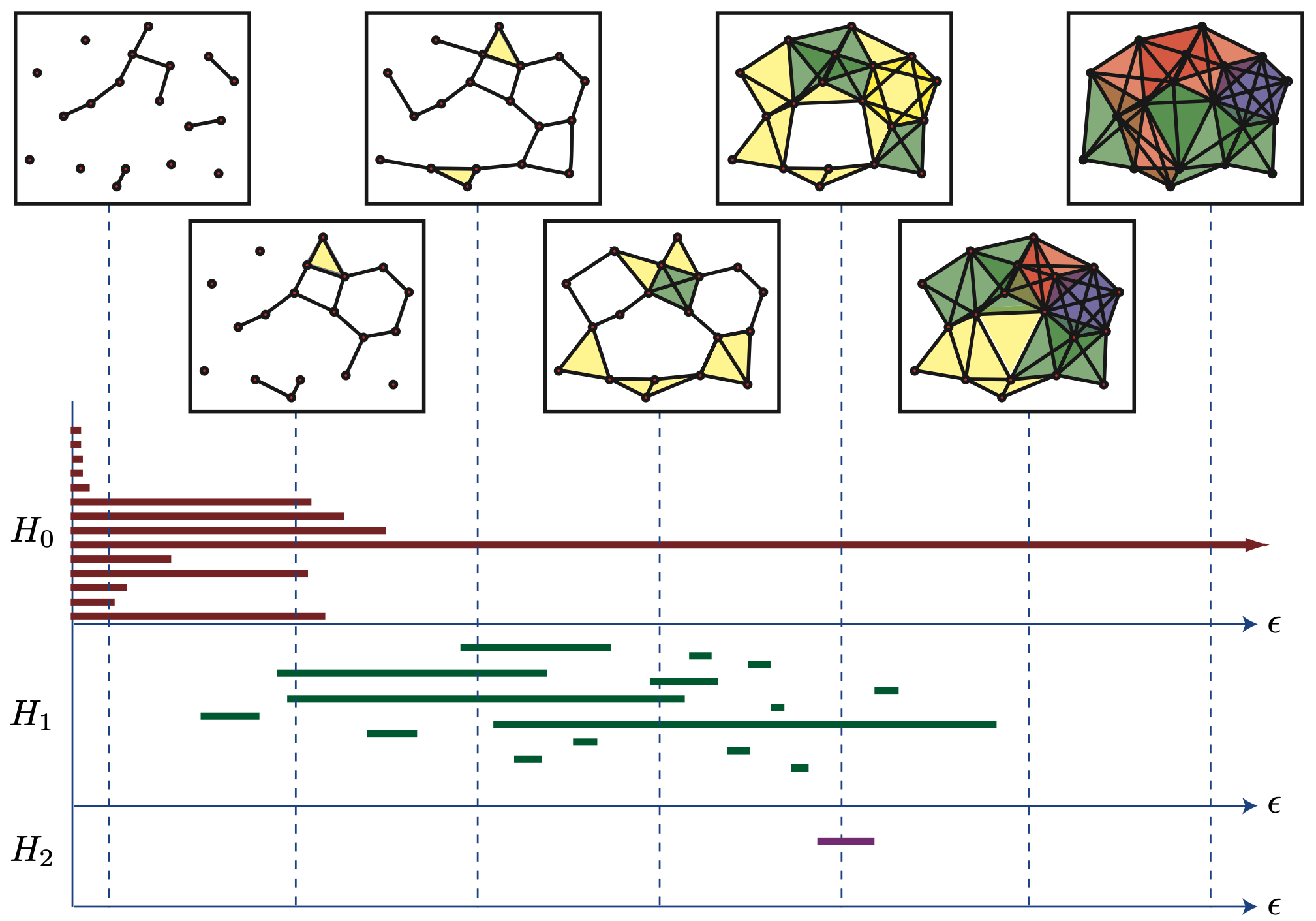}
\caption{This figure is adapted from Ref.~\cite{ghrist2008barcodes}. It demonstrates barcodes, a faithful representation of the persistent homology (PH) results, showing how topological features emerge and persist as the parameter $\epsilon$ increases. The lines in the barcodes are categorized by their homology group ($H_0$, $H_1$, $H_2$), with each line representing a homology class. The left endpoint marks the feature’s birth at $\epsilon_i$, while the right endpoint indicates its death at $\epsilon_j$, with the length representing its persistence. Each line corresponds to either a birth-death pair $(\epsilon_i,\epsilon_j)$ or a birth-persistence pair $(\epsilon_i,\epsilon_j - \epsilon_i)$, where the persistence is the lifespan of the feature. The number of lines intersecting a vertical line at any $\epsilon$ represents the number of $H_k$ topological features at that scale, corresponding to the Betti number $\beta_k$. This paper focuses on $k=0,1,2$.}
\label{fig:persistent_homology_process}
\end{figure}

\begin{figure*}[htbp]
\includegraphics[width=1\textwidth]{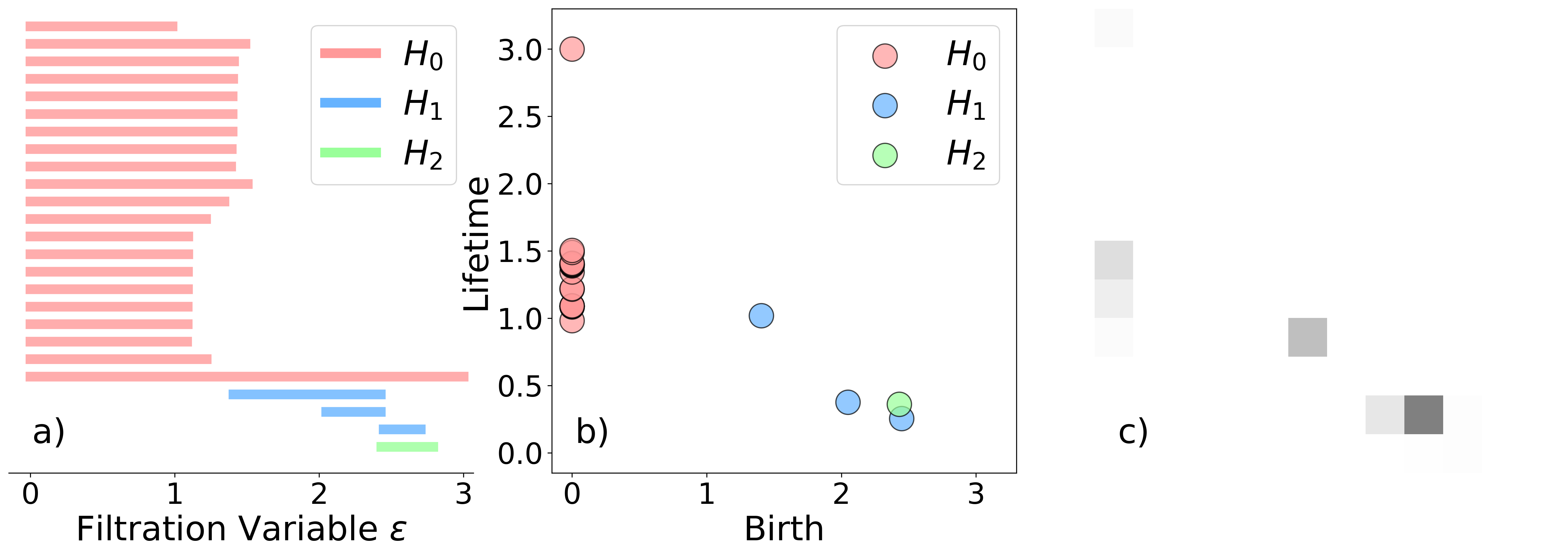}
\caption{This figure shows the relationship between barcodes, persistence diagrams (PD), and persistence images (PI): a) shows barcodes, b) is the PD, and c) is the PI. PD maps barcode points to a 2D Cartesian system, and PI smooths these points using kernel-density estimation (KDE, see Eq.~\ref{eq:mapping}), compressing varying PDs into fixed-size images for machine learning (ML) tasks.}
\label{fig:barcodes}
\end{figure*}

Fig.~\ref{fig:persistent_homology_process} and Fig.~\ref{fig:barcodes} illustrate how persistent homology (PH) is applied to point cloud data, leading to equivalent representations like barcodes~\cite{ghrist2008barcodes}, persistence diagrams (PD)~\cite{edelsbrunner2002topological,zomorodian2004computing}, and the smoothed representation as persistence images (PI)~\cite{adams2017persistence}.

\textit{Persistent Homology} (PH) analyzes topological features by constructing a sequence of VR complexes $\mathrm{VR}(P, \epsilon)$ as the distance threshold $\epsilon$ varies. Tracking changes in homology groups reveals features such as connected components ($H_0$), loops ($H_1$), and cavities ($H_2$). The filtration variable~\cite{edelsbrunner2002topological} $\epsilon$ determines the scale at which the topological structure is built. As $\epsilon$ increases, the VR complexes evolve, capturing different features at various scales. Betti numbers~\cite{betti1870sopra,poincare1895analysis} ($\beta_k$) count the number of $k$-dimensional features, where $\beta_0$ counts connected components, $\beta_1$ counts cycles, and $\beta_2$ counts cavities.

Topological features are ``born" when they firstly appear in the complex. For example, a new connected component ($H_0$) is born when a point first appears as $\epsilon$ increases. Similarly, a cycle ($H_1$) is born when a closed cycle (or loop) forms, and a cavity ($H_2$) is born when an enclosed space emerges.

Topological features are ``died" when they vanish. For connected components ($H_0$), death occurs when two components merge. For cycles ($H_1$), death happens when the cycle is filled, meaning the cycle becomes the boundary of a 2-simplex (such as a triangle), effectively eliminating the open space enclosed by the cycle. For cavities ($H_2$), death occurs when the cavity is fully enclosed by higher-dimensional simplices, becoming the boundary of a 3-dimensional structure. In general, the death of a topological feature in homology group $H_k$ happens when the feature becomes the boundary of a higher-dimensional simplex in $H_{k+1}$. Filling the feature with lower-dimensional simplices (such as adding a line segment inside a triangle) will not cause the $H_k$ feature to die.

A \textit{filtration chain}~\cite{edelsbrunner2002topological} is a sequence of nested complexes $\emptyset = K_0 \subseteq K_1 \subseteq \ldots \subseteq K_n = \mathrm{VR}(P, \epsilon_n)$, with $\epsilon_0 < \epsilon_1 < \ldots < \epsilon_n$, and the homology groups satisfy $H_k(K_0) \rightarrow H_k(K_1) \rightarrow \ldots \rightarrow H_k(K_n)$. Each $k$-dimensional homology class~\cite{poincare1895analysis} $H_k$ is born at some $K_i$ and dies at some $K_j$, forming birth-death pairs $(\epsilon_i, \epsilon_j)$ or equivalent birth-persistence pairs $(\epsilon_i, \epsilon_j - \epsilon_i)$.

A \textit{barcode}~\cite{ghrist2008barcodes} is a set of intervals $\{ [\epsilon_i, \epsilon_j) \}$, where each interval represents the lifespan of a homology class, with $\epsilon_i$ as the birth time and $\epsilon_j$ as the death time. As shown in Fig.~\ref{fig:persistent_homology_process}, the barcodes of the point cloud $P$ are a faithful representation of the results of the persistent homology analysis of $P$.

A \textit{Persistence Diagram} (PD)~\cite{edelsbrunner2002topological,zomorodian2004computing} is the set of birth-lifespan pairs $PD_k = \{ (\epsilon_i, \epsilon_j - \epsilon_i) \}$, where each element indicates the birth time and the lifespan of a homology class. It is obtained by mapping each interval from the barcode onto a 2D Cartesian coordinate system, where each interval is represented as a point, as shown in Fig.~\ref{fig:barcodes}a) and~\ref{fig:barcodes}b). The PDs for $H_0$, $H_1$, and $H_2$ are often combined as $PD = \cup_{k} PD_k, \forall k = 0, 1, 2$.

A \textit{Persistence Image} (PI)~\cite{adams2017persistence} uses kernel density estimation (KDE)~\cite{parzen1962estimation,davis2011remarks} to convert points from a PD into a smooth distribution, producing a fixed-size image, as demonstrated in Fig.~\ref{fig:barcodes}c). Each point $(\epsilon_i, \epsilon_j - \epsilon_i)$ in the PD is denoted as $(b, d)$, where $b = \epsilon_i$ is the birth time and $d = \epsilon_j - \epsilon_i$ is the persistence. Assuming the PI has a resolution of $\mathfrak{M} \times \mathfrak{N}$, the pixel value at position $(\mathfrak{m}, \mathfrak{n})$, where $\mathfrak{m} \in [1, \mathfrak{M}]$ and $\mathfrak{n} \in [1, \mathfrak{N}]$, is computed by:

\begin{equation}
PI(\mathfrak{m}, \mathfrak{n}) = \sum_{k \in PD} \left[ \Phi \left( \frac{x_{\mathfrak{m}} - b_k}{\sigma} \right) \Phi \left( \frac{y_{\mathfrak{n}} - d_k}{\sigma} \right) w(b_k, d_k) \right]
\label{eq:mapping}
\end{equation}

\noindent where $\sigma$ is the standard deviation of the Gaussian kernel, $\omega(b,d)=d/\max(d)$ represents the weight and $\Phi(z)$ is the cumulative distribution function (CDF) of the standard normal distribution~\cite{gauss1877theoria}. Finally, the PI is flattened into a vector for use in ML tasks.

In this article, we used the Ripser~\cite{bauer2021ripser} package to compute persistent homology (PH), and its time complexity for obtaining barcodes from a point cloud with $N$ points, including the computation of $H_0$, $H_1$, and $H_2$ homology classes, is $O(N^4)$ in the worst case~\cite{bauer2021ripser}.

\subsubsection{PH Descriptors}
\label{subsec:descriptors}

\textit{Local Characterization.} The local environment of a particle is characterized by its neighborhood (surrounding region). For particle $p_i$, its neighborhood is denoted as $\mathbb{L}^{(P)}_i(r) = \{ a \in \mathbb{R}^3 \mid \| p_i - a \|_2 \leq r \}$, where $r > 0$.

To analyze the local environment of a particle $p \in P$, we compute the PH for a sequence of increasing radii $r_1 < r_2 < \cdots < r_{\mathcal{N}}$, generating point clouds $\mathfrak{P}^{(P)}_{p,q}$ for each corresponding neighborhood. Each point cloud $\mathfrak{P}^{(P)}_{p,q}$ has a corresponding PI vector $\bold{V}^{(P)}_{p,q}$. Intuitively, a smaller-radius neighborhood has a stronger impact on local characteristics, while a larger-radius neighborhood contributes less. Therefore, the multi-scale PI feature vector $\bold{I}^{(P)}_{p}$ for particle $p$ is then obtained by adding the PI vectors with exponential decay as follows:

\begin{equation}
    \bold{I}^{(P)}_{p} = \sum_{q=1}^{\mathcal{N}} \bold{V}^{(P)}_{p,q} \cdot e^{-\tau (r_q-r_1)}
    \label{equ:PI_local}
\end{equation}

\textit{Global Characterization.} For the entire system, PH is applied directly to $P$, resulting in the global PI vector:

\begin{equation}
    \bold{I}^{(P)}_{\text{entire}} = \bold{Vec}(P)
    \label{equ:PI_global}
\end{equation}

The global PI vector captures the overall structure of the system, while local PI vectors provide multi-scale information by expanding neighborhood radii with an exponential decay in influence for more distant regions. Together, these vectors establish a unified mathematical framework that links the global structures with the local environments of individual particles.

\subsection{Traditional Metrics}
\label{sec:traditional}

Here, we present the methods for labeling or classifying the three phases (liquid, amorphous solids, and crystalline) of a multi-particle system constrained by the Lennard-Jones interactions within traditional physics.

\subsubsection{Mean Square Displacement (MSD)}

The mean square displacement (MSD)~\cite{einstein1905molekularkinetischen} describes the translation diffusion of the particles. In an $\mathcal{N}$-body system, for each particle $i$, its position at time $t$ (initial time is denoted as $t_0$) can be represented by $\overrightarrow{r_i}$. The MSD is defined as:

\begin{equation*}
    \text{MSD}(t) = \langle |\overrightarrow{r_i}(t)-\overrightarrow{r_i}(t_0)|^2 \rangle = \frac{1}{\mathcal{N}} \sum_{i=1}^{\mathcal{N}} |\overrightarrow{r_i}(t)-\overrightarrow{r_i}(t_0)|^2
\end{equation*}

In crystals, particles are arranged in an orderly structure with minimal mobility. The MSD shows slight fluctuations initially but stabilizes over longer timescales, reflecting local vibrations. In liquids, particles exhibit high mobility, and the MSD increases approximately linearly with time, indicating free diffusion. Specifically, amorphous solids lie between these states, with the MSD showing nonlinear growth as particles start confined locally and gradually stabilize or increase slowly, indicating moderate mobility.

Generally, MSD can effectively distinguish between solids and liquids and capture the crystallization of amorphous solid, but it cannot differentiate low-mobility amorphous solids from crystals.

\subsubsection{Bond-orientational Order Parameters}

\label{subsec:Bond_orientational_Order_Parameters}

The Steinhardt's local bond order parameter~\cite{steinhardt1983bond} $q_l(i)$ quantifies the order of local atomic structure. Each $q_l(i)$ is a vector with components $q_{lm}(i)$ defined by spherical harmonics $Y_{lm}$:

\begin{equation*}
q_{lm}(i)=\frac{1}{N_b(i)}\sum_{j=1}^{N_b(i)}Y_{lm}(\theta_{ij},\phi_{ij})
\end{equation*}

\noindent where $N_b(i)$ is the number of neighbors of particle $i$, and $\theta_{ij}$, $\phi_{ij}$ are the angles of neighbor $j$ relative to $i$.

The ten Wolde's order parameter~\cite{ten1995numerical,jungblut2016pathways} refines this by defining connectivity $S_{ij}$ between neighbors $i$ and $j$:

\begin{equation*}
S_{ij}=\sum_{m=-l}^{l}q_{lm}(i)\cdot q_{lm}(j)
\end{equation*}

Particles $i$ and $j$ are connected if $S_{ij}>c_q=0.5$. A particle is classified as crystal-like if it has at least $N_c=8$ connections. Based on this, the number of crystal particles in the system can be counted as:

\begin{equation*}
    N_{\text{solid}} = \sum_{i=1}^{N} \Theta\left( \sum_{j=1}^{N_b(i)} \Theta(S_{ij} - c_q) - (N_c-1) \right)
\end{equation*}

\noindent where $\Theta$ is the Heaviside step function~\cite{heaviside2003electromagnetic}:

\begin{equation*}
\Theta(x) = 
\begin{cases}
1, & \text{if } x \geq 0 \\
0, & \text{if } x < 0
\end{cases}
\end{equation*}

The global bond order parameter $Q_l$ averages $|q_l(i)|$ over all solid-like particles $N_{\text{solid}}$:

\begin{equation*}
Q_l=\frac{1}{N_{\text{solid}}}\sum_{i\in \text{solid}}|q_l(i)|
\end{equation*}

\noindent where $|q_l(i)|=\left(\sum_{m=-l}^{l}|q_{lm}(i)|^2\right)^{1/2}$. The value of $Q_l$ reflects the degree of order in the solid regions.

In a uniform system with $N$ particles where $Q_l$ can sufficiently represent the overall degree of order, the following relationship can be established:

\begin{equation*}
N_{\text{solid}} \approx N \times Q_l
\end{equation*}

It is evident because, from a statistical perspective, $Q_l$ can be viewed as an average measure of the probability of being ordered. Thus, $Q_l$ effectively serves as an estimate of the fraction of ordered particles in the system.

Typically, we set $l=6$ and use $Q_6$ to represent the average degree of order in the system, distinguishing between ordered phases (crystals) and disordered phases (liquids and amorphous solids). This choice of $l=6$ is particularly effective because it captures the sixfold symmetry typical of many crystalline structures and provides strong discriminative power between ordered and disordered phases~\cite{steinhardt1983bond}.

Combining the structural information presented in $Q_6$ and the mobility shown in MSD allow us to distinguish between liquids, amorphous solids, and crystals.

\subsubsection{Identifying Rearrangements}
\label{subsec:Identifying_Rearrangements}

To characterize the tendency of particle rearrangement, we define the positional change $p_{\text{hop},p}(t)$ of a given particle $p$ within a time window $[t-t_{R/2}, t+t_{R/2}]$ \cite{candelier2010spatiotemporal, smessaert2013distribution}. If $p_{\text{hop},p}(t)$ exceeds a threshold $p_c$ during this interval, the particle is labeled as “soft”; otherwise, it is labeled as “hard” \cite{schoenholz2016structural}. Specifically, $p_{\text{hop},p}(t)$ is defined by:

\begin{equation}
    p_{\text{hop},p}(t)=\sqrt{\langle (\bold{r}_p-\langle \bold{r}_p \rangle_B)^2 \rangle_A \cdot \langle (\bold{r}_p-\langle \bold{r}_p \rangle_A)^2 \rangle_B}
    \label{equ:p_hop}
\end{equation}

\noindent where $\bold{r}_p$ is the position vector of particle $p$, and $\langle . \rangle_A$ and $\langle . \rangle_B$ represent averages over the intervals $A=[t-t_{R/2},t]$ and $B=[t,t+t_{R/2}]$, respectively.

We set $t_{R/2}=20$ and $p_c=0.1$ based on the assumption that most particles in a liquid state are expected to exhibit “soft” behavior. The detailed explanation of threshold selection is provided in the Appendix~\ref{app:select_of_p_hop}.

\subsection{Machine Learning (ML) Techniques}

\subsubsection{Basic Concepts}
\label{subsec:ml_def_basic_conc}

Machine learning (ML) can be employed to investigate the relationship between data instances and target variables. In this article, ML specifically refers to interpretable discriminative classification tasks with handcrafted feature engineering.

Firstly, data instances are transformed into a feature matrix through feature engineering, denoted as $\mathbf{X} = g(\mathbf{x})$, where $\mathbf{x}$ is a vector representing all instances and $g$ represents the process of feature engineering. Each instance is mapped to a corresponding feature vector, forming the feature matrix $\mathbf{X}$, where the $i$-th row represents the $i$-th instance and the $j$-th column corresponds to the $j$-th feature.

Based on this representation, ML is applied to fit a composite function $\mathbf{y} = f(g(\mathbf{x})) = f(\mathbf{X})$, where $\mathbf{y}$ is the target property vector, and $f$ is the ML model that captures the relationship between the feature matrix $\mathbf{X}$ and $\mathbf{y}$. Here, we choose support vector machines (SVM)~\cite{cortes1995support} as $f$, not only because of their interpretability but also due to the convenience of constructing a scalar field based on the directed distance from samples to the decision hyperplane in the input space when using a linear kernel. This allows for a more intuitive characterization of the target properties of the samples.

\subsubsection{Principal Component Analysis (PCA)}
\label{subsec:pca}

In our research, the feature matrix $\bold{X}$ is constructed by concatenating the vectors $\bold{I}^{(P)}_{p}$ (defined in Eq.~\ref{equ:PI_local}) and $\bold{I}^{(P)}_{\text{entire}}$ (defined in Eq.~\ref{equ:PI_global}) corresponding to each sample into a single matrix.

Typically, $\mathbf{X}$ contains a large number of columns corresponding to multiple predictors, making dimensionality reduction necessary. Principal Component Analysis (PCA)~\cite{pearson1901liii} is employed to achieve this by applying a linear transformation that generates a new set of uncorrelated principal components. These components are ranked according to their explained variance, enabling the selection of a smaller subset while preserving most of the original information. Since the importance of each principal component is determined by its associated eigenvalue, ranking the eigenvalues quantifies their relative contribution to the total variance in the data.

The steps for performing PCA on $\textbf{X}$ with an explained variance proportion of $z$ are as follows:

Firstly, we standardize $\bold{X}$ to $\bold{X}_{\text{std}}$, ensuring each feature has a mean of 0 and variance of 1 using the formula $\bold{X}_{\text{std}} = (\bold{X} - \mu) / \sigma$, where $\mu$ is the mean and $\sigma$ the standard deviation.

Next, we need to compute the covariance matrix $\bold{C} = (\bold{X}_{\text{std}}^T \bold{X}_{\text{std}}) / (n-1)$, where $n$ is the number of samples, to capture linear relationships between features. We then perform eigen decomposition on $\bold{C}$, solving $\text{det}(\bold{C} - \lambda \bold{I}) = 0$ to find eigenvalues $\lambda_1, \lambda_2, \dots, \lambda_m$ and their corresponding eigenvectors, sorting the eigenvalues in descending order and retaining the first $k$ components such that the cumulative explained variance ratio $\sum_{i=1}^{k} \lambda_i / \sum_{i=1}^{m} \lambda_i \geq z$.

Finally, we construct the projection matrix $\bold{P}$ using the top $k$ eigenvectors and project the standardized data as $\bold{X}_{\text{PCA}} = \bold{X}_{\text{std}} \cdot \bold{P}$, reducing dimensionality while preserving essential information, which aids in data compression, feature extraction, and noise reduction.

Note that, in the testing stage, it is essential to apply the same PCA transformation used during model selection and training. Firstly, it maintains a consistent feature space between training and testing, as the model was trained using the PCA-defined space from the training sets. Secondly, it prevents data leakage because recalculating PCA on the test sets could unintentionally incorporate test data characteristics into the model. Consistent PCA transformations are therefore essential for an unbiased evaluation of model performance.

For all ML tasks discussed in this article, the feature matrix undergoes PCA by default, which will not be elaborated further. For simplicity, we will use $\bold{X}$ to represent $\bold{X}_{\text{PCA}}$ in the following sections.

\subsubsection{ML Models}

This article uses two kinds of Support Vector Machines (SVM)~\cite{cortes1995support} as the ML model $f$:

\vspace{0.25cm}

\paragraph{SVM with Linear Kernel}
\begin{itemize}
    \item Decision function:
    \begin{equation}
        f(\bold{X}) = \bold{X} \cdot \bold{w} + u
        \label{eq:linear_svm}
    \end{equation}
    where $\bold{w}$ is the weight vector, $u$ is the bias term.
    \item Objective function:
    \begin{equation*}
        \mathop{\min}_{\bold{w},b} \frac{1}{2} ||\bold{w}||_2^2 + C \sum_{i=1}^N \mathop{\max} (0, 1-y_i (\bold{X}_{(i)} \cdot \bold{w} + u))
    \end{equation*}
    where $C$ is the regularization parameter that controls the trade-off between the regularization term and the loss term.
    \item Hyperparameters: $C \in [0.001,1000]$, log-scaled.
\end{itemize}

\vspace{0.25cm}
\paragraph{SVM with RBF Kernel}
\begin{itemize}
    \item RBF kernel:
    \begin{equation*}
        K(\bold{X}_{(i)},\bold{X}_{(j)}) = \mathop{\exp}(-\gamma || \bold{X}_{(i)} - \bold{X}_{(j)} ||^2)
    \end{equation*}
    where $\gamma$ is a parameter that controls the extent of the influence of each data point on the similarity measure.

    \item Decision function:
    \begin{equation*}
        f(\bold{X}) = \sum_{i=1}^N \alpha_i y_i K(\bold{X},\bold{X}_{(i)}) + u
    \end{equation*}
    where $\alpha_i$ are Lagrange multipliers quantifying the importance of $X_{(i)}$ within the decision function and $u$ is the bias term.

    \item Objective function:
    \begin{equation*}
        \mathop{\max}_{\alpha} \left[ \sum_{i=1}^N \alpha_i - \frac{1}{2} \sum_{i=1}^N \sum_{j=1}^N \alpha_i \alpha_j y_i y_j K(\bold{X}_{(i)},\bold{X}_{(j)}) \right]
    \end{equation*}
    subject to the constraints:
    \begin{equation*}
        \sum_{i=1}^N \alpha_i y_i = 0 \text{ and } 0 \leq \alpha_i \leq C, \forall i.
    \end{equation*}
    Here, $C$ is a regularization parameter, and $\alpha_i$ are Lagrange multipliers relative to each sample.

    \item Hyperparameters: (1) $C \in [0.001,1000]$, log-scaled; (2) $\gamma \in [0.001,1000]$, log-scaled.
\end{itemize}

In this research, we set the hyperparameter optimization as follows: 10-fold cross-validation for validation, Matthews Correlation Coefficient (MCC))~\cite{matthews1975comparison} and accuracy as evaluation metrics, and a Bayesian optimizer~\cite{snoek2012practical} with 200 steps using Expected Improvement per second plus (EIps) as the acquisition function. For the multi-class strategy (only for Task 2), we try the methods of One-vs-One and One-vs-All at the same time, and select the best-performing one.

\subsubsection{Scalar Field Trained by Linear SVM}

\begin{figure}[htbp!]
\includegraphics[width=0.5\textwidth]{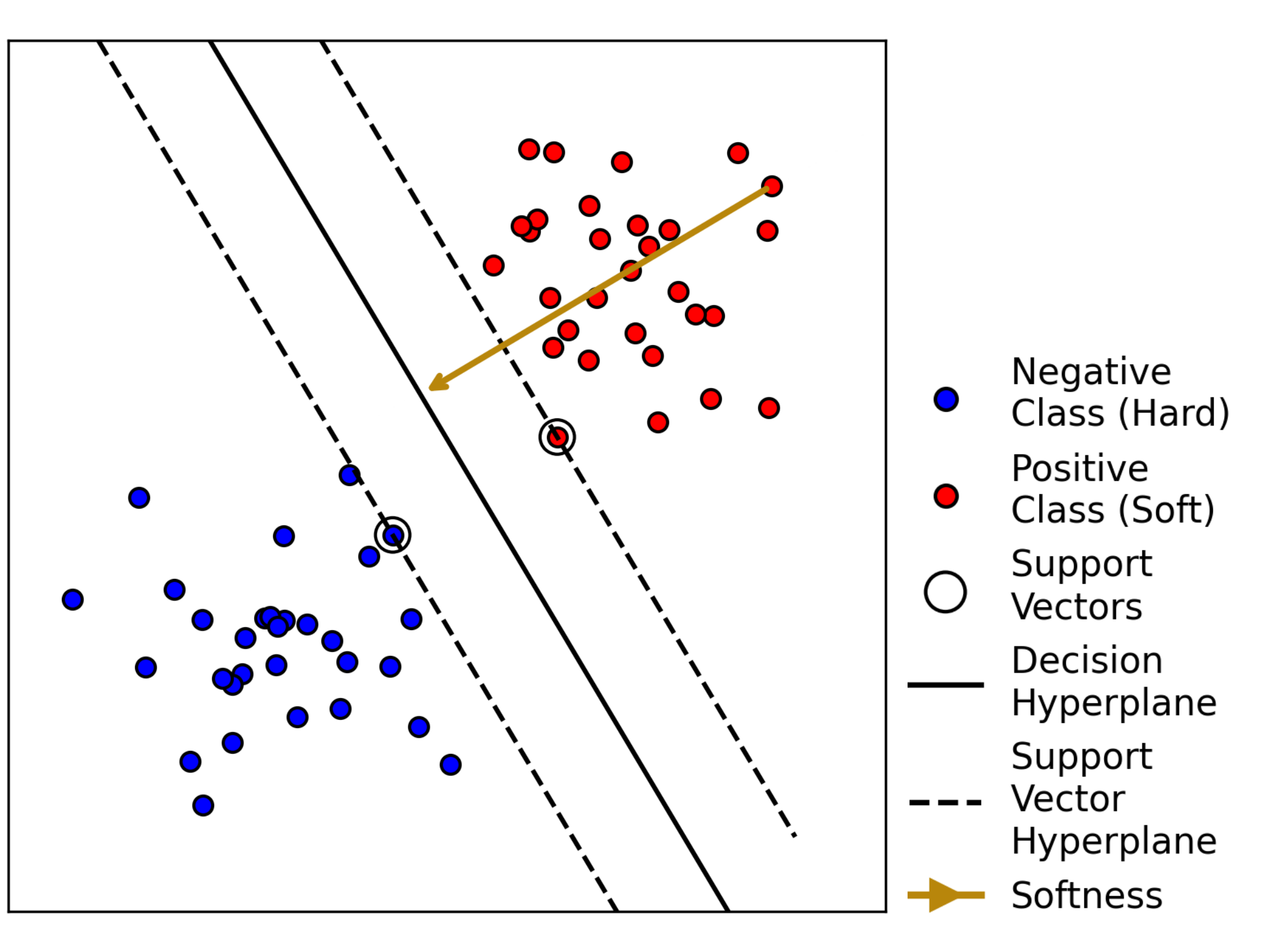}
\caption{A schematic diagram of a linear SVM, illustrating the decision hyperplane, support vectors, support vector hyperplanes, positive and negative samples, and the directed distance from a certain sample to the decision hyperplane. For convenience, this figure only illustrates the idealized two-dimensional case without loss of generality.}
\label{fig:svm}
\end{figure}

As shown in Fig.~\ref{fig:svm}, when explicitly choosing a linear SVM, the training objective is to determine an optimal decision hyperplane that maximizes the margin between the two classes. According to Eq.~\ref{eq:linear_svm}, the function of this decision hyperplane is:

\begin{equation}
    \bold{X} \cdot \bold{w} + u - \bold{y} = 0
\end{equation}

This hyperplane is determined by a subset of data points closest to it, known as support vectors, which are marked with circles on the dashed lines in Fig.\ref{fig:svm}. The support vectors not only define the placement of the decision boundary (the solid line in Fig.\ref{fig:svm}) but also establish two support vector hyperplanes (the symmetric dashed lines in Fig.~\ref{fig:svm}). These hyperplanes are parallel to the decision boundary, forming the margins between the two classes and ensuring that all other data points remain outside this margin.

Furthermore, the directed distance from a sample point to the decision hyperplane, which forms a scalar field, can be used to quantify the confidence of the sample belonging to a particular class, as illustrated by the solid yellow line with an arrow in Fig.~\ref{fig:svm}.

In Sec.~\ref{subsec:Identifying_Rearrangements}, we introduced a binary classification system for particles, using Eq.\ref{equ:p_hop} to distinguish between soft particles, which are prone to rearrangement, and hard particles, which are more resistant to it. When the sample consists of particles, the directed distance of a corresponding point from the classification hyperplane represents the confidence level in the classification of a particle as ``soft" or ``hard". A positive value indicates a soft particle, while a negative value corresponds to a hard particle. The magnitude of this value quantifies the tendency of a particle to undergo rearrangement.

Following the method by Cubuk~\textit{et al.}~\cite{cubuk2015identifying,schoenholz2016structural}, we define the directed distance from the corresponding point of the $p$-th particle to the decision hyperplane at time $t$ as the \textit{Local Softness} of the particle at time $t$:

\begin{equation}
    \mathcal{s}_{(p)}(t) = \frac{f(\bold{X}_{(p)}(t))}{||\bold{w}||_2} = \frac{\bold{X}_{(p)}(t) \cdot \bold{w} + u}{||\bold{w}||_2} 
    \label{label:softness}
\end{equation}

In essence, this approach represents the evolution of the tendency for particle rearrangement as a temporal scalar field~\cite{schoenholz2016structural}. 

For a system with $\mathscr{N}$ particles, the overall fluidity trend of the system can be obtained by summing the Local Softness of all particles and taking the average. We define this averaged scalar field as \textit{Global Softness}, which quantifies the overall fluidity trend of the system at time step $t$:

\begin{equation}
    \mathcal{S}(t) = \frac{1}{\mathscr{N}} \sum_{p=1}^{\mathscr{N}} \frac{\bold{X}_{(p)}(t) \cdot \bold{w} + u}{||\bold{w}||_2}
    \label{label:softness}
\end{equation}

Note that Global Softness is not our original method but an application of the SVM decision hyperplane.

\subsubsection{Shapley Values}
\label{subsec:shapley_values}

To determine which of the homology classes, $H_0$ (connected components), $H_1$ (cycles), or $H_2$ (cavities), plays the dominant role in the classification, we employed Shapley Values~\cite{shapley1953value,winter2002shapley} to quantify the contribution of each homology class.

In detail, we used 7 combinations of $H_0$, $H_1$, and $H_2$ (i.e., $H_0$, $H_1$, $H_2$, $H_0H_1$, $H_0H_2$, $H_1H_2$, $H_0H_1H_2$) to generate persistence images, flattened them into vectors, and trained models to obtain accuracies. For comparison, we applied PCA to retain 98\% of the variance and used an SVM with an RBF kernel. Then, we calculate the marginal contribution of each homology class in various combinations as follow:

\begin{equation*}
    \phi_i = \sum_{S \subseteq C \setminus \{i\}} \frac{|S|!(|C|-|S|-1)!}{|C|!} (v(S \cup \{i\}) - v(S))
\end{equation*}

\noindent where $C$ represents all homology classes (${H_0, H_1, H_2}$), $S$ is a subset of $C$ excluding $i$, $v(S)$ is the classification accuracy of subset $S$, and $\phi_i$ is the Shapley value for homology class $i$. The Shapley values are normalized to sum to $1$ for easy comparison.

\subsection{Descriptive Statistics on PH (non-ML approach)}

\label{subsec:def_descr_si_t}

We define a mapping from the results of PH on point cloud data to a real number in the range $[0,+\infty)$, referred to as the Separation Index (SI).

\textit{Separation Index.} For a temporal point cloud $P$ with $N$ points evolving over time step $t$, we obtain its PD at time step $t$, i.e., $\{(b_{H_i}^j(t), d_{H_i}^j(t))\}$, where $i \in \{0,1,2\}$, $j \in [1,\beta^{\text{max}}_i] \cap \mathbb{N}$, and $\beta^{\text{max}}_i$ is the maximum Betti number for the homology class $H_i$ over the range of $\epsilon$. The PD is generated by plotting the birth-persistence pairs $(b_{H_i}^j(t), d_{H_i}^j(t))$ corresponding to each homology class in a 2D Cartesian coordinate system. We define a new non-parametric metric, the Separation Index (SI), based on descriptive statistics to quantify the clarity of the boundary between clusters of $H_1$ and $H_2$ on the PD:

\begin{equation}
    \text{SI}(t) = \frac{\left|\text{mean}\left(\{d_{j_1}^{H_1}(t)\}_{j_1=1}^{\beta_1^{\max}}\right) - \text{mean}\left(\{d_{j_2}^{H_2}(t)\}_{j_2=1}^{\beta_2^{\max}}\right)\right|}{\text{std}\left(\{d_{j_1}^{H_1}(t)\}_{j_1=1}^{\beta_1^{\max}}\right) + \text{std}\left(\{d_{j_2}^{H_2}(t)\}_{j_2=1}^{\beta_2^{\max}}\right)}
    \label{label:si}
\end{equation}

The process of increasing $\epsilon$ from 0 can be interpreted as the emergence and disappearance of topological features. As defined in Eq.~\ref{label:si}, the subscript $j$ denotes the index of each topological feature within the homology group, corresponding to each birth-persistence pair in the barcode data. The Betti number $\beta_i^{\max}$ indicates the total number of topological features in $H_i$, meaning the total number of birth-persistence pairs in dimension $i$. The time step $t$ captures the evolution of the system at different time points. Ultimately, SI quantifies the separation between the point clusters corresponding to different homology classes in the PD by calculating the sum of mean differences of the $(b,d)$ pairs in $H_1$ and $H_2$ to the sum of their standard deviations.

\begin{figure}[htbp]
\includegraphics[width=0.48\textwidth]{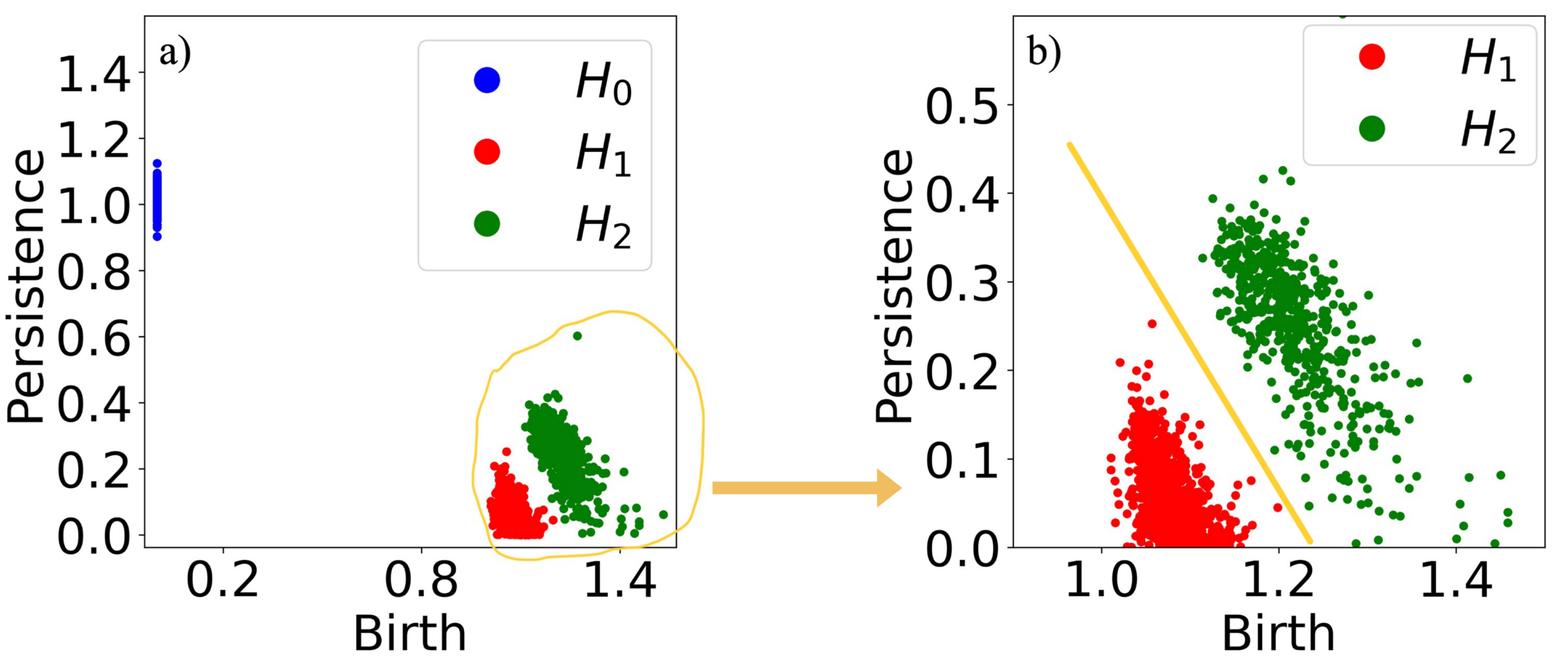}
\caption{a) is the persistence diagram (PD) for a crystal ($\text{Traj}(T^{(1)}_5)$ at time step $t=900$), of which the homology classes of $H_0$ are distributed along a line near the y-axis, while the $H_1$ and $H_2$ classes are located within the yellow circle and magnified in b). The yellow solid line in b) divides the point sets of $H_1$ and $H_2$ into two disjoint parts. Inspired by this, the Separation Index (SI) is defined to measure the clarity of the boundary of the point sets for $H_1$ and $H_2$.}
\label{fig:visual_si}
\end{figure}

As shown in Fig.~\ref{fig:visual_si}, SI provides an intuitive visual interpretation. Specifically, the PD of a homogeneous Lennard-Jones (LJ) system, generated by Protocol 1, as defined in Sec.~\ref{sec:md_protocols}, typically displays a characteristic pattern: $H_0$ persistence points align along a vertical line on the left, while $H_1$ and $H_2$ persistence points form two distinct but adjacent clusters in the lower-right corner. The value of SI quantifies the clarity of the boundary between these $H_1$ and $H_2$ clusters, providing a measure of particle ordering within the analyzed region. The figure illustrates the distribution of these clusters and their boundary, with the SI calculated from the mean difference and combined standard deviation of the clusters. 

Notably, SI is a purely mathematical and non-ML metric derived from the descriptive statistics of persistent homology, providing a complementary perspective to traditional physical methods. Besides, SI is applicable to both local and global characterizations, as the objects they describe are essentially point clouds.

\color{black}

\section{Dataset Generation}

\label{sec:data_set_generation}

In our study, molecular dynamics (MD) simulations~\cite{alder1957phase,rahman1964correlations} serve as the core technique. To begin with, we utilized the MD simulations to generate multiple trajectories. Then, we sampled these trajectories to obtain system-level samples (a single frame from a trajectory) and particle-level samples (the local environment of an individual particle within a single frame of the trajectory). Finally, using these samples, we constructed different datasets for both ML and non-ML tasks.

We provide a comprehensive summary in Table~\ref{tab:protocols}, detailing the molecular dynamics (MD) simulation protocols, system categories, and simulation dynamics. Additionally, we outline potential structural transitions observed in the trajectories, classification of groups, extracted datasets, corresponding state sets, their respective tasks, and the figures generated from these datasets for reference.

\begin{table*}[t!]
    \centering
    \renewcommand{\arraystretch}{1.3}
    \setlength{\tabcolsep}{4pt}
    \begin{tabular}{|>{\centering\arraybackslash}m{2.45cm}|
                    >{\centering\arraybackslash}m{2.8cm}|
                    >{\centering\arraybackslash}m{2.8cm}|
                    >{\centering\arraybackslash}m{2.8cm}|
                    >{\centering\arraybackslash}m{2.8cm}|
                    >{\centering\arraybackslash}m{2.45cm}|}
        \hline
        \textbf{MD Protocol} & \multicolumn{4}{c|}{\textbf{Protocol 1}} & \textbf{Protocol 2} \\
        \hline
        \textbf{Group} & \multicolumn{2}{c|}{Group 1} & \multicolumn{2}{c|}{Group 2} & Group 3 \\
        \hline
        \textbf{Set} & Set 1 & Set 3 & Set 2 & Set 4 & Not applicable \\
        \hline
        \textbf{Object} & Particle & System & Particle & System & Not applicable \\
        \hline
        \textbf{Feature matrix} & $\bold{X}_1$ & $\bold{X}_3$ & $\bold{X}_2$ & $\bold{X}_4$ & Not applicable \\
        \hline
        \textbf{Scale} & Local & Global & Local & Global & Not applicable \\
        \hline
        \textbf{Property and/or status} & Rearrangement Tendency & Global Phase Structure & Rearrangement Tendency & Global Phase Structure & Not applicable \\
        \hline
        \textbf{State set} & Soft or hard & Crystal, amorphous or liquid & Soft or hard & Crystal, amorphous or liquid & Amorphous or liquid \\
        \hline
        \multirow{4}{*}{\textbf{ML task}} & \multicolumn{2}{c|}{Training} & \multicolumn{2}{c|}{Testing} & Not applicable \\
        \cline{2-6}
        & Task 1 & Task 2 & Task 1 & Task 2 & Not applicable \\
        \cline{2-6}
        & \multirow{2}{*}{Not applicable} & \multirow{2}{*}{Not applicable} & \multirow{2}{*}{Global Softness} & \multirow{2}{*}{Not applicable} & \multirow{2}{*}{Global Softness} \\
        &  &  &  &  &  \\

        &  &  & Fig.~\ref{fig:four_rep_group_2}a) and~\ref{fig:four_rep_group_2}b) &  & Fig.~\ref{fig:four_rep_group_2}c) and~\ref{fig:four_rep_group_2}d) \\
        \hline
        \multirow{2}{*}{\textbf{Non-ML task}} & Separation Index & Separation Index & Separation Index & Separation Index & Separation Index \\
        & Fig.~\ref{fig:separation_index_all}a) and~\ref{fig:betti_numbers}a) & Fig.~\ref{fig:separation_index_all}b) and~\ref{fig:betti_numbers}b) & Fig.~\ref{fig:separation_index_all}a) and~\ref{fig:betti_numbers}a) & Fig.~\ref{fig:separation_index_all}b),~\ref{fig:betti_numbers}b),~\ref{fig:four_rep_group_2}a) and~\ref{fig:four_rep_group_2}b) & Fig.~\ref{fig:four_rep_group_2}c) and~\ref{fig:four_rep_group_2}d) \\
        \hline
        \textbf{System} & \multicolumn{4}{c|}{LJ system} & KA system \\
        \hline
        \textbf{Dynamics} & \multicolumn{4}{c|}{Isothermal evolution} & Linear quenching \\
        \hline
        \textbf{Potential Structural Transition} & \multicolumn{4}{m{11.2cm}<{\centering}|}{(1) Crystallization from liquids, (2) Crystallization from amorphous solids, (3) Incomplete crystallization from liquids to amorphous solids, or (4) Remaining liquid phases} & (1) Vitrification, or (2) Remaining liquid phases \\
        \hline
    \end{tabular}
    \caption{Summary of MD simulation protocols, system categories, simulation dynamics, potential structural transitions, group classifications, extracted datasets, state sets, corresponding tasks, and the figures generated from these datasets. Note that ``Not applicable" means that the content has no substantive meaning or is not relevant.}
    \label{tab:protocols}
\end{table*}

\subsection{Protocols of MD Simulation}
\label{sec:md_protocols}

This study involves two MD simulation protocols: Protocol 1 focuses on the isothermal evolution of the LJ system, while Protocol 2 addresses the KA mixture during a linear quenching process. Both protocols utilized a particle count of $\mathscr{N}=864$ to balance the demands and computational efficiency. In the simulation, we use Lennard-Jones reduced units~\cite{hirschfelder1964molecular}, and denote each time step as $100 \times (0.2*(m*\sigma^2/\epsilon)^{1/2})$.

\textit{Protocol 1}. For LJ systems, the parameters of the Lennard-Jones potentials are: $\epsilon = 1$, $\sigma = 1$, cutoff radius $r_c = 3.5\sigma$ and mass $m = 1 M$. We have adopted a tail correction as the truncation scheme. The computational protocol we have followed begins with a linearly quench of the liquid from $T_{\text{init}} = 1.25$ to a given temperature $T_{\text{final}}$ in 20 steps. Then, we perform a $1000$ steps equilibration at temperature $T_{\text{final}}$. These simulations have been conducted within the NPT ensemble with an isotropic pressure $P=5.68 (\epsilon/{\sigma^3})$, enforced via a chain of five thermostats coupled to a Nosé–Hoover barostat~\cite{hoover1996kinetic} (which accounts for the Martyna-Tobias-Klein correction~\cite{martyna1994constant}). The damping parameter of the barostat is $0.5t^{*}$, where $t^{*} = 0.002 (m * \sigma^2 /\epsilon)^{1/2}$. For convenience, we denote the trajectory corresponding to its temperature $T$ as $\text{Traj}(T)$. 
All these trajectories were generated using the same initial positions and velocities for each particle. These settings are the same as those adopted in Ref.~\cite{wang2024graph}, and previously work shown~\cite{blow2021seven} these settings allow us to observe crystal nucleation within a timescale accessible via unbiased MD simulations.

\textit{Protocol 2}. For KA mixtures~\cite{kob1995testing,pedersen2018phase}, we use the parameter setting: $\epsilon_{AA} = \sigma_{AA} = 1$, $\epsilon_{AB} = 1.5$, $\sigma_{AB} = 0.8$, $\epsilon_{BB} = 0.5$, $\sigma_{BB} = 0.88$. We started from $T_{\text{init}} = 1.25$ and quenched the system to the final temperature $T_{\text{final}}$. The details of these MD simulations are identical to those utilised for the homogeneous LJ system including the system size - 864 particles in total, with the proportion of B particles as $\chi_B = 0.2$. We generated two trajectories within the time step of $1000$, where one time step equals to $100 \times (0.2*(m*\sigma^2/\epsilon)^{1/2})$ in LJ reduced units.

\subsection{Group Division}
\label{subsec:data_group_division}

We generated trajectories for three groups according to the protocols provided in Sec.~\ref{sec:md_protocols}. Specifically, Group 1 and Group 2 used Protocol 1, while two trajectories of Group 3 followed Protocol 2.

Group 1 and 2 considered 20 corresponding $T_{\text{final}}$ values, denoted as $T_1,T_2,\dots,T_{20}$, which form a monotonically increasing sequence with equal intervals (see Table~\ref{tab:temperature_list} in Appendix~\ref{app:extra_data}). The trajectory corresponding to $T_j$ is denoted as $\text{Traj}(T_j)$, where $j\in[1,20]\cap\mathbb{N}$. Group $1$ includes temperatures with odd indices, i.e, the trajectories corresponding to $T^{(1)}_i=T_{2i-1}$; conversely, Group $2$ includes temperatures with even indices, i.e, the trajectories corresponding to $T^{(2)}_i=T_{2i}$, where $i \in [1,10] \cap \mathbb{N}$. Group 3 considered two trajectories, with their corresponding final $T_{\text{final}}$ values being: $T_{\text{final}} = 0.5$ and $T_{\text{final}} = 1.0$. We denote these two trajectories as $\text{Traj}_{\text{KA}}^{(1)}$ and $\text{Traj}_{\text{KA}}^{(2)}$, respectively.

In ML tasks, Group 1 was used for training, and Group 2 was used for testing. In non-ML tasks, Groups 1 and 2 were jointly used for descriptive statistics of barcodes. Specifically, the data presented in Table~\ref{tab:ml_result}, Table~\ref{tab:importance}, Fig.~\ref{fig:separation_index_all}, and Fig.~\ref{fig:betti_numbers} were derived from Groups 1 and 2, while Fig.~\ref{fig:four_rep_group_2}a), Fig.~\ref{fig:four_rep_group_2}b), and Fig.~\ref{fig:group_2_faded} were generated by testing the model trained on data from Group 1 with Group 2. The inclusion of Group 3 was solely due to the need for simulating the vitrification of supercooled liquids, which could not be captured from the isothermal evolution trajectories of Groups 1 and 2. Consequently, Fig.~\ref{fig:four_rep_group_2}c) and Fig.~\ref{fig:four_rep_group_2}d) were produced by testing the model trained on data from Group 1 with data from Group 3.

\subsection{Sample Labelling}
\label{subsec:labelling}

Sample labelling is the assignment of a state set to its corresponding sample. In our study, we focus on the influence of particle rearrangement tendencies on the global phase structure. Thus, the state set of individual particles is \{soft, hard\}, while the global system exhibits a state set of \{liquid, crystal, amorphous\}.

The rationale behind this approach lies in the effectiveness of selecting \{soft, hard\} as the particle state set (or target property) in capturing the fundamental microscopic characteristics of particles. Previous studies~\cite{cubuk2015identifying,schoenholz2016structural,ganapathi2021structure} have demonstrated that machine learning models trained on particle rearrangement properties (i.e., soft/hard) as target variables are highly effective in exploring structural rearrangements~\cite{cubuk2015identifying,schoenholz2016structural}, dense packing structures~\cite{cubuk2015identifying,schoenholz2016structural}, non-exponential relaxation~\cite{schoenholz2016structural}, glassy dynamics~\cite{schoenholz2016structural}, and crystallization~\cite{ganapathi2021structure}. Representing these intrinsic properties (i.e., soft/hard) as a scalar field offers a robust framework for tracking their temporal evolution and serves as a powerful tool for investigating collective particle behavior at the system level~\cite{cubuk2015identifying,schoenholz2016structural}. In contrast, defining particle states solely based on structural properties or symmetry, rather than employing the set of \{soft, hard\}, fails to accurately capture particle activity and, as a result, limits the ability to effectively study macroscopic phenomena~\cite{cubuk2015identifying,schoenholz2016structural}. Therefore, setting the particle state set as \{soft, hard\} and the global system state set as \{liquid, crystal, amorphous\} must be done simultaneously.

We use MD simulations to obtain the data required for our research. Naturally, the data is stored in the form of simulation trajectories. The smallest accessible unit in a trajectory is a frame, and within each frame, the smallest accessible unit is the position and velocity vectors of an individual particle in the system at the corresponding time of this frame.

Each trajectory is essentially a time series composed of frame-by-frame data, describing the changes in the coordinates of each particle in the system over time. Our study involves two types of samples: particle samples and system samples. A system sample refers to a specific frame in a trajectory, while a particle sample refers to a specific particle within a frame of a trajectory.

The particle samples are labelled according to Eq.~\ref{equ:p_hop}, which is used to characterize whether the particles will undergo rearrangement within a specific time interval. The labels for particle samples are then derived from these $p_{\text{hop},p}(t)$ values, with the label for particle $p$ in the system at time step $t$ in the trajectory $\text{Traj}(T)$ denoted as $y^{(T)}_{p,t} \in \{ \text{soft}, \text{hard} \}$.

The system samples are jointly labelled by MSD and $Q_6$. Firstly, MSD is used to distinguish between liquids and low-mobility solids. In liquids, the MSD shows a linear increase with a significant slope, indicating high particle mobility, whereas in crystals and amorphous solids, the MSD remains steady over time, reflecting particle stability. Note that the MSD of an amorphous material undergoing crystallization will be greater than 0. The difference between two types of solids, amorphous and crystalline, lies in whether long-range order (LRO) is established, and $Q_6$ is used as its estimator. Note that both rapid linear quenching of a liquid and incomplete crystallization caused by frustration can lead to the formation of an amorphous state. The former typically suppresses crystallization, resulting in $Q_6$ values close to 0, while the latter produces nonzero $Q_6$ values that are constant but still significantly smaller than those of crystals. Therefore, we set a threshold of $0.5$ to determine whether the system has crystallized. When $Q_6$ exceeds this value, the system is considered to be crystallized.

For system samples, we use MSD and $Q_6$ for phase labeling, denoting the label for trajectory $\text{Traj}(T)$ at time step $t$ as $y^{(T)}_{\text{entire},t} \in \{\text{liquid}, \text{crystal}, \text{amorphous}\}$. All data are listed in Table~\ref{tab:phase_labels} of Appendix~\ref{app:extra_data} for verification.

\subsection{Dataset Construction}
\label{subsec:dataset}

For convenience, we denote the point cloud corresponding to the frame in $\text{Traj}(T)$ at time step $t$ as $P(T,t)$. We randomly select 15,000 balanced positive and negative samples from $\{ ( \bold{I}^{(P(T_i^{(1)},t))}_{p}, y^{(T_i^{(1)})}_{p,t} ) \}$ to create Set 1, and 30,000 balanced samples from $\{ ( \bold{I}^{(P(T_i^{(2)},t))}_{p}, y^{(T_i^{(2)})}_{p,t} ) \}$ to create Set 2, where $t = 100 + 100 \times (k-1), k \in [1,9] \cap \mathbb{N}, i \in [1,10] \cap \mathbb{N}$, $p \in [1,N] \cap \mathbb{N}$, and $N = 864$ is the particle number of each system. Besides, we set $\{ ( \bold{I}^{(P(T^{(1)}_i,t))}_{\text{entire}}, y_{\text{entire},t}^{(T^{(1)}_i)} ) \}$ as Set 3 and $\{ ( \bold{I}^{(P(T^{(2)}_i,t))}_{\text{entire}}, y_{\text{entire},t}^{(T^{(2)}_i)} ) \}$ as Set 4, where $t = 100 + 25 \times (k-1), k \in [1, 33] \cap \mathbb{N}$, and $i \in [1,10] \cap \mathbb{N}$.

That is to say, Set 1 and Set 2 are particle-level sample sets with 15,000 and 30,000 balanced positive and negative samples, respectively. Set 3 and Set 4 are based on global system descriptors for classifying phases and structures. Set 1 and Set 3 are drawn from Group 1, while Set 2 and Set 4 are from Group 2.

\color{black}

\section{Experiments and Results}
\label{sec:results}

In this article, we conducted the following three computational experiments:

\textbf{1. Applying PH descriptors to interpretable classification tasks:} We demonstrated that PH-based descriptors are effective for interpretable classification tasks using SVMs. Moreover, we showed that a single variable is insufficient for soft-hard classification of particles but can successfully achieve the three-phase classification of the system. (Sec.~\ref{subsec:machine_learning}).

\textbf{2. Analyzing the Separation Index (SI):} We computed the SI values and observed that as the neighborhood radius increases, the SI distribution of soft and hard particles gradually approaches that of liquid and solid systems. Moreover, the SI of global systems can serve as an approximate surrogate for explicitly mapping a single variable to global phase structures. (Sec.~\ref{subsec:results_non_ml_appr}).

\textbf{3. Tracking the temporal evolution of four metrics:} We analyzed the variations of SI and global softness along trajectories and compared them with $Q_6$ and MSD (Sec.~\ref{subsec:metrics_on_trajs}).

\subsection{Machine Learning (ML)}
\label{subsec:machine_learning}

\subsubsection{Task Description}
\label{subsec:ml_task}

This article aims to establish a relationship between local structure and global properties within a unified mathematical framework. As a case study, we examine the relationship between particle mobility trends and the global phase structure of the system to evaluate the effectiveness of two types of descriptors.

The motivation is clear: in liquids, most particles tend to exhibit high mobility, whereas in solids, the opposite is true.

Therefore, we conducted two classification tasks. Task 1 assesses the local descriptor $\bold{I}^{(P)}_{p}$ (Eq.~\ref{equ:PI_local}) to determine whether a particle has a tendency for rearrangement (soft or hard), while Task 2 examines the global descriptor $\bold{I}^{(P)}_{\text{entire}}$ (Eq.~\ref{equ:PI_global}) to classify phases throughout the entire system (liquid, crystal, or amorphous). Note that the relationship between data generation, dataset construction, and ML tasks is listed in Table~\ref{tab:protocols}.

The ML process follows the steps outlined below and is applicable to both Task 1 and Task 2:

\vspace{0.1cm}

1. We constructed four datasets, namely Set 1, Set 2, Set 3, and Set 4, according to the description in Sec.~\ref{subsec:dataset} (or the summary provided in Table~\ref{tab:protocols}). Each set corresponds to its respective feature matrix, denoted as $\bold{X}_1$, $\bold{X}_2$, $\bold{X}_3$, and $\bold{X}_4$.

\vspace{0.1cm}

2. We performed PCA on $\mathbf{X_1}$, $\mathbf{X}_2$, $\mathbf{X}_3$, and $\mathbf{X}_4$, ensuring that the explained variance ratio is $z$, and obtain the corresponding dimension-reduced matrices. Note that $\mathbf{X}_2$ should undergo the same PCA transformation as $\mathbf{X}_1$, while $\mathbf{X}_4$ should undergo the same PCA transformation as $\mathbf{X}_3$, as required in Sec.~\ref{subsec:pca}. For convenience, we retain the original notation for the matrices after PCA transformation.

\vspace{0.1cm}

3. We trained our model using SVM, with its decision function uniformly expressed as $\bold{y} = f(\bold{X})$, evaluating both linear~\cite{cortes1995support} and RBF kernels~\cite{boser1992training}. Here, $\bold{X}$ denotes the feature matrix corresponding to the training sets, i.e., $\mathbf{X}_1$ or $\mathbf{X}_3$, while $\bold{y}$ represents the label vector formed by concatenating $y^{(T)}_{p,t}$ (Task 1) or $y^{(T)}_{\text{entire},t}$ (Task 2) for each sample in the respective set. For Task 1, we trained the model on Set 1 and tested it on Set 2, while for Task 2, we trained on Set 3 and tested on Set 4.

Through this process, we not only derived the classification mappings for Task 1 and Task 2, $f: \mathbf{X} \to \mathbf{y}$, but also obtained the corresponding performance metrics, as reported in Table~\ref{tab:ml_result}.

\subsubsection{ML Results and Analysis}
\label{subsec:ml_results}

The ML results are presented in Table~\ref{tab:ml_result}. As described in Appendix~\ref{app:parameter_selection_pi}, the feature matrix $\bold{X}$ initially contains 1,600 columns, corresponding to a $\mathfrak{m} \times \mathfrak{n}$ grid with $\mathfrak{m} = \mathfrak{n} = 40$, resulting in 1,600 predictors. However, applying PCA with an explained variance ratio of $z = 98\%$ reduces the number of predictors to 188 for Task 1 and 122 for Task 2. Regardless of the kernel function used, Task 1 consistently achieves an accuracy above 86\%, while Task 2 exceeds 95\%. Overall, the linear kernel performs slightly better than the RBF kernel, though the difference remains marginal.

\begin{table}[ht!]
\caption{
The Performance of Models with Persistent Homology in Two ML Tasks.}
\begin{ruledtabular}
\begin{tabular}{>{\bfseries}l >{\bfseries}l cc cc}
\multicolumn{2}{c}{\textrm{}} & \multicolumn{2}{c}{\textrm{Task 1 (Local)}} & \multicolumn{2}{c}{\textrm{Task 2 (Global)}} \\
\cline{3-4} \cline{5-6}
\textrm{Models} & \textrm{Metrics} & \multicolumn{2}{c}{\textrm{Variance}} & \multicolumn{2}{c}{\textrm{Variance}} \\
 &  & \textrm{98\%} & \textrm{50\%} & \textrm{98\%} & \textrm{45\%} \\
\hline
\multirow{2}{*}{\textrm{SVM (Linear)}} & \textrm{Accuracy} & 86.1\% & 84.2\% & 96.1\% & 94.2\% \\
 & \textrm{MCC} & 0.723 & 0.686 & 0.763 & 0.763 \\
\multirow{2}{*}{\textrm{SVM (RBF)}} & \textrm{Accuracy} & 86.1\% & 82.4\% & 95.5\% & 91.5\% \\
 & \textrm{MCC} & 0.723 & 0.649 & 0.760 & 0.715 \\
\multicolumn{2}{c}{\textrm{Number of Predictors}} & 188 & 7 & 122 & 1 \\
\end{tabular}
\end{ruledtabular}
\label{tab:ml_result}
\end{table}

As shown in Table~\ref{tab:ml_result}, further reducing the explained variance ratio $z$ reveals that at least 7 predictors are required to prevent a significant decline in the accuracy of Task 1. Additionally, we evaluated scenarios with the number of predictors ranging from 1 to 7, with detailed results provided in Table~\ref{tab:pca_task_1_to_1} of Appendix~\ref{app:extra_data}. Our analysis indicates that at least 3 predictors are necessary for the accuracy of Task 1 to be significantly higher than that of random selection, confirming that a single predictor is insufficient. This suggests that mapping particles to a binary classification (soft or hard) using a single variable is not feasible.

In contrast, Task 2 can be effectively performed using a single variable without a significant loss in accuracy. This suggests that mapping the entire system to a three-phase classification (crystal, amorphous, or liquid) based on a single variable is feasible. However, deriving such a mapping analytically remains still challenging.

\subsubsection{Feature Importance Analysis (FIA)}
\label{susbec:key_factor_task_2}

Here, we address two key questions: first, the significance of homology classes of different dimensions in the two classification tasks; and second, in the three-phase classification task of Task 2, which column in the feature matrix plays a decisive role—specifically, which variable serves as the primary determining factor.

Firstly, we use the Shapley values introduced in Sec.~\ref{subsec:shapley_values} to evaluate the importance of different homology classes, specifically $H_0$, $H_1$ and $H_2$, in two classification tasks. The results listed in Table~\ref{tab:importance} indicate that the significance of $H_0$ is slightly lower than that of $H_1$ and $H_2$, with $H_1$ and $H_2$ together contributing over 70\% of the total importance in both tasks.

\begin{table}[h]
\caption{The Contribution of Each Homology Class to the Classification in Task 1 and 2.}
\begin{ruledtabular}
\begin{tabular}{>{\bfseries}l ddd}
\multicolumn{1}{c}{} &
\multicolumn{1}{c}{\textrm{$\phi_0$ ($H_0$)}} &
\multicolumn{1}{c}{\textrm{$\phi_1$ ($H_1$)}} &
\multicolumn{1}{c}{\textrm{$\phi_2$ ($H_2$)}} \\
\hline
\textrm{Task 1 (Local)} & 28.08\% & 35.29\% & 36.63\% \\
\textrm{Task 2 (Global)} & 30.02\% & 35.05\% & 34.93\% \\
\end{tabular}
\end{ruledtabular}
\label{tab:importance}
\end{table}

Secondly,  in Task 2, a single predictor is sufficient to achieve an accuracy of 94.2\%, suggesting that the three-phase classification of a multi-particle system is largely dictated by this single critical predictor. We conclude that this variable corresponds to the 1389th element of the vector $\bold{I}_{\text{entire}}^{(P)}$, with the reasoning process detailed in Appendix~\ref{app:details_fia}. This single variable, in conjunction with the SVM mapping $f$, effectively characterizes the global phase structure.

\subsection{Descriptive Statistics on PH (non-ML approach)}
\label{subsec:results_non_ml_appr}

In Sec.~\ref{subsec:def_descr_si_t}, we introduced a mapping from the results of the PH analysis to a single real number in the range $[0, +\infty)$, specifically the Separation Index (SI), as defined in Eq.\ref{label:si}. Since both the neighborhood of a particle and the entire system can be represented as point clouds, the SI applies to both local and global structures, mapping each to a single real number.

\begin{figure}[htbp]
\includegraphics[width=0.5\textwidth]{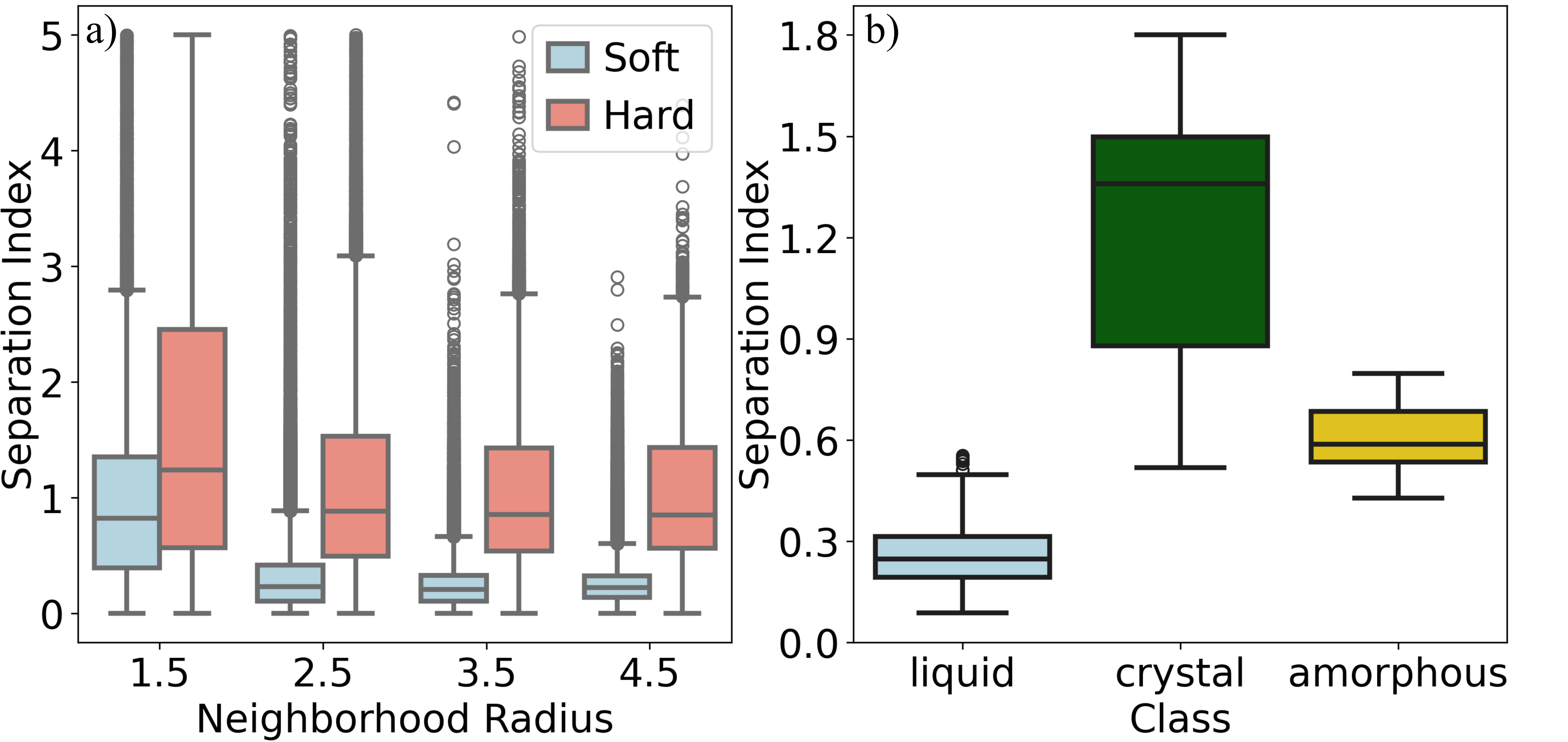}
\caption{Box plots of the Seperation Index (SI) of a) all neighborhoods with multiple radii of particle samples at time step $t=100+100 \times (k-1), k \in [1,10] \cap \mathbb{N}$, and b) all systems sampled at time step $t=100+25 \times (k-1), k \in [1,33] \cap \mathbb{N}$, from each trajectory in Group 1 and 2.}
\label{fig:separation_index_all}
\end{figure}

\begin{figure}[htbp]
\includegraphics[width=0.5\textwidth]{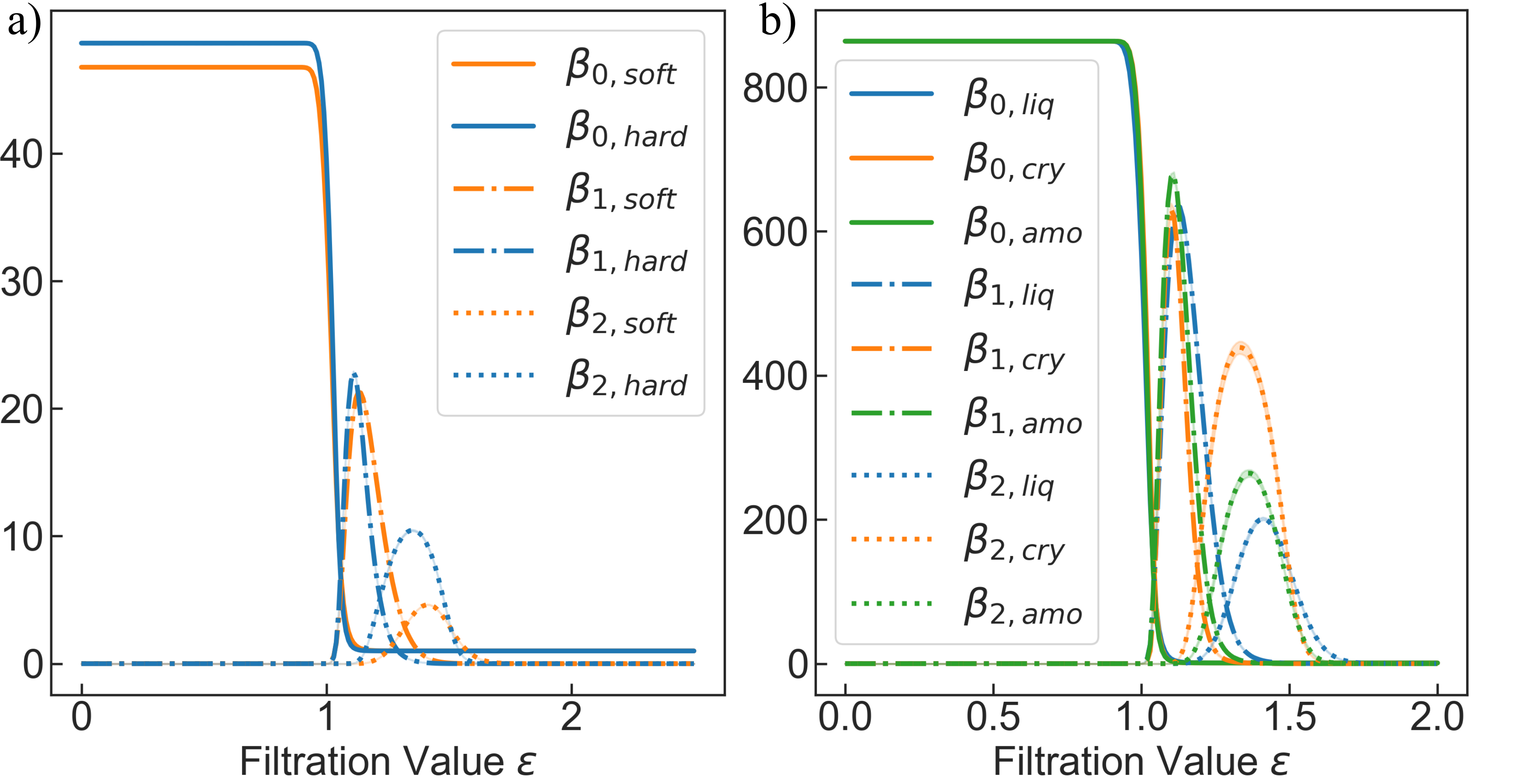}
\caption{The variation of average Betti numbers with $\epsilon$ for a) all neighborhoods with a radius $r=2.5$ of particle samples at time step $t=100+100 \times (k-1), k \in [1,10] \cap \mathbb{N}$, and b) all systems sampled at time step $t=100+25 \times (k-1), k \in [1,33] \cap \mathbb{N}$, from each trajectory in Group 1 and 2.}
\label{fig:betti_numbers}
\end{figure}

We analyzed the SI (Fig.\ref{fig:separation_index_all}) and Betti numbers (Fig.\ref{fig:betti_numbers}) of neighborhoods around two types of particles and systems in different phases across various times and trajectories in Groups 1 and 2.

The SI can be seen as an analytical approximation for the mapping trained by ML. For Task 2 (three-phase classification), while its accuracy is slightly lower, it can still achieve near-perfect classification between liquids and solids (Fig.~\ref{fig:separation_index_all}b)). Additionally, the overlapping regions between amorphous and crystalline phases lie outside the interquartile range (Fig.~\ref{fig:separation_index_all}b)), indicating that SI can, in most cases, achieve reasonably accurate three-phase classification. Although SI sacrifices some accuracy, it significantly enhances interpretability. For Task 1, although we have rigorously demonstrated that a single variable is insufficient for fully classifying particles on \{soft, hard\}, the interquartile ranges of the SI for soft and hard particles begin to separate as the neighborhood radius increases (Fig.\ref{fig:separation_index_all}a)). Simultaneously, their distribution patterns progressively resemble those of liquids and solids (compare Fig.\ref{fig:separation_index_all}a) to Fig.~\ref{fig:separation_index_all}b)).

Therefore, beyond providing an approximate analytical mapping for three-phase classification, SI plays a more crucial role as a bridge, enabling the analysis of the relationship between the soft-hard property of a particle and the global phase structure within a unified framework. This analysis is visualized in Fig.~\ref{fig:separation_index_all}.

Although hard particles tend to be located in symmetric environments, the tendency for rearrangement and the preservation of order are fundamentally distinct properties. Although symmetry is related to particle rigidity, other factors—such as local packing density and stress distribution—can also significantly influence rearrangement behavior. Therefore, $q_6$, which primarily measures six-fold symmetry, is not a reliable predictor of particle rearrangements, as demonstrated in Appendix~\ref{appendix:exchange_exp}.

Firstly, for the entire systems, as depicted in Fig.~\ref{fig:separation_index_all}b), crystals exhibit the highest SI values, followed by amorphous solids, with liquids having the lowest. The distinct and non-overlapping interquartile ranges (IQRs) of SI for these three phases allow for effective differentiation in most cases. Notably, the highest SI for liquids is below the lowest SI for crystals, ensuring perfect classification between the two. As shown in Fig.~\ref{fig:betti_numbers}b), the mean $\beta_2$ peak is distinct for each phase: crystals ($\beta_{2,\text{cry}}$) show the highest peak, amorphous phases ($\beta_{2,\text{amo}}$) slightly lower, and liquids ($\beta_{2,\text{liq}}$) the lowest.
This is expected because crystals, with their highly symmetric and periodic structures, exhibit long-range order (LRO), resulting in concentrated and distinct topological features, especially in $H_1$ (cycles) and $H_2$ (cavities). These factors contribute to the highest SI values. In contrast, amorphous solids, while lacking long-range order (LRO), exhibit some short-range order (SRO), resulting in SI and mean $\beta_2$ values lower than those of crystals but higher than those of liquids.

Specifically, the mean $\beta_2$ primarily reflects the number of two-dimensional cavities in the system. In crystals, the highly ordered particle arrangement forms well-defined, stable cavities, resulting in a concentrated peak in the mean $\beta_2$. In amorphous solids, although long-range periodicity is absent, localized short-range order (SRO) creates a more scattered cavity distribution, resulting in a broader $\beta_2$ peak. In the case of liquids, with even less structural coherence, the distribution of cavities is highly random, resulting in the lowest and most diffuse $\beta_2$ peak. These differences in topological features directly reflect the degree of structural organization in each phase: crystals exhibit the most ordered and interconnected structures, followed by amorphous solids, and then liquids.

\begin{figure}[htbp]
\includegraphics[width=0.48\textwidth]{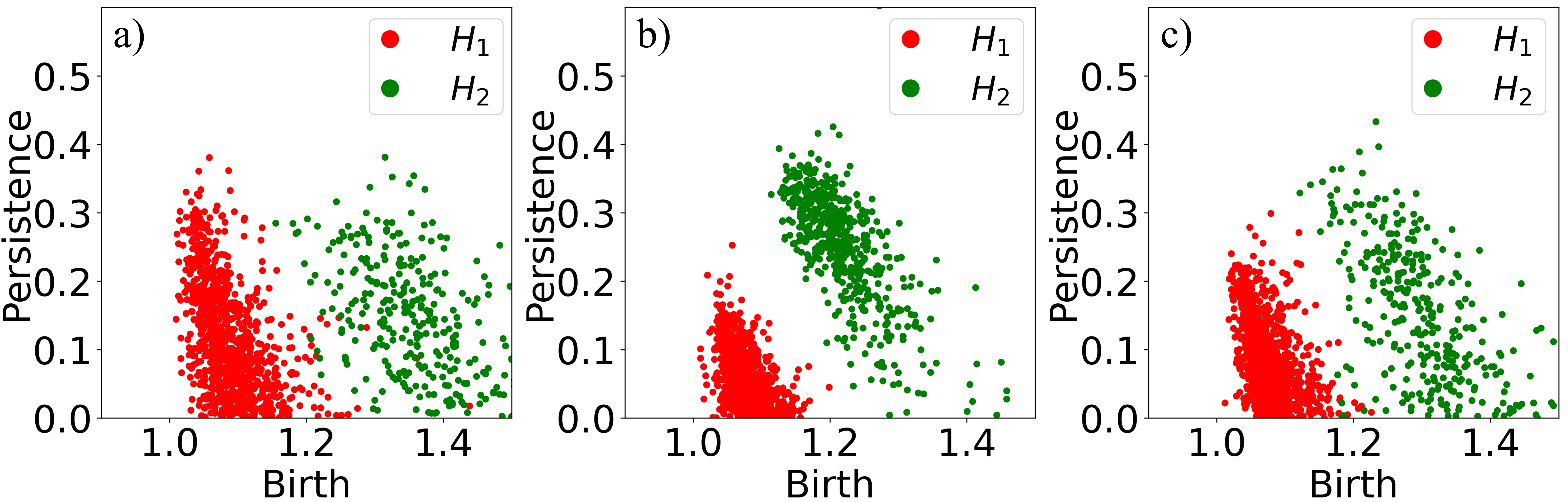}
\caption{The $H_1$ and $H_2$ points for the entire systems from selected trajectories as follow: a) for liquid ($\text{Traj}(T^{(2)}_9)$ at time step $t=900$); b) for crystal ($\text{Traj}(T^{(1)}_5)$ at time step $t=900$); c) for amorphous ($\text{Traj}(T^{(2)}_1)$ at time step $t=200$).}
\label{fig:si_global_phase}
\end{figure}

Fig.~\ref{fig:si_global_phase} illustrates the concept of SI by showing the clarity of boundaries in the clusters of $H_1$ and $H_2$ points across different phases. Clearer cluster boundaries correspond to higher SI values, indicating a more ordered structure. Note that the degree of order here is relative. In the liquids (Fig.~\ref{fig:si_global_phase}a)), the clusters of $H_1$ and $H_2$ points are diffuse with indistinct boundaries, resulting in the lowest SI value. The amorphous solids (Fig.~\ref{fig:si_global_phase}c)) shows the boundaries of moderate clarity, yielding SI values between those of the liquids and crystals. In contrast, the crystals (Fig.~\ref{fig:si_global_phase}b)) has the most clearly defined boundaries, corresponding to the highest SI value.

Secondly, for particle environments, as illustrated in Fig.~\ref{fig:separation_index_all}a), hard particles exhibit higher SI values than soft particles in their neighborhoods. As the neighborhood radius $r$ increases, the IQRs of the neighborhoods of hard and soft particles diverge, and the number of outliers decreases. This suggests that small-scale structures progressively merge as $r$ increases, leading to a more coherent understanding of the global structure. The neighborhoods of hard particles tend to exhibit SI distributions typical of solids, whereas those of soft particles show more liquid-like characteristics, as seen in Fig.~\ref{fig:separation_index_all}b). Furthermore, as shown in Fig.~\ref{fig:betti_numbers}a), the mean $\beta_2$ of hard-particle neighborhoods peaks higher than that of soft-particle neighborhoods. When $\epsilon \in [0,1]$, the mean $\beta_0$ for hard-particle neighborhoods is greater, indicating that these neighborhoods contain more particles. This is expected because hard particles generally reside in environments with strong symmetry and SRO, where particles are densely packed. This symmetry and dense packing enhance resistance to rearrangement, giving hard particles solid-like properties. In contrast, soft particles are found in less symmetric or ordered environments, exhibiting more liquid-like behavior.

Physically, these topological differences can be understood by considering the role of cavities (mean $\beta_2$) and connected components (mean $\beta_0$). The higher mean $\beta_2$ peaks in hard-particle neighborhoods reflect well-defined cavities resulting from tight, symmetric packing, indicative of solid-like stability. Besides, higher mean $\beta_0$ values in hard-particle neighborhoods also indicate greater local connectivity, reflecting a higher number of neighboring particles. In contrast, soft-particle neighborhoods, with more dispersed and irregular cavity structures, indicate environments where particles are less tightly bound, making these neighborhoods more liquid-like. The lower local connectivity in these environments, indicated by mean $\beta_0$, further emphasizes their liquid-like mobility.

\begin{figure}[htbp]
\includegraphics[width=0.4\textwidth]{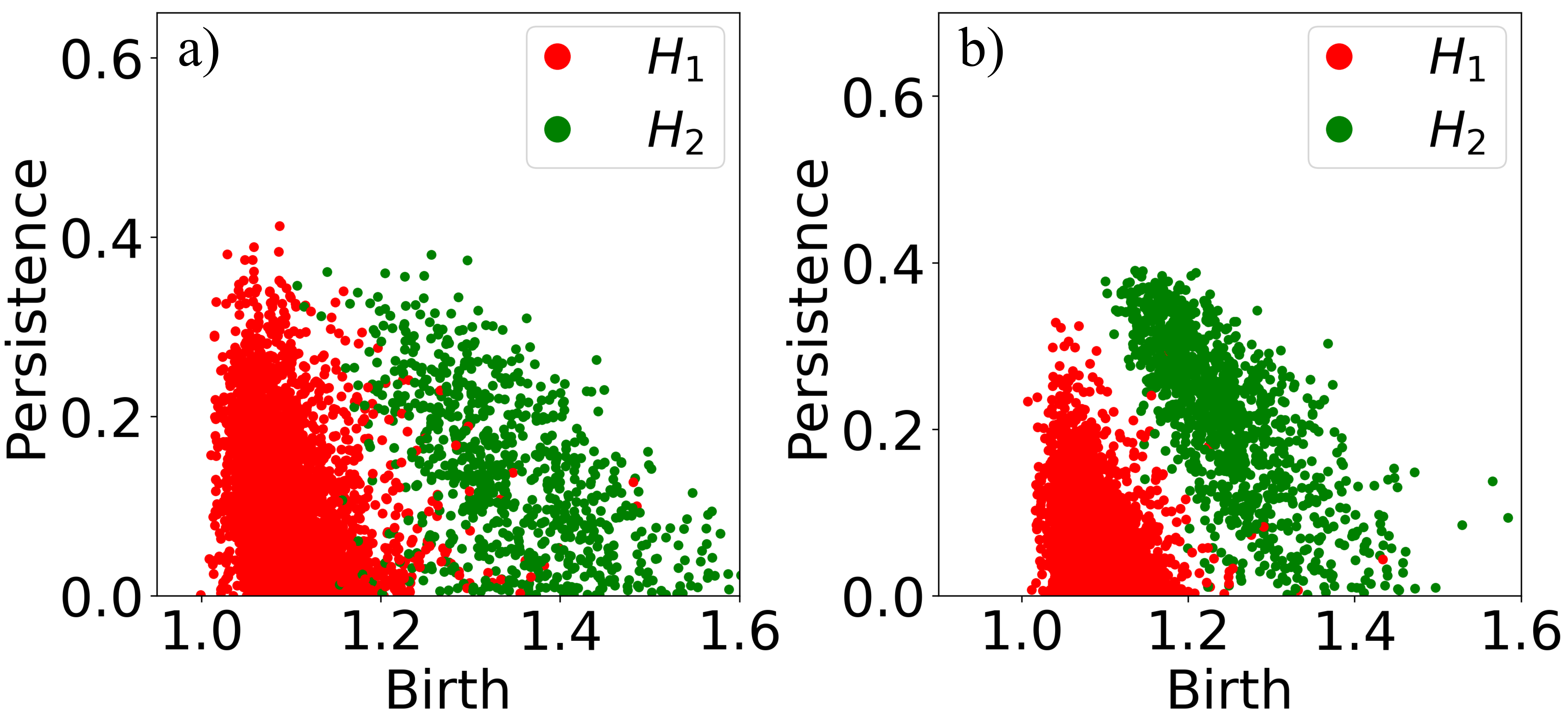}
\caption{The $H_1$ and $H_2$ points of neighborhoods with a radius $r=2.5$, after randomly sampling 100 soft particles and 100 hard particles from from each trajectory in Group 1 and 2 at time step $t=100+100 \times (k-1), k \in [1,10] \cap \mathbb{N}$ for a) soft particles, and b) hard particles.}
\label{fig:si_particle_environment}
\end{figure}

Fig.~\ref{fig:si_particle_environment} shows the $H_1$ and $H_2$ points in the neighborhoods of soft and hard particles. For soft particles (Fig.~\ref{fig:si_particle_environment}a)), the $H_1$ and $H_2$ clusters nearly merge, yielding lower SI values, which indicate low symmetry and disorder. In contrast, hard particles (Fig.~\ref{fig:si_particle_environment}b)) exhibit relatively clear cluster boundaries, resulting in higher SI values that reflect a more symmetric and ordered environment. In comparison with Fig.~\ref{fig:si_global_phase}, the neighborhoods of soft particles resemble liquid characteristics, while those of hard particles are more solid-like.

In crystalline environments, the local symmetry and SRO around hard particles lead to higher $\beta_2$, as stable cavities form in their surroundings. With a small radius $r$, the neighborhood around a hard particle may exhibit properties similar to those in amorphous materials, with evident SRO and local symmetry. However, as the neighborhood radius increases, periodicity, global symmetry, and LRO begin to emerge, transitioning the topological features from local to global. This shift results in increased $\beta_2$ and SI values as more extensive, higher-dimensional structures develop, reflecting the mechanism through which LRO and global symmetry are established. In contrast, soft particles, especially those in liquids, due to their higher fluidity and lack of such ordered structures, exhibit lower $\beta_2$ and SI values, indicative of the randomness and instability of the topological features in liquid-like environments.

\subsection{Metrics on Trajectories}
\label{subsec:metrics_on_trajs}

Tracking the evolution of various metrics along molecular dynamics trajectories is a widely used approach for studying how target properties evolve over time under controlled conditions. In this study, we leverage this method to identify the onset of phase transitions using our proposed metrics, such as SI and Global Softness, while also providing a comparison with traditional metrics like $Q_6$ and MSD.

\begin{figure}[htbp!]
\includegraphics[width=0.5\textwidth]{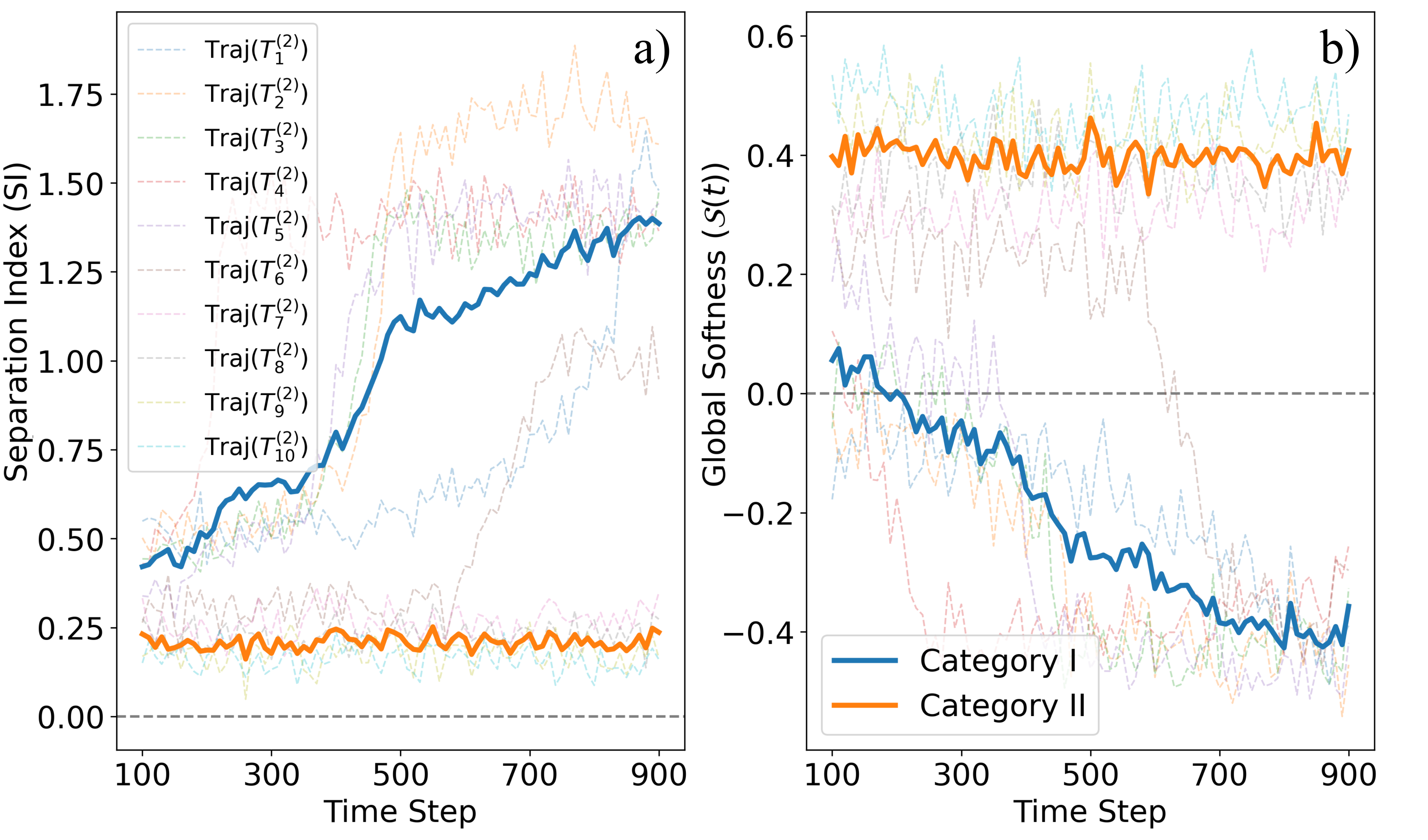}
\caption{The a) Separation Index $\text{SI}(t)$ and b) Global Softness $\mathcal{S}(t)$, where $t=100+10 \times (i-1)$, $i \in [1,81] \cap \mathbb{N}$, for trajectories in Group 2, denoted as $\text{Traj}(T^{(2)}_k), k \in [1,10] \cap \mathbb{N}$. These 10 trajectories are divided into two categories: those with $k \leq 6$ (Category I) crystallized, while those with $k \geq 7$ (Category II) remained liquid. The metrics over time for each trajectory are shown as faded dashed lines, with bold lines indicating the average in each category—blue for Category I and orange for Category II.}
\label{fig:group_2_faded}
\end{figure}

\begin{figure*}[htbp!]
\includegraphics[width=1\textwidth]{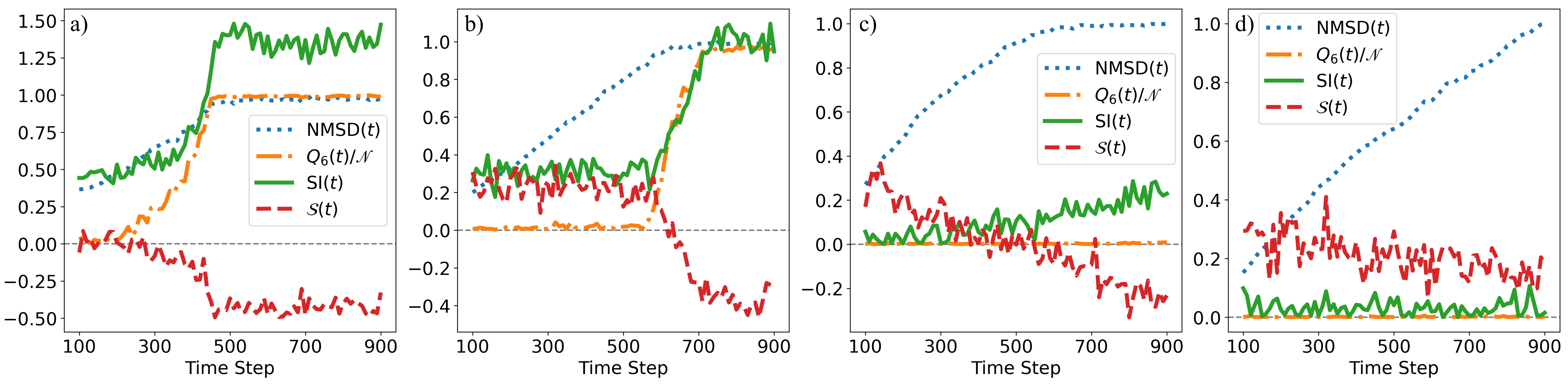}
\caption{The Normalized MSD (abbreviated as $\text{NMSD}(t)$, normalizing the MSD into $[0,1]$), $Q_6(t)/\mathscr{N}$, $\text{SI}(t)$, and $\mathcal{S}(t)$ where $t=100+10 \times (i-1)$, $i \in [1,81] \cap \mathbb{N}$ on the selected trajectories a) $\text{Traj}(T_3^{(2)})$ containing a whole transition from amorphous to crystal; b) $\text{Traj}(T_6^{(2)})$ containing a whole transition from liquid to crystal; c) $\text{Traj}_{\text{KA}}^{(1)}$ containing a process of vitrification from a supercooled liquid; d) $\text{Traj}_{\text{KA}}^{(2)}$ remaining liquid phase during quenching.}
\label{fig:four_rep_group_2}
\end{figure*}

We tracked the evolution in four metrics including normalized MSD, $Q_6(t)/\mathscr{N}$, $\text{SI}(t)$ and $\mathcal{S}(t)$ over time on ten trajectories in test group (Group 2), i.e., $\text{Traj}(T^{(2)}_k), k\in [1,10] \cap \mathbb{N}$, as well as two trajectories in Group 3, i.e.,  $\text{Traj}_{\text{KA}}^{(1)}$ and $\text{Traj}_{\text{KA}}^{(2)}$. The data for all metrics are provided in Fig.~\ref{fig:metrics_group_2} of Appendix~\ref{app:extra_data} for reference.

The trajectories corresponding to the test set of Group 2 are divided into two categories: Category I, with $k \leq 6$, undergoes complete crystallization from an amorphous or liquid state, while Category II, with $k \geq 7$, remains in the liquid phase. In Fig.~\ref{fig:group_2_faded}, the blue solid line shows the average of SI (Fig.~\ref{fig:group_2_faded}a)) and Global Softness (Fig.~\ref{fig:group_2_faded}b)) for each trajectory in Category I, while the orange solid line shows the same for Category II. In Category I, SI rises sharply as Global Softness decreases, whereas in Category II, SI stays low and stable, with Global Softness remaining at a high positive level. This distinction clearly differentiates crystallized states from non-crystallized ones. Notably, the Global Softness was calculated using a linear SVM explaining 98\% of the explained variance, trained in Task 1. In contrast, other three metrics are non-ML based.

In Fig.~\ref{fig:four_rep_group_2}, without loss of generality, we report the evolution of four metrics over time for four representative trajectories: (i) $\text{Traj}(T_3^{(2)})$, which undergoes a complete transition from amorphous to crystalline (Fig.~\ref{fig:four_rep_group_2}a)); (ii) $\text{Traj}(T_6^{(2)})$, which undergoes a complete crystallization from liquid (Fig.~\ref{fig:four_rep_group_2}b)); (iii) $\text{Traj}_{\text{KA}}^{(1)}$, depicting a glass transition from a liquid (Fig.~\ref{fig:four_rep_group_2}c)); and (iv) $\text{Traj}_{\text{KA}}^{(2)}$, which remains in the liquid phase during quenching (Fig.~\ref{fig:four_rep_group_2}d)).

The order parameter $Q_6$ captures the global ordering in the system, particularly six-fold symmetry, and is effective for detecting crystallization, especially from a higher-energy phase (liquid or amorphous) to a more ordered crystalline phase. In liquid or amorphous states, $Q_6$ values are typically lower, indicating minimal structural symmetry. However, as the system crystallizes, $Q_6$ rises sharply, signaling the emergence of LRO. As local particle order increases and aligns with the global structure, $Q_6$ gradually rises. A sharp increase or spike in $Q_6$ during this transition (Fig.~\ref{fig:four_rep_group_2}a) and Fig.~\ref{fig:four_rep_group_2}b)) clearly indicates the formation of LRO in crystals.

However, $Q_6$ lacks sufficient sensitivity to subtle variations in SRO and thus cannot effectively distinguish between liquid and amorphous solids (see Fig.~\ref{fig:four_rep_group_2}c)). Therefore, $Q_6$ needs to be combined with MSD to effectively differentiate these three phases (liquids, amorphous solids, and crystals). Particle mobility in the system is quantified by MSD, which measures the average displacement of particles over a specified time interval. In the liquids, particles move freely, and MSD increases rapidly in a linear manner, indicating high mobility. After crystallization, the MSD stabilizes, indicating that particles are restricted to fixed positions (see Fig.~\ref{fig:four_rep_group_2}b)). Notably, in the amorphous solids that are undergoing crystallization, the MSD also increases, but more gradually than in the liquids. This slower increase reflects the gradual ordering of local structures, as opposed to random diffusion (see Fig.~\ref{fig:four_rep_group_2}a)).

We propose that $\text{SI}(t)$ is highly effective for distinguishing among liquids, amorphous solids, and crystals. In Fig.~\ref{fig:four_rep_group_2}a), $\text{SI}(t)$ increases steadily during the crystallization of the amorphous phase, indicating a transition from a disordered to an ordered structure. Similarly, Fig.~\ref{fig:four_rep_group_2}b) shows a sharp rise in $\text{SI}(t)$ during liquid crystallization, indicating a shift of the system to a solid state. In Fig.~\ref{fig:four_rep_group_2}c), $\text{SI}(t)$ captures the increase in SRO as the system quenches from liquid to amorphous, indicated by the upward turn in the curve. In contrast, $Q_6$ fails to capture these subtle changes and remains close to zero. Finally, Fig.~\ref{fig:four_rep_group_2}d) shows that $\text{SI}(t)$ remains stable during quenching in the liquids, with minimal changes in SRO. Notably, $\text{SI}(t)$ demonstrates exceptional sensitivity in capturing transitions between two disordered phases (e.g., liquid and amorphous), as consistent with the results observed in Fig.~\ref{fig:separation_index_all}b). This sensitivity significantly surpasses that of traditional order parameters such as $Q_6$, which exhibit minimal or no ability to distinguish between disordered phases.

In addition, $\mathcal{S}(t)$ captures the fluidity trend of the system, offering insights into both its current state and potential future behavior. Note that, $\mathcal{S}(t)$ was trained on the training set (Group 1, LJ system) and applied into the test set (Group 2, LJ system) and Group 3 (KA system). When applied to Group 2, a negative $\mathcal{S}(t)$ indicates a trend towards solidification, while a positive value suggests the system is likely to remain fluid. By correlating with potential energy in the system, $\mathcal{S}(t)$ captures both current structural changes and the potential evolution of fluidity over time. As the system transitions from an amorphous phase (see Fig.~\ref{fig:four_rep_group_2}a)) or a liquid phase (see Fig.~\ref{fig:four_rep_group_2}b)) to a crystalline phase, a sharp drop in $\mathcal{S}(t)$  indicates a substantial reduction in fluidity and the stabilization of the crystalline phase. When applied to Group 3 (KA mixture, see Fig.~\ref{fig:four_rep_group_2}c) and Fig.~\ref{fig:four_rep_group_2}d)), $\mathcal{S}(t)$ required a correction to accurately reflect the positive and negative relationships (see Fig.~\ref{fig:correctness} of Appendix~\ref{app:correctness}), as the differences in atomic configurations caused systematic shifts in the absolute scale of fluidity. However, it only serves as a reference, and its shift does not impact the presentation of the underlying principles. Therefore, we ignore exploring the correction mechanism as it is not the key factor. If the purpose is solely to track the fluidity on a single trajectory, this correction to align the zero point is not necessary. On the contrary, as mentioned at the beginning of this paragraph, when applied to Group 2 (LJ system), $\mathcal{S}(t)$ required no such adjustment (see Fig.~\ref{fig:four_rep_group_2}b)), indicating that its positive and negative relationships can be directly used without any corrections.

Overall, our proposed SI and Global Softness not only effectively detect phase transitions but also reflect the harmonious integration of ML and non-ML approaches within our PH-based framework.

\section{Conclusion}
\label{sec:conclusion}

We employed persistent homology (PH) to develop a unified mathematical framework for both local and global characterization in disordered systems. This approach produces high-performance descriptors for interpretable machine learning (ML) and offers deep insights into the structure-function relationships in these systems. It reveals crucial links between local particle environments and global structures, predicting particle rearrangement and classifying phases in a universal framework. In three-phase classification, the SVM trained using ML methods can achieve near-perfect accuracy. In contrast, our proposed Separation Index (SI) sacrifices some accuracy to enhance the interpretability of the three-phase classification model. Our proposed model effectively captures the differences in order among liquid, amorphous, and crystalline phases in multi-particle systems, providing an explanation for the mechanism of long-range order formation. Besides, SI, in conjunction with Global Softness, achieves precise detection of phase transitions through non-ML and ML pathways, respectively. Furthermore, complexity science suggests that global behavior in a system arises from nonlinear interactions among local components, rather than from simple linear addition. Our unified framework fundamentally avoids the issue of nonlinear adding, while treating local and global characteristics as homogeneous representations. This offers a novel perspective for research in complexity science.

We acknowledge that there are certain limitations to our approach as it currently stands. Firstly, computational complexity limits its scalability and cost-effectiveness when applied to large datasets. Secondly, while the method effectively captures topological features, its applicability may be limited in QSAR tasks that depend on geometric properties like metrics and curvature. Thirdly, although the framework performs well in specific multi-particle systems constrained by Lennard-Jones potentials, its generalizability to other disordered systems and diverse dynamical processes remains unverified. Finally, regarding the Separation Index (SI) and its relationship to order and symmetry, current validation is computational, lacking rigorous proof of how SI quantitatively reflects these physical characteristics.

Our future work will naturally focus on addressing these limitations. Firstly, improving computational efficiency to handle systems with over $10^5$ particles will be a key objective in scaling our approach. Secondly, we will consider combining our PH-based approach with computational conformal geometry~\cite{gu2002computing,gu2003global,gu2010fundamentals} to better accommodate targets requiring precise geometric detail, such as defects and polycrystals, and tasks like atomic-scale defect reconstruction or complex surface structuring. Thirdly, we plan to explore dynamical processes beyond phase transitions, including defect migration and reconstruction, particle diffusion and aggregation, and phase separation with multicomponent interactions. Lastly, we aim to rigorously establish how SI captures order and symmetry in the system, and develop its mathematical foundation in relation to underlying physics.

The code related to this article is available at \url{https://github.com/anwanguow/PH_structural}.

\acknowledgments

We thank Prof. Robert MacKay FRS from Warwick Mathematics Institute for the helpful discussions and suggestions.

\appendix

\section{Parameter Selection}
\label{app:parameter_selection}

\subsection{The parameter selection for the PI}
\label{app:parameter_selection_pi}

\begin{figure}[htbp!]
\includegraphics[width=0.48\textwidth]{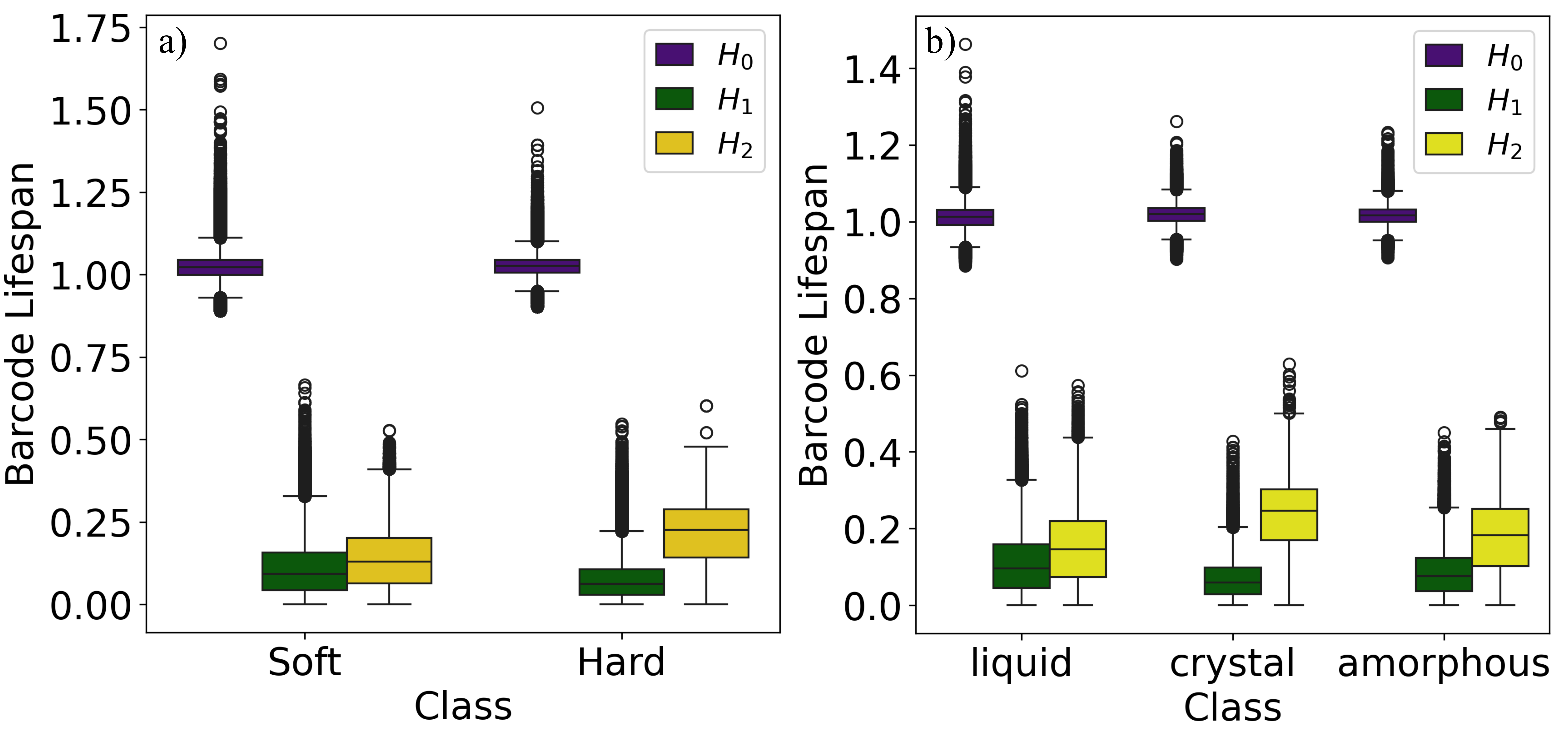}
\caption{Box plots of the barcode lifespan (persistence) of a) all neighborhoods with a radius $r=2.5$ of particle samples at time step $t=100+100 \times (k-1), k \in [1,10] \cap \mathbb{N}$, and b) all systems sampled at time step $t=100+25 \times (k-1), k \in [1,33] \cap \mathbb{N}$, from each trajectory in Group 1 and 2.}
\label{fig:box_lifespan}
\end{figure}

Here, the parameters we need to set are \texttt{minBD}, \texttt{maxBD}, $\tau$, $\mathfrak{M}$, $\mathfrak{N}$, and $\sigma$.

The parameters \texttt{minBD} and \texttt{maxBD} are used to filter out features with very short or excessively long persistence. Based on the analysis in Fig.~\ref{fig:box_lifespan}, all $H_1$ and $H_2$ features have lifetimes below 0.75, and most $H_0$ features are below 1.5. Therefore, setting \texttt{maxBD} to 1.5 is sufficient to capture all relevant features. To avoid numerical errors, \texttt{minBD} is set to -0.1.

\begin{figure}[htbp!]
\includegraphics[width=0.3\textwidth]{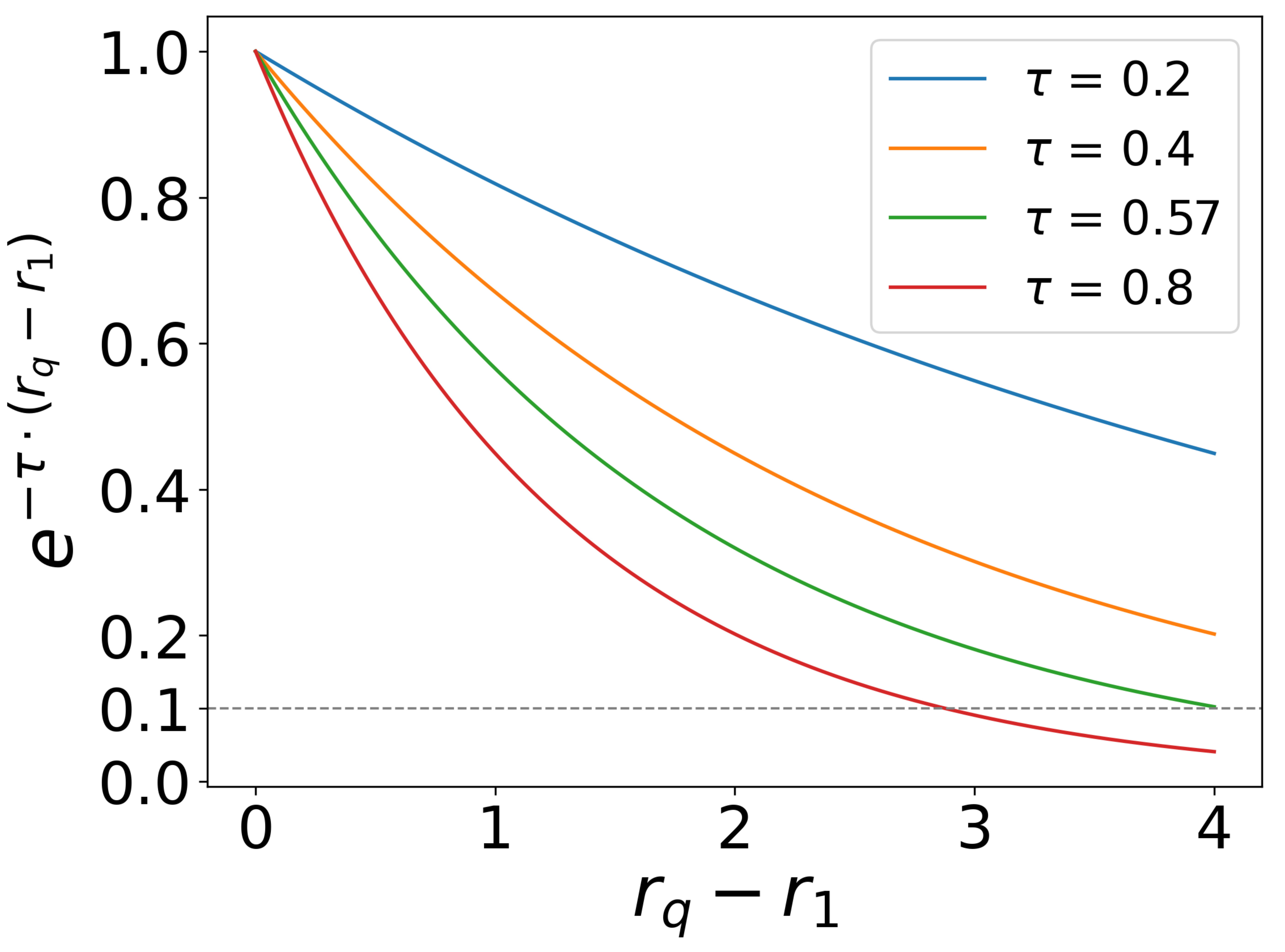}
\caption{The four curves in different color respectively represent the change in $e^{-\tau (r_q-r_1)}$ as $r_q-r_1$ increases from 0 to 4 (or, $r_q$ increases from 1 to 5) under the premise of taking different values of $\tau$. The gray horizontal dashed line is $y=0.1$.}
\label{fig:decay}
\end{figure}

In the formula $\bold{I}^{(P)}_{p} = \sum_{q=1}^{\mathcal{N}} \bold{V}^{(P)}_{p,q} \cdot e^{-\tau (r_q - r_1)}$, the decay factor $\tau$ is chosen to balance the rate of decay, ensuring $\bold{V}_{p,q}^{(P)}$ decays from 100\% to 10\% as $r_q$ increases from 1 to 5. Based on the decay curves in Fig.~\ref{fig:decay}, we set $\tau = 0.57$ as the most appropriate value.

\begin{figure}[htbp!]
\includegraphics[width=0.48\textwidth]{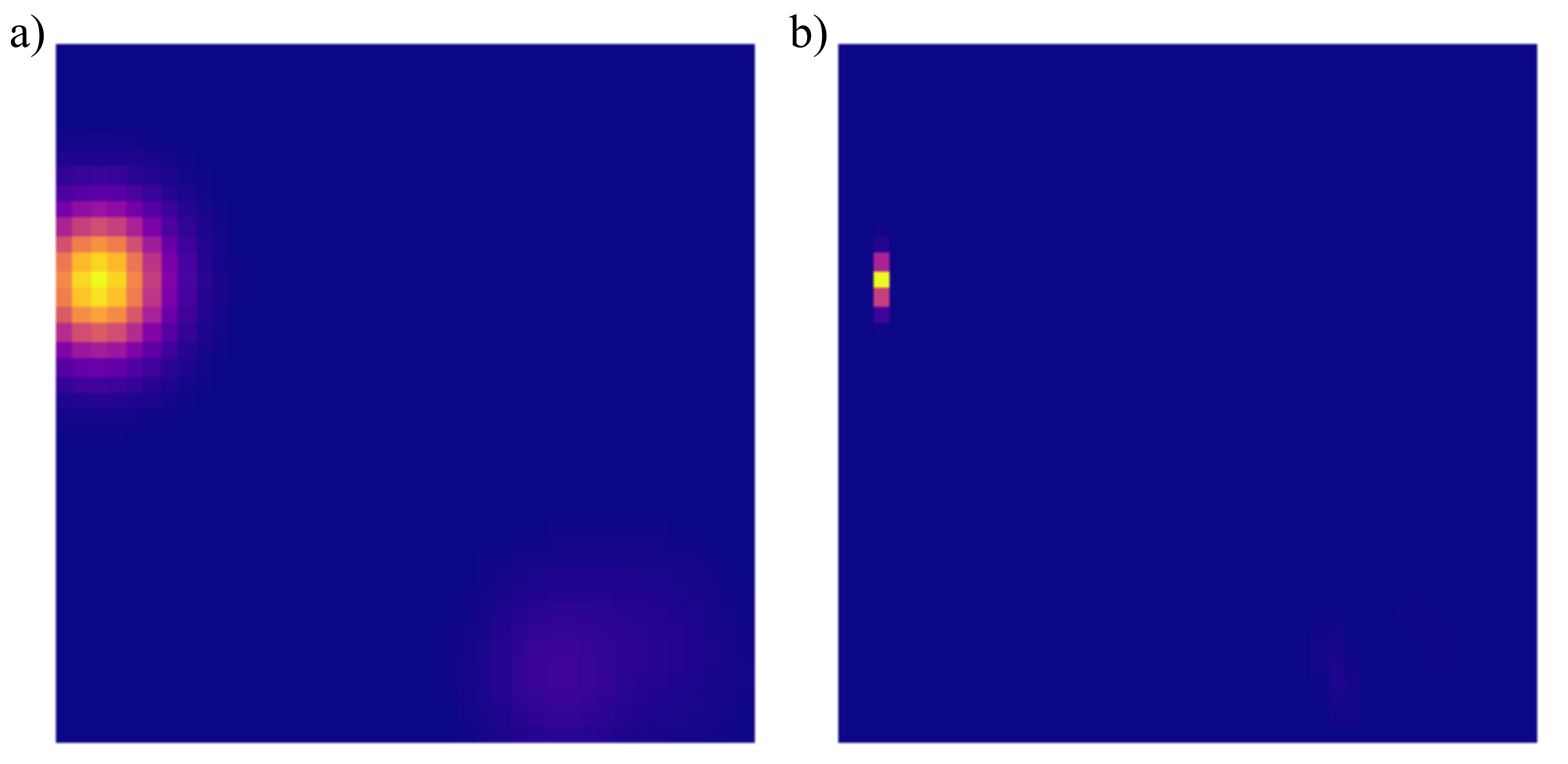}
\caption{The examples for a) an unclear PI, with $\tau$ = 0.1, and b) a clear PI, with $\tau = 0.002$, which are both generated from the frame corresponding to $\text{Traj}(T_4^{(2)})$ at time step $t = 100$. }
\label{fig:clear_pi}
\end{figure}

To balance computational efficiency and model performance, we set $\mathfrak{M}=\mathfrak{N}=40$ initially and treat $\sigma$ as a hyperparameter. The selection of $\sigma$ follows a two-stage process: (1) coarse screening; (2) fine-tuning.

\textit{Coarse Screening:} In this stage, we visually inspected the clarity of the PI generated with different values of $\sigma$. A smaller $\sigma$ preserves more details of topological features, while a larger $\sigma$ smooths the image, reducing the impact of individual points, as shown in Fig.~\ref{fig:clear_pi}. By observing the degree of blurring in the PI, we quickly identified a suitable range of $\sigma$ between 0.001 and 0.01.

\begin{table}[h]
\caption{
Performance comparison of different values of $\tau \in [0.001,0.01]$ in Task 2.}
\begin{ruledtabular}
\begin{tabular}{>{\bfseries}lcccc}
\multicolumn{1}{c}{\textbf{SVM (98\% Var)}} & \multicolumn{2}{c}{\textrm{Linear}} & \multicolumn{2}{c}{\textrm{RBF Kernel}} \\
\hline
\textbf{$\sigma$} & \textrm{ACC} & \textrm{MCC} & \textrm{ACC} & \textrm{MCC} \\
\hline
\textrm{0.001} & 95.5\% & 0.760 & 94.8\% & 0.750 \\
\textrm{0.002} & 94.5\% & 0.730 & 95.5\% & 0.760 \\
\textrm{0.003} & 95.5\% & 0.764 & 95.5\% & 0.764 \\
\textrm{0.004} & 95.5\% & 0.764 & 95.5\% & 0.764 \\
\textrm{0.005} & 95.8\% & 0.757 & 95.2\% & 0.757 \\
\textrm{0.006} & 95.2\% & 0.743 & 95.5\% & 0.760 \\
\textrm{0.007} & 96.1\% & 0.763 & 95.2\% & 0.753 \\
\textrm{0.008} & 95.5\% & 0.764 & 95.2\% & 0.753 \\
\textrm{0.009} & 95.8\% & 0.767 & 95.5\% & 0.760 \\
\textrm{0.010} & 96.1\% & 0.763 & 95.5\% & 0.760 \\
\end{tabular}
\end{ruledtabular}
\label{tab:tau_selection}
\end{table}

\textit{Fine-Tuning:} In this stage, we treated $\sigma$ as a hyperparameter and refined it through iterative training and validation. Using both two kinds of SVMs, we found that while $\sigma$ had minimal impact on overall performance, as listed in Table~\ref{tab:tau_selection}. The model achieved the optimal performance with $\sigma = 0.01$.

\subsection{The Parameter Selection for Rearrangement Tendency (Task 1)}
\label{app:select_of_p_hop}

\begin{figure}[htbp!]
\includegraphics[width=0.5\textwidth]{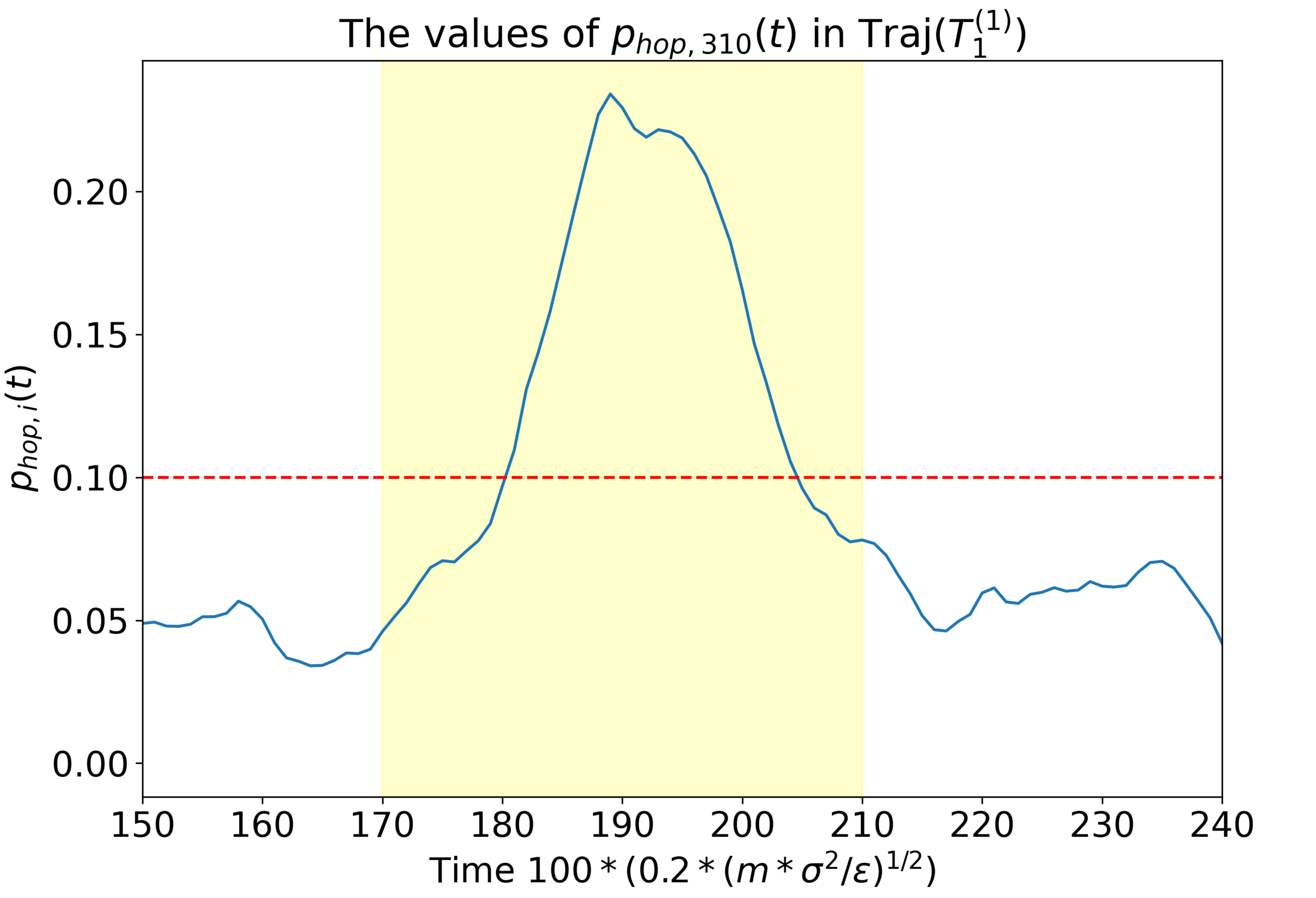}
\caption{The evolution of $p_{\text{hop}}$ for a selected particle (No.310) over a period of time under the lowest temperature (i.e., $T^{(1)}_1$). The yellow region represents the interval with the time length of $2t_{R/2}$, encompassing a complete rearrangement, which means $p_{hop}$ rises above $p_c$ and then falls back below $p_c$. The red-dotted line is the selected $p_c=0.1$.}
\label{fig:p_hop_trend}
\end{figure}

\begin{figure}[htbp!]
\includegraphics[width=0.5\textwidth]{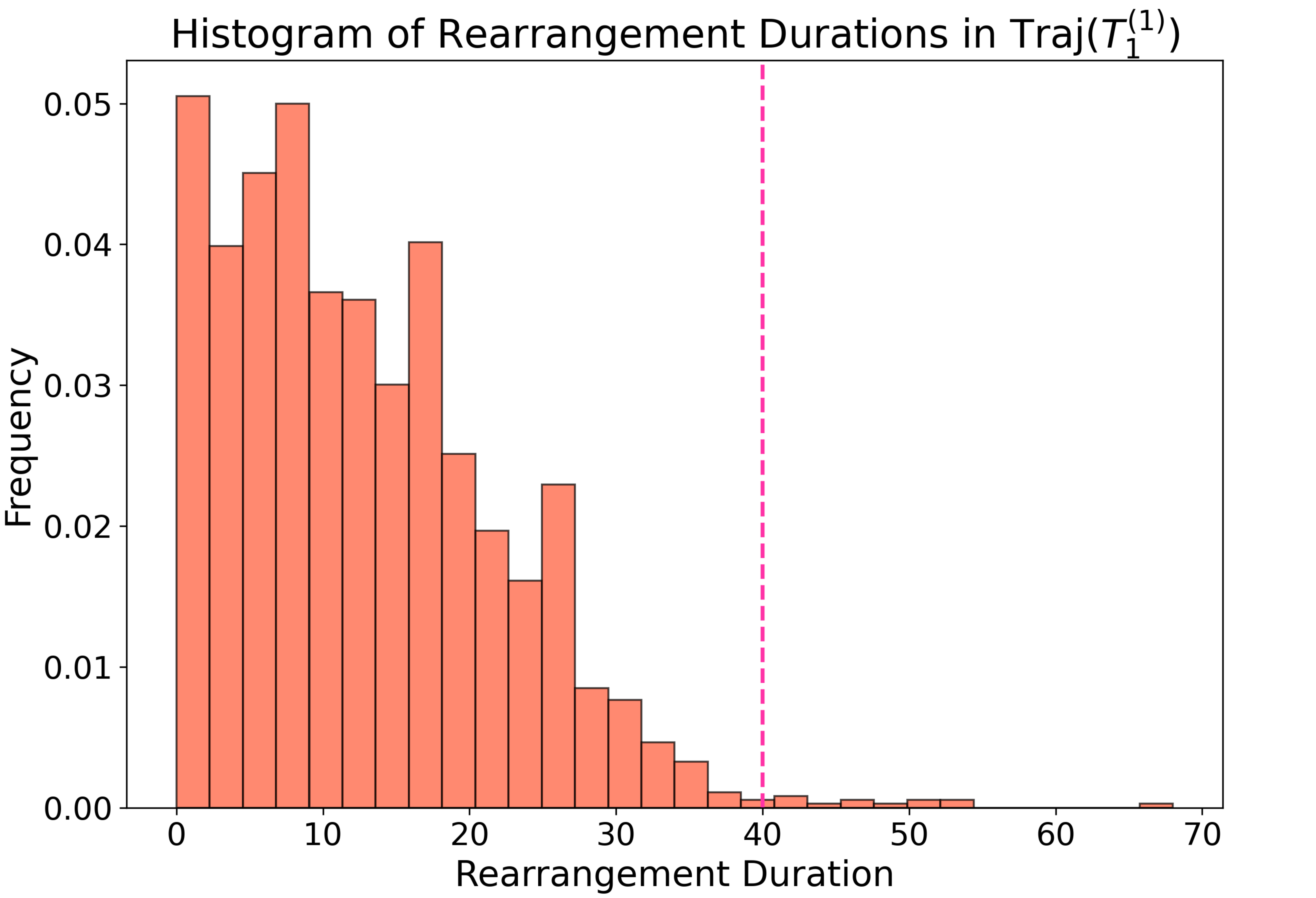}
\caption{The histogram of rearrangement duration in $\text{Traj}(T^{(1)}_1)$, corresponding to the lowest temperature and the system with the lowest mobility. The red-dotted line is the selected $t_{R}=40$, which equals to $2t_{R/2}$.}
\label{fig:histogram_p}
\end{figure}

\begin{figure}[t!]
\includegraphics[width=0.5\textwidth]{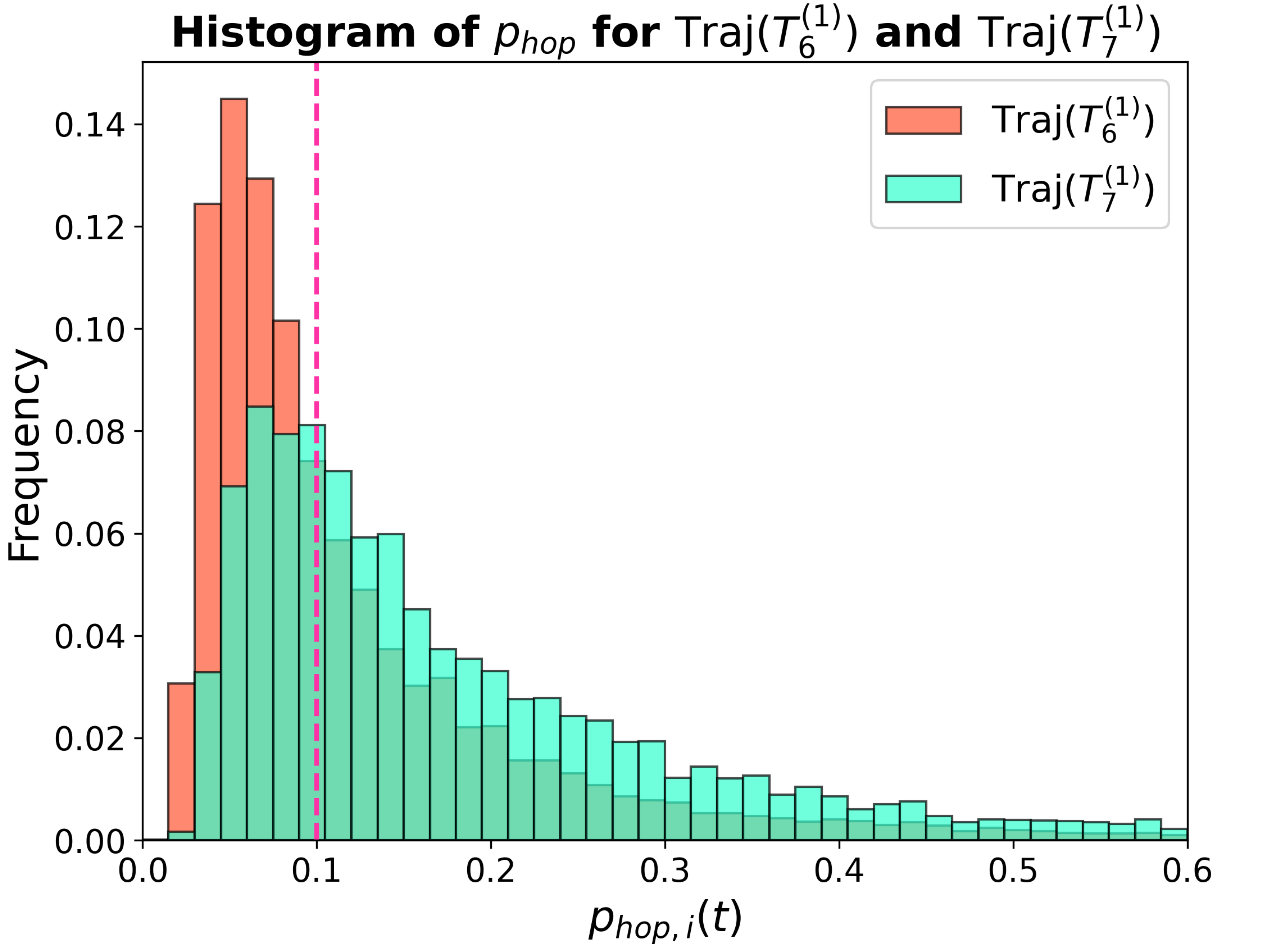}
\caption{The histogram of the value of $p_{\text{hop}}$ in $\text{Traj}(T^{(1)}_6)$ and $\text{Traj}(T^{(1)}_7)$. The onset temperature $T^{(1)}_c$ of crystallization separates $T^{(1)}_6$ and $T^{(1)}_7$, i.e., $T^{(1)}_6 < T^{(1)}_c < T^{(1)}_7$. The red-dotted line is the selected $p_c=0.1$. The reason for setting $p_c=0.1$ is when $p_{\text{hop}} < p_c$, the frequency of $p_{\text{hop}}$ in $\text{Traj}(T^{(1)}_6)$ is larger than that in $\text{Traj}(T^{(1)}_7)$; when $p_{\text{hop}} \geq p_c$, the frequency of $p_{\text{hop}}$ in $\text{Traj}(T^{(1)}_6)$ is smaller than that in $\text{Traj}(T^{(1)}_7)$. $p_c=0.1$ is the value at which the relative size of two frequency of $\text{Traj}(T^{(1)}_6)$ and $\text{Traj}(T^{(1)}_7)$ reverse.}
\label{fig:histogram}
\end{figure}

Here, the parameters we need to set are $t_{R/2}$ and $p_c$, aligned with Ref.~\cite{wang2024graph}. Fig.~\ref{fig:p_hop_trend},~\ref{fig:histogram_p}, and~\ref{fig:histogram} are all adapted from Ref.~\cite{wang2024graph}. The settings is based on the statistics on Group 1 (Set 1), and we cannot use the data in the Group 2 (Set 2) at this stage to avoid the data leakage.

To select $t_{R/2}$, we use the lowest temperature $T^{(1)}_1$ as a reference because it corresponds to the system's lowest mobility. The interval $[t - t_{R/2}, t + t_{R/2}]$ should encompass as many complete rearrangements as possible at this low mobility, as shown in Fig.~\ref{fig:p_hop_trend}. The yellow region marks a full rearrangement, where $p_{\text{hop}}$ rises above and then falls below $p_c$. As seen in Fig.~\ref{fig:histogram_p}, most rearrangements at this reference temperature occur within $t_R = 40$, so we set $t_{R/2} = 20$.

As shown in Fig.~\ref{fig:histogram}, we chose $p_c = 0.1$ because it effectively separates the majority of particles in two trajectories above and below the crystallization onset temperature, i.e., $T^{(1)}_6$ and $T^{(1)}_7$. The $p_{\text{hop},i}(t)$ values are mostly significantly greater or less than $p_c = 0.1$, respectively. Specifically, $T^{(1)}_6$ is the highest temperature at which crystallization occurs, while $T^{(1)}_7$ is the lowest temperature where the liquid phase is maintained, both of which are critical for this analysis. By plotting the histograms of $p_{\text{hop}}$ in $\text{Traj}(T^{(1)}_6)$ and $\text{Traj}(T^{(1)}_7)$, we determined that $p_c = 0.1$ is the value at which the relative sizes of the two distributions reverse.

\section{Correctness of Global Softness}
\label{app:correctness}

\begin{figure*}[ht!]
\includegraphics[width=0.63\textwidth]{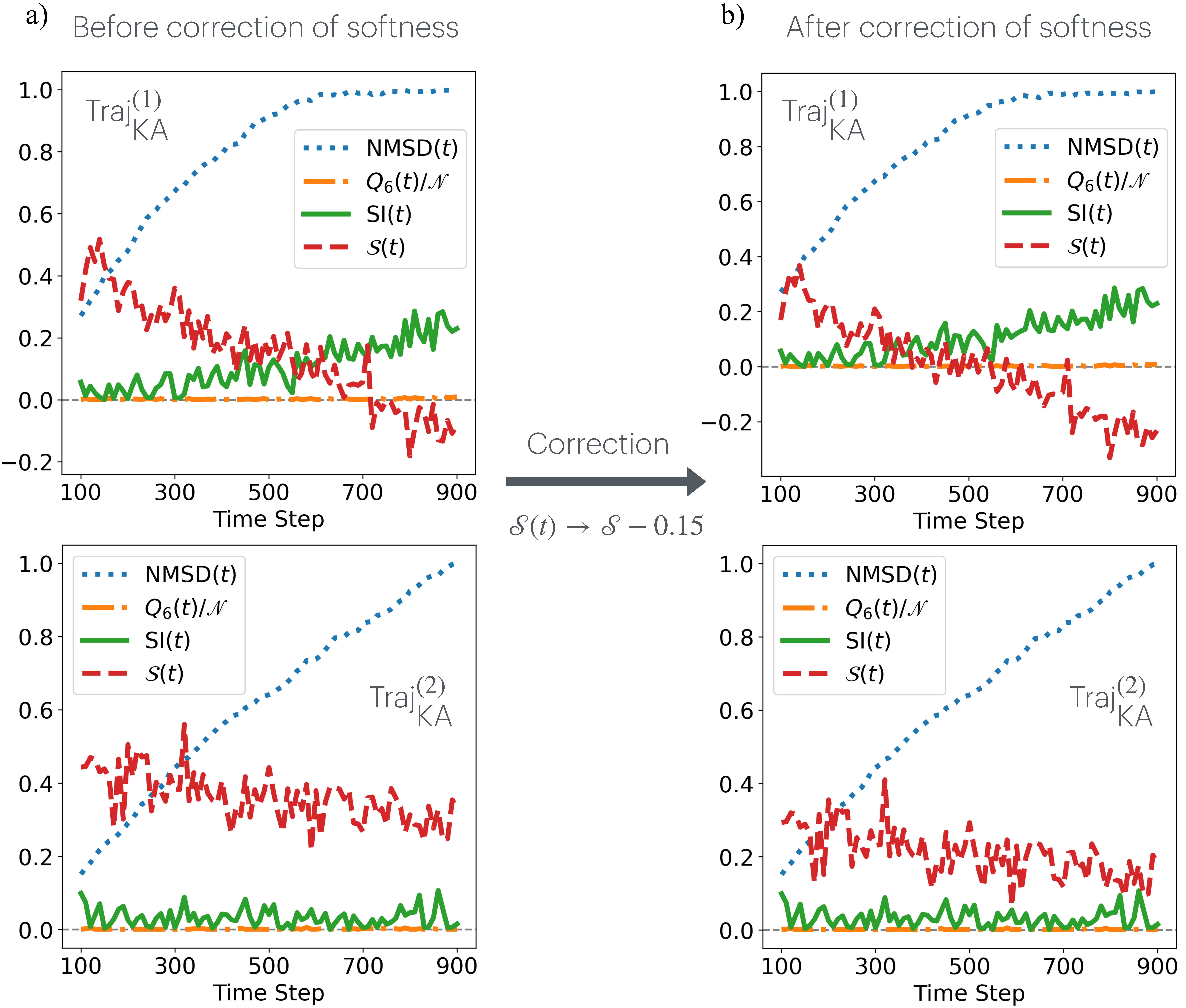}
\caption{The global softness $\mathcal{S}(t)$ (red lines) shown in a) is before the correction, while the softness shown in b) is after the correction by a downward shift of $0.15$.}
\label{fig:correctness}
\end{figure*}

Fig.~\ref{fig:correctness} demonstrate the correction by a downward shift of 0.15 when applying the model to the KA systems. After correction, the positive or negative sign of Global Softness can indicate the class of fluidity trend.

\section{Details of Feature Importance Analysis (FIA)}

\label{app:details_fia}

The single decisive predictor mentioned in Sec.~\ref{susbec:key_factor_task_2}, corresponding to the largest eigenvalue from the PCA, is the 1389th element of $\bold{I}_{\text{entire}}^{(P)}$, located at pixel (28, 34) in the PI. Given the PI dimensions of $40 \times 40$ and birth and persistence ranges of $[-0.1, 1.5]$, each pixel has a resolution of $0.04$. We can get that the pixel $(28, 34)$ is with birth range of $[1.02, 1.06]$ and persistence ranges of  and $[0.1, 0.14]$ in the PD, respectively. The feature importance, based on the eigenvalues of the principal components, revealed that this key predictor corresponds to the 1389th column of the original feature matrix, which is the 1389th element of the vector $\bold{I}_{\text{entire}}^{(P)}$.

\section{comparative Analysis Based on the $q_6$-SI Swapped State Sets}

\label{appendix:exchange_exp}

\begin{figure}[ht!]
\includegraphics[width=0.5\textwidth]{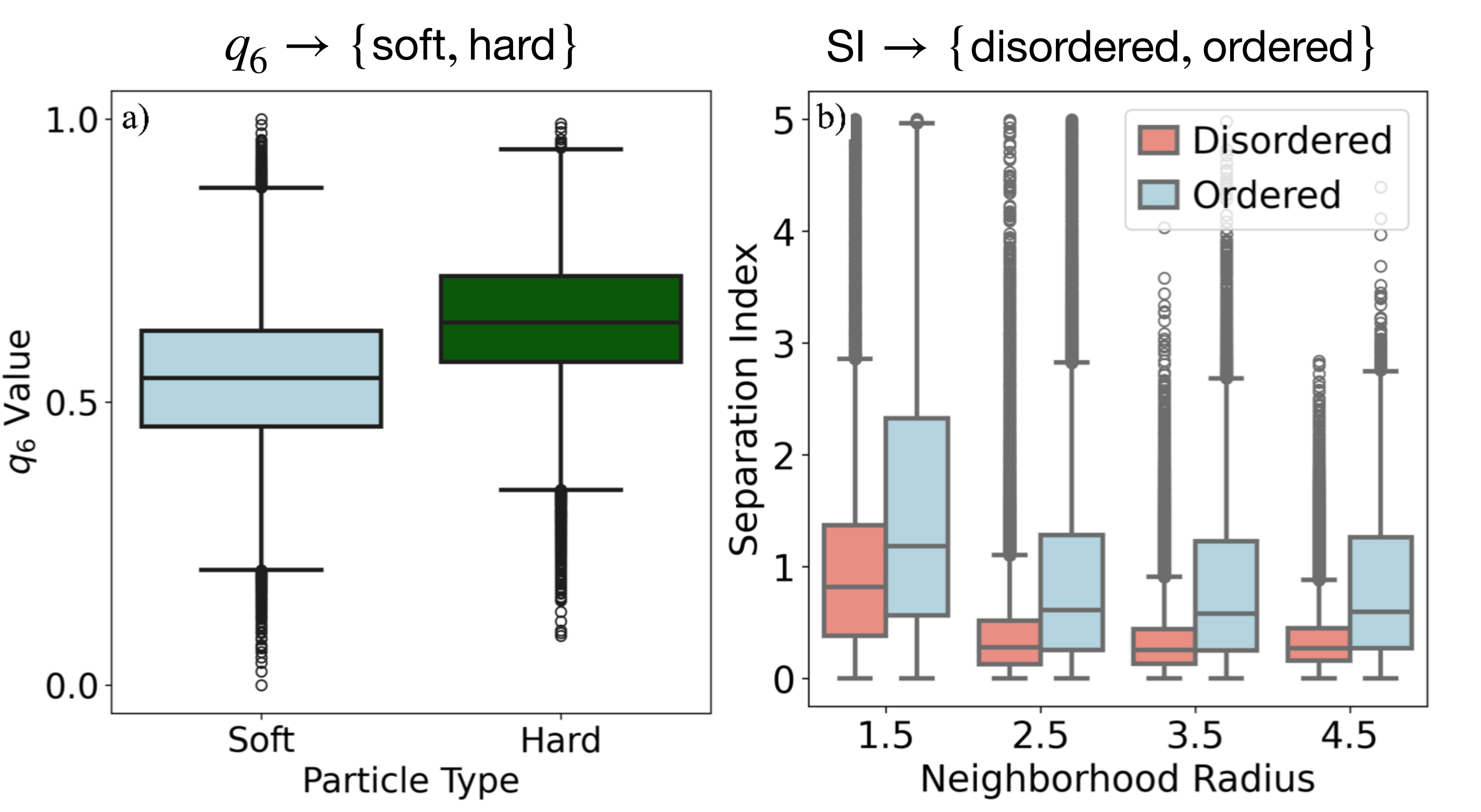}
\caption{Distribution of $q_6$ over the state set \{soft, hard\} and the Separation Index (SI) over the state set \{ordered, disordered\}. Note that the dataset used here is the same as in the Fig.~\ref{fig:separation_index_all}a), with only new metrics and labels applied.}
\label{fig:crossover}
\end{figure}

To begin with, we examined the local bond order parameter $q_6$, which characterizes the six-fold symmetry of the local particle environment (defined in Sec.~\ref{subsec:Bond_orientational_Order_Parameters}), to assess its ability to distinguish particles with different rearrangement tendencies. Specifically, we evaluated the performance of $q_6$ in differentiating between the state set of $\{\text{soft}, \text{hard}\}$, as reported in Fig.~\ref{fig:crossover}a). Here, the states of $\{\text{soft}, \text{hard}\}$ are determined by $p_{\text{hop}}$ (see Eq.~\ref{equ:p_hop}).

Furthermore, we investigated the effectiveness of the Separation Index (SI) in distinguishing the degree of six-fold symmetry preservation in the local particle environment. In particular, we analyzed how SI differentiates between the states $\{\text{ordered}, \text{disordered}\}$, with the results reported in Fig.~\ref{fig:crossover}b). It is important to note that the states $\{\text{ordered}, \text{disordered}\}$ are classified according to the values of $q_6$, which reflect the level of local structural order around the particles.

As shown in Fig.~\ref{fig:crossover}a), aside from the slightly lower average $q_6$ value observed for soft particles compared to hard particles, no other significant non-trivial patterns are evident. Thus, $q_6$ is not an effective predictor of particle rearrangement tendencies, nor does it serve as a reliable criterion for distinguishing between soft and hard particles. While the results suggest that hard particles exhibit, on average, a higher degree of local symmetry, this difference alone is insufficient to justify its use as a classification metric.

In Fig.~\ref{fig:crossover}b), although the average SI differs between ordered and disordered particles, the lower bound of the IQR for ordered particles does not separate significantly from the upper bound of the IQR for disordered particles as the neighborhood radius increases. Instead, it does not exhibit any clear non-trivial trends. Therefore, while particles in ordered environments generally display higher average SI values, this distinction also falls short of providing a robust basis for classification.

The particle rearrangement tendency and the degree of six-fold symmetry preservation in the local environment represent entirely different properties and cannot be used interchangeably as classification criteria. Overall, SI exhibits slightly better performance than $q_6$.

\section{Extra Data}
\label{app:extra_data}

Table~\ref{tab:temperature_list} provides the values of the 20 monotonically increasing temperatures mentioned in Sec.~\ref{subsec:data_group_division}.

\begin{table}[h]
    \centering
    \resizebox{0.45\textwidth}{!}{
    \begin{tabular}{c c c c c c c c c c}
        \toprule
        $T_1$ & $T_2$ & $T_3$ & $T_4$ & $T_5$ & $T_6$ & $T_7$ & $T_8$ & $T_9$ & $T_{10}$ \\
        0.50 & 0.53 & 0.55 & 0.57 & 0.60 & 0.63 & 0.65 & 0.67 & 0.70 & 0.73 \\
        \midrule
        $T_{11}$ & $T_{12}$ & $T_{13}$ & $T_{14}$ & $T_{15}$ & $T_{16}$ & $T_{17}$ & $T_{18}$ & $T_{19}$ & $T_{20}$ \\
        0.75 & 0.77 & 0.80 & 0.83 & 0.85 & 0.88 & 0.90 & 0.93 & 0.95 & 0.97 \\
        \bottomrule
    \end{tabular}
    }
    \caption{The 20 different temperatures used in Protocol 1 for generating LJ systems.}
    \label{tab:temperature_list}
\end{table}

\begin{table}[ht!]
\caption{\textcolor{black}{Classification performance of RBF kernel SVM after PCA with different explained variances in Task 1.}}
\begin{ruledtabular}
\begin{tabular}{cccc}
\textrm{VAR Explained} & \textrm{Num of Predictors} & \textrm{Accuracy} & \textrm{MCC} \\
\hline
98\% & 188 & 84.2\% & 0.686 \\
50\% & 7 & 82.4\% & 0.649 \\
48\% & 6 & 81.3\% & 0.626 \\
45\% & 5 & 80.7\% & 0.614 \\
40\% & 4 & 80.3\% & 0.607 \\
35\% & 3 & 79.5\% & 0.589 \\
30\% & 2 & 51.2\% & 0.030 \\
23\% & 1 & 50.3\% & 0.007 \\
\end{tabular}
\end{ruledtabular}
\label{tab:pca_task_1_to_1}
\end{table}

As demonstrated in Table.~\ref{tab:pca_task_1_to_1}, in Task 1, after applying PCA with different variance explained ratios to the feature matrix, we assessed the classification performance of an SVM with an RBF kernel. The results indicate that at least 3 variables are required to characterize the hard-or-soft property of the particles with an accuracy significantly higher than random selection. Furthermore, to achieve an accuracy loss within 2\% compared to the baseline (the first row of Table~\ref{tab:pca_task_1_to_1}), at least 7 variables are necessary.

\begin{figure*}[ht!]
\includegraphics[width=1\textwidth]{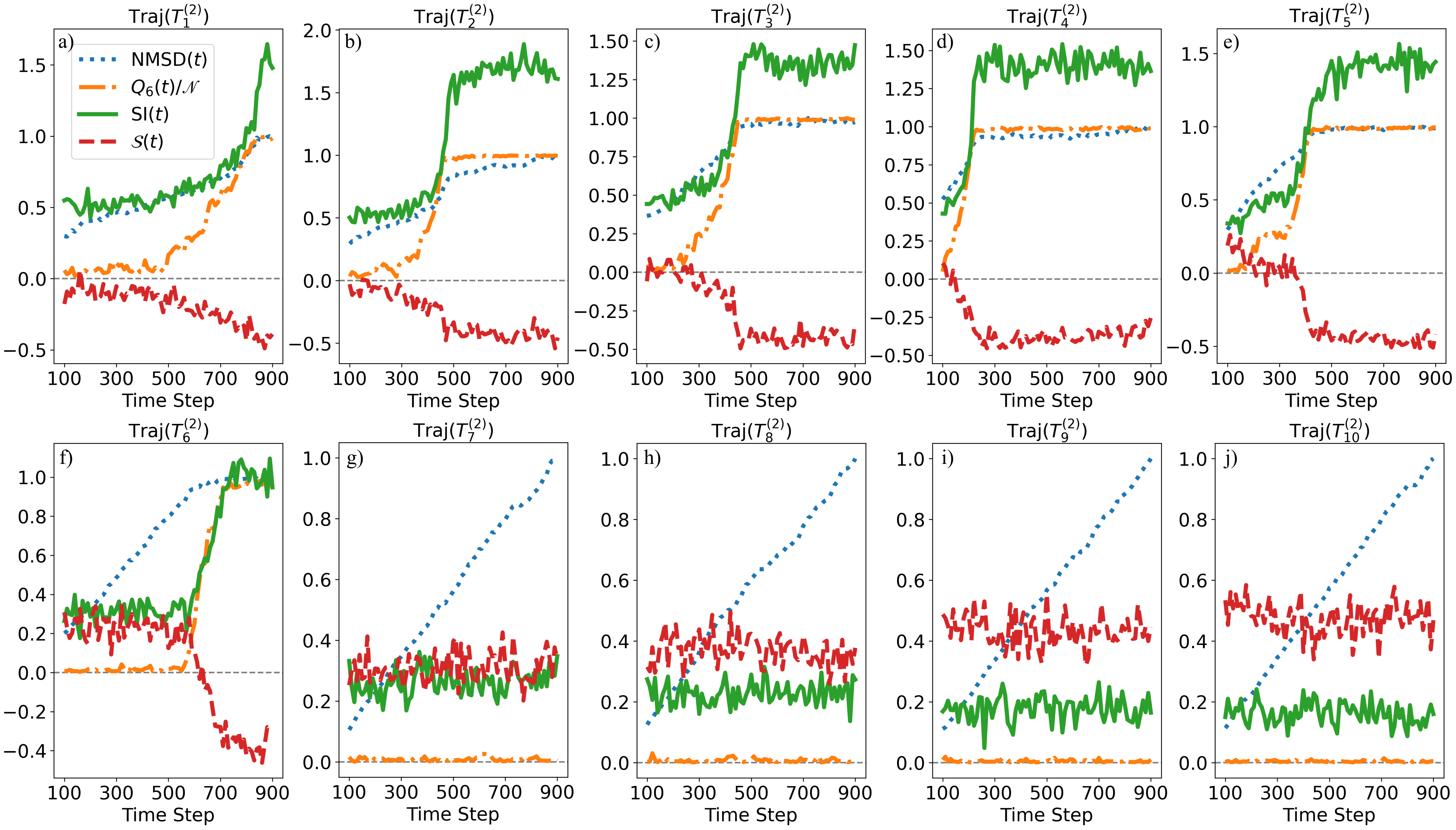}
\caption{The Normalized MSD (abbreviated as $\text{NMSD}(t)$, normalizing the MSD into $[0,1]$), $Q_6(t)/\mathscr{N}$, $\text{SI}(t)$, and $\mathcal{S}(t)$ where $t=100+10 \times (i-1)$, $i \in [1,81] \cap \mathbb{N}$ on all trajectories on Group 2 (Testing Group), containing a) $\text{Traj}(T_1^{(2)})$, b) $\text{Traj}(T_2^{(2)})$, c) $\text{Traj}(T_3^{(2)})$, d) $\text{Traj}(T_4^{(2)})$, e) $\text{Traj}(T_5^{(2)})$, f) $\text{Traj}(T_6^{(2)})$, g) $\text{Traj}(T_7^{(2)})$, h) $\text{Traj}(T_8^{(2)})$, i) $\text{Traj}(T_9^{(2)})$, and j) $\text{Traj}(T_{10}^{(2)})$.}
\label{fig:metrics_group_2}
\end{figure*}

The datasets for Task 2 are listed in Table~\ref{tab:phase_labels}, and the evolution of all four metrics over time for each trajectory in Group 2 (testing group) are listed in Fig.~\ref{fig:metrics_group_2}.

\begin{table*}[ht!]
\footnotesize
\centering
\caption{The labels, i.e., $y_{\text{entire},t}^{(T^{(1)}_i)}$ for Set 3 (sampled from Group 1) and $y_{\text{entire},t}^{(T^{(2)}_i)}$ for Set 4 (sampled from Group 2). In this table, 0 represents liquid, 1 represents crystal, and 2 represents amorphous. The $i$-th row represents trajectory $\text{Traj}(T_{i}^{(1 \text{ or } 2)})$, and the $j$-th column represents the sample point at $t = 100 + 25 \times (j-1)$.}
{\small
\begin{tabular}{|c|c|c|c|c|c|c|c|c|c|c|c|c|c|c|c|c|c|c|c|c|c|c|c|c|c|c|c|c|c|c|c|c|}
\hline
\multicolumn{33}{|c|}{\textbf{Set 3 (sampled from Group 1)}} \\
\hline
2 & 2 & 2 & 2 & 2 & 2 & 2 & 2 & 2 & 2 & 2 & 2 & 2 & 2 & 2 & 2 & 2 & 2 & 2 & 2 & 2 & 2 & 2 & 2 & 2 & 2 & 2 & 2 & 2 & 2 & 2 & 2 & 2 \\
\hline
2 & 2 & 2 & 2 & 2 & 2 & 2 & 2 & 2 & 2 & 1 & 1 & 1 & 1 & 1 & 1 & 1 & 1 & 1 & 1 & 1 & 1 & 1 & 1 & 1 & 1 & 1 & 1 & 1 & 1 & 1 & 1 & 1 \\
\hline
2 & 2 & 2 & 2 & 2 & 2 & 1 & 1 & 1 & 1 & 1 & 1 & 1 & 1 & 1 & 1 & 1 & 1 & 1 & 1 & 1 & 1 & 1 & 1 & 1 & 1 & 1 & 1 & 1 & 1 & 1 & 1 & 1 \\
\hline
2 & 2 & 2 & 2 & 1 & 1 & 1 & 1 & 1 & 1 & 1 & 1 & 1 & 1 & 1 & 1 & 1 & 1 & 1 & 1 & 1 & 1 & 1 & 1 & 1 & 1 & 1 & 1 & 1 & 1 & 1 & 1 & 1 \\
\hline
0 & 0 & 0 & 0 & 0 & 0 & 0 & 0 & 0 & 0 & 0 & 0 & 0 & 0 & 0 & 0 & 0 & 0 & 0 & 0 & 0 & 0 & 1 & 1 & 1 & 1 & 1 & 1 & 1 & 1 & 1 & 1 & 1 \\
\hline
0 & 0 & 0 & 0 & 0 & 0 & 0 & 0 & 0 & 0 & 0 & 0 & 0 & 0 & 0 & 0 & 0 & 0 & 0 & 0 & 0 & 0 & 0 & 0 & 0 & 0 & 0 & 0 & 0 & 0 & 0 & 0 & 0 \\
\hline
0 & 0 & 0 & 0 & 0 & 0 & 0 & 0 & 0 & 0 & 0 & 0 & 0 & 0 & 0 & 0 & 0 & 0 & 0 & 0 & 0 & 0 & 0 & 0 & 0 & 0 & 0 & 0 & 0 & 0 & 0 & 0 & 0 \\
\hline
0 & 0 & 0 & 0 & 0 & 0 & 0 & 0 & 0 & 0 & 0 & 0 & 0 & 0 & 0 & 0 & 0 & 0 & 0 & 0 & 0 & 0 & 0 & 0 & 0 & 0 & 0 & 0 & 0 & 0 & 0 & 0 & 0 \\
\hline
0 & 0 & 0 & 0 & 0 & 0 & 0 & 0 & 0 & 0 & 0 & 0 & 0 & 0 & 0 & 0 & 0 & 0 & 0 & 0 & 0 & 0 & 0 & 0 & 0 & 0 & 0 & 0 & 0 & 0 & 0 & 0 & 0 \\
\hline
0 & 0 & 0 & 0 & 0 & 0 & 0 & 0 & 0 & 0 & 0 & 0 & 0 & 0 & 0 & 0 & 0 & 0 & 0 & 0 & 0 & 0 & 0 & 0 & 0 & 0 & 0 & 0 & 0 & 0 & 0 & 0 & 0 \\
\hline
\multicolumn{33}{|c|}{\textbf{Set 4 (sampled from Group 2)}} \\
\hline
2 & 2 & 2 & 2 & 2 & 2 & 2 & 2 & 2 & 2 & 2 & 2 & 2 & 2 & 2 & 2 & 2 & 2 & 2 & 2 & 2 & 2 & 2 & 1 & 1 & 1 & 1 & 1 & 1 & 1 & 1 & 1 & 1 \\
\hline
2 & 2 & 2 & 2 & 2 & 2 & 2 & 2 & 2 & 2 & 2 & 2 & 2 & 1 & 1 & 1 & 1 & 1 & 1 & 1 & 1 & 1 & 1 & 1 & 1 & 1 & 1 & 1 & 1 & 1 & 1 & 1 & 1 \\
\hline
2 & 2 & 2 & 2 & 2 & 2 & 2 & 2 & 2 & 2 & 2 & 2 & 1 & 1 & 1 & 1 & 1 & 1 & 1 & 1 & 1 & 1 & 1 & 1 & 1 & 1 & 1 & 1 & 1 & 1 & 1 & 1 & 1 \\
\hline
2 & 2 & 2 & 1 & 1 & 1 & 1 & 1 & 1 & 1 & 1 & 1 & 1 & 1 & 1 & 1 & 1 & 1 & 1 & 1 & 1 & 1 & 1 & 1 & 1 & 1 & 1 & 1 & 1 & 1 & 1 & 1 & 1 \\
\hline
0 & 0 & 0 & 0 & 0 & 0 & 0 & 0 & 0 & 0 & 0 & 1 & 1 & 1 & 1 & 1 & 1 & 1 & 1 & 1 & 1 & 1 & 1 & 1 & 1 & 1 & 1 & 1 & 1 & 1 & 1 & 1 & 1 \\
\hline
0 & 0 & 0 & 0 & 0 & 0 & 0 & 0 & 0 & 0 & 0 & 0 & 0 & 0 & 0 & 0 & 0 & 0 & 0 & 0 & 0 & 1 & 1 & 1 & 1 & 1 & 1 & 1 & 1 & 1 & 1 & 1 & 1 \\
\hline
0 & 0 & 0 & 0 & 0 & 0 & 0 & 0 & 0 & 0 & 0 & 0 & 0 & 0 & 0 & 0 & 0 & 0 & 0 & 0 & 0 & 0 & 0 & 0 & 0 & 0 & 0 & 0 & 0 & 0 & 0 & 0 & 0 \\
\hline
0 & 0 & 0 & 0 & 0 & 0 & 0 & 0 & 0 & 0 & 0 & 0 & 0 & 0 & 0 & 0 & 0 & 0 & 0 & 0 & 0 & 0 & 0 & 0 & 0 & 0 & 0 & 0 & 0 & 0 & 0 & 0 & 0 \\
\hline
0 & 0 & 0 & 0 & 0 & 0 & 0 & 0 & 0 & 0 & 0 & 0 & 0 & 0 & 0 & 0 & 0 & 0 & 0 & 0 & 0 & 0 & 0 & 0 & 0 & 0 & 0 & 0 & 0 & 0 & 0 & 0 & 0 \\
\hline
0 & 0 & 0 & 0 & 0 & 0 & 0 & 0 & 0 & 0 & 0 & 0 & 0 & 0 & 0 & 0 & 0 & 0 & 0 & 0 & 0 & 0 & 0 & 0 & 0 & 0 & 0 & 0 & 0 & 0 & 0 & 0 & 0 \\
\hline
\end{tabular}
}
\label{tab:phase_labels}
\end{table*}

\clearpage


\begin{thebibliography}{102}%
\makeatletter
\providecommand \@ifxundefined [1]{%
 \@ifx{#1\undefined}
}%
\providecommand \@ifnum [1]{%
 \ifnum #1\expandafter \@firstoftwo
 \else \expandafter \@secondoftwo
 \fi
}%
\providecommand \@ifx [1]{%
 \ifx #1\expandafter \@firstoftwo
 \else \expandafter \@secondoftwo
 \fi
}%
\providecommand \natexlab [1]{#1}%
\providecommand \enquote  [1]{``#1''}%
\providecommand \bibnamefont  [1]{#1}%
\providecommand \bibfnamefont [1]{#1}%
\providecommand \citenamefont [1]{#1}%
\providecommand \href@noop [0]{\@secondoftwo}%
\providecommand \href [0]{\begingroup \@sanitize@url \@href}%
\providecommand \@href[1]{\@@startlink{#1}\@@href}%
\providecommand \@@href[1]{\endgroup#1\@@endlink}%
\providecommand \@sanitize@url [0]{\catcode `\\12\catcode `\$12\catcode `\&12\catcode `\#12\catcode `\^12\catcode `\_12\catcode `\%12\relax}%
\providecommand \@@startlink[1]{}%
\providecommand \@@endlink[0]{}%
\providecommand \url  [0]{\begingroup\@sanitize@url \@url }%
\providecommand \@url [1]{\endgroup\@href {#1}{\urlprefix }}%
\providecommand \urlprefix  [0]{URL }%
\providecommand \Eprint [0]{\href }%
\providecommand \doibase [0]{https://doi.org/}%
\providecommand \selectlanguage [0]{\@gobble}%
\providecommand \bibinfo  [0]{\@secondoftwo}%
\providecommand \bibfield  [0]{\@secondoftwo}%
\providecommand \translation [1]{[#1]}%
\providecommand \BibitemOpen [0]{}%
\providecommand \bibitemStop [0]{}%
\providecommand \bibitemNoStop [0]{.\EOS\space}%
\providecommand \EOS [0]{\spacefactor3000\relax}%
\providecommand \BibitemShut  [1]{\csname bibitem#1\endcsname}%
\let\auto@bib@innerbib\@empty
\bibitem [{\citenamefont {Zhang}\ \emph {et~al.}(2019)\citenamefont {Zhang}, \citenamefont {Zhao}, \citenamefont {Chen}, \citenamefont {Huang}, \citenamefont {Dong}, \citenamefont {Chen}, \citenamefont {Liu}, \citenamefont {Wang}, \citenamefont {Yan}, \citenamefont {Li} \emph {et~al.}}]{zhang2019tuning}%
  \BibitemOpen
  \bibfield  {author} {\bibinfo {author} {\bibfnamefont {J.}~\bibnamefont {Zhang}}, \bibinfo {author} {\bibfnamefont {Y.}~\bibnamefont {Zhao}}, \bibinfo {author} {\bibfnamefont {C.}~\bibnamefont {Chen}}, \bibinfo {author} {\bibfnamefont {Y.-C.}\ \bibnamefont {Huang}}, \bibinfo {author} {\bibfnamefont {C.-L.}\ \bibnamefont {Dong}}, \bibinfo {author} {\bibfnamefont {C.-J.}\ \bibnamefont {Chen}}, \bibinfo {author} {\bibfnamefont {R.-S.}\ \bibnamefont {Liu}}, \bibinfo {author} {\bibfnamefont {C.}~\bibnamefont {Wang}}, \bibinfo {author} {\bibfnamefont {K.}~\bibnamefont {Yan}}, \bibinfo {author} {\bibfnamefont {Y.}~\bibnamefont {Li}}, \emph {et~al.},\ }\bibfield  {title} {\bibinfo {title} {Tuning the coordination environment in single-atom catalysts to achieve highly efficient oxygen reduction reactions},\ }\href@noop {} {\bibfield  {journal} {\bibinfo  {journal} {Journal of the American Chemical Society}\ }\textbf {\bibinfo {volume} {141}},\ \bibinfo {pages} {20118} (\bibinfo {year} {2019})}\BibitemShut {NoStop}%
\bibitem [{\citenamefont {Li}\ \emph {et~al.}(2020)\citenamefont {Li}, \citenamefont {Rong}, \citenamefont {Zhang}, \citenamefont {Wang},\ and\ \citenamefont {Li}}]{li2020modulating}%
  \BibitemOpen
  \bibfield  {author} {\bibinfo {author} {\bibfnamefont {X.}~\bibnamefont {Li}}, \bibinfo {author} {\bibfnamefont {H.}~\bibnamefont {Rong}}, \bibinfo {author} {\bibfnamefont {J.}~\bibnamefont {Zhang}}, \bibinfo {author} {\bibfnamefont {D.}~\bibnamefont {Wang}},\ and\ \bibinfo {author} {\bibfnamefont {Y.}~\bibnamefont {Li}},\ }\bibfield  {title} {\bibinfo {title} {Modulating the local coordination environment of single-atom catalysts for enhanced catalytic performance},\ }\href@noop {} {\bibfield  {journal} {\bibinfo  {journal} {Nano Research}\ }\textbf {\bibinfo {volume} {13}},\ \bibinfo {pages} {1842} (\bibinfo {year} {2020})}\BibitemShut {NoStop}%
\bibitem [{\citenamefont {Cowley}(1965)}]{cowley1965short}%
  \BibitemOpen
  \bibfield  {author} {\bibinfo {author} {\bibfnamefont {J.}~\bibnamefont {Cowley}},\ }\bibfield  {title} {\bibinfo {title} {Short-range order and long-range order parameters},\ }\href@noop {} {\bibfield  {journal} {\bibinfo  {journal} {Physical Review}\ }\textbf {\bibinfo {volume} {138}},\ \bibinfo {pages} {A1384} (\bibinfo {year} {1965})}\BibitemShut {NoStop}%
\bibitem [{\citenamefont {Capella}\ and\ \citenamefont {Krzywicki}(1978)}]{capella1978unitarity}%
  \BibitemOpen
  \bibfield  {author} {\bibinfo {author} {\bibfnamefont {A.}~\bibnamefont {Capella}}\ and\ \bibinfo {author} {\bibfnamefont {A.}~\bibnamefont {Krzywicki}},\ }\bibfield  {title} {\bibinfo {title} {Unitarity corrections to short-range order: Long-range rapidity correlations},\ }\href@noop {} {\bibfield  {journal} {\bibinfo  {journal} {Physical Review D}\ }\textbf {\bibinfo {volume} {18}},\ \bibinfo {pages} {4120} (\bibinfo {year} {1978})}\BibitemShut {NoStop}%
\bibitem [{\citenamefont {Ferrari}\ \emph {et~al.}(2023)\citenamefont {Ferrari}, \citenamefont {K{\"o}rmann}, \citenamefont {Asta},\ and\ \citenamefont {Neugebauer}}]{ferrari2023simulating}%
  \BibitemOpen
  \bibfield  {author} {\bibinfo {author} {\bibfnamefont {A.}~\bibnamefont {Ferrari}}, \bibinfo {author} {\bibfnamefont {F.}~\bibnamefont {K{\"o}rmann}}, \bibinfo {author} {\bibfnamefont {M.}~\bibnamefont {Asta}},\ and\ \bibinfo {author} {\bibfnamefont {J.}~\bibnamefont {Neugebauer}},\ }\bibfield  {title} {\bibinfo {title} {Simulating short-range order in compositionally complex materials},\ }\href@noop {} {\bibfield  {journal} {\bibinfo  {journal} {Nature Computational Science}\ }\textbf {\bibinfo {volume} {3}},\ \bibinfo {pages} {221} (\bibinfo {year} {2023})}\BibitemShut {NoStop}%
\bibitem [{\citenamefont {Fuller}(1959)}]{fuller1959hydrogen}%
  \BibitemOpen
  \bibfield  {author} {\bibinfo {author} {\bibfnamefont {W.}~\bibnamefont {Fuller}},\ }\bibfield  {title} {\bibinfo {title} {Hydrogen bond lengths and angles observed in crystals},\ }\href@noop {} {\bibfield  {journal} {\bibinfo  {journal} {The Journal of Physical Chemistry}\ }\textbf {\bibinfo {volume} {63}},\ \bibinfo {pages} {1705} (\bibinfo {year} {1959})}\BibitemShut {NoStop}%
\bibitem [{\citenamefont {Geisinger}\ \emph {et~al.}(1985)\citenamefont {Geisinger}, \citenamefont {Gibbs},\ and\ \citenamefont {Navrotsky}}]{geisinger1985molecular}%
  \BibitemOpen
  \bibfield  {author} {\bibinfo {author} {\bibfnamefont {K.}~\bibnamefont {Geisinger}}, \bibinfo {author} {\bibfnamefont {G.}~\bibnamefont {Gibbs}},\ and\ \bibinfo {author} {\bibfnamefont {A.}~\bibnamefont {Navrotsky}},\ }\bibfield  {title} {\bibinfo {title} {A molecular orbital study of bond length and angle variations in framework structures},\ }\href@noop {} {\bibfield  {journal} {\bibinfo  {journal} {Physics and Chemistry of Minerals}\ }\textbf {\bibinfo {volume} {11}},\ \bibinfo {pages} {266} (\bibinfo {year} {1985})}\BibitemShut {NoStop}%
\bibitem [{\citenamefont {Laskowski}\ \emph {et~al.}(1993)\citenamefont {Laskowski}, \citenamefont {Moss},\ and\ \citenamefont {Thornton}}]{laskowski1993main}%
  \BibitemOpen
  \bibfield  {author} {\bibinfo {author} {\bibfnamefont {R.~A.}\ \bibnamefont {Laskowski}}, \bibinfo {author} {\bibfnamefont {D.~S.}\ \bibnamefont {Moss}},\ and\ \bibinfo {author} {\bibfnamefont {J.~M.}\ \bibnamefont {Thornton}},\ }\bibfield  {title} {\bibinfo {title} {Main-chain bond lengths and bond angles in protein structures},\ }\href@noop {} {\bibfield  {journal} {\bibinfo  {journal} {Journal of molecular biology}\ }\textbf {\bibinfo {volume} {231}},\ \bibinfo {pages} {1049} (\bibinfo {year} {1993})}\BibitemShut {NoStop}%
\bibitem [{\citenamefont {Shirley}\ \emph {et~al.}(1995)\citenamefont {Shirley}, \citenamefont {Hoffmann},\ and\ \citenamefont {Mastryukov}}]{shirley1995approach}%
  \BibitemOpen
  \bibfield  {author} {\bibinfo {author} {\bibfnamefont {W.~A.}\ \bibnamefont {Shirley}}, \bibinfo {author} {\bibfnamefont {R.}~\bibnamefont {Hoffmann}},\ and\ \bibinfo {author} {\bibfnamefont {V.~S.}\ \bibnamefont {Mastryukov}},\ }\bibfield  {title} {\bibinfo {title} {An approach to understanding bond length/bond angle relationships},\ }\href@noop {} {\bibfield  {journal} {\bibinfo  {journal} {The Journal of Physical Chemistry}\ }\textbf {\bibinfo {volume} {99}},\ \bibinfo {pages} {4025} (\bibinfo {year} {1995})}\BibitemShut {NoStop}%
\bibitem [{\citenamefont {Hohenberg}(1967)}]{hohenberg1967existence}%
  \BibitemOpen
  \bibfield  {author} {\bibinfo {author} {\bibfnamefont {P.~C.}\ \bibnamefont {Hohenberg}},\ }\bibfield  {title} {\bibinfo {title} {Existence of long-range order in one and two dimensions},\ }\href@noop {} {\bibfield  {journal} {\bibinfo  {journal} {Physical Review}\ }\textbf {\bibinfo {volume} {158}},\ \bibinfo {pages} {383} (\bibinfo {year} {1967})}\BibitemShut {NoStop}%
\bibitem [{\citenamefont {White}\ and\ \citenamefont {Geballe}(2013)}]{white2013long}%
  \BibitemOpen
  \bibfield  {author} {\bibinfo {author} {\bibfnamefont {R.~M.}\ \bibnamefont {White}}\ and\ \bibinfo {author} {\bibfnamefont {T.~H.}\ \bibnamefont {Geballe}},\ }\href@noop {} {\emph {\bibinfo {title} {Long Range Order in Solids: Solid State Physics}}}\ (\bibinfo  {publisher} {Elsevier},\ \bibinfo {year} {2013})\BibitemShut {NoStop}%
\bibitem [{\citenamefont {Yang}(1962)}]{yang1962concept}%
  \BibitemOpen
  \bibfield  {author} {\bibinfo {author} {\bibfnamefont {C.~N.}\ \bibnamefont {Yang}},\ }\bibfield  {title} {\bibinfo {title} {Concept of off-diagonal long-range order and the quantum phases of liquid he and of superconductors},\ }\href@noop {} {\bibfield  {journal} {\bibinfo  {journal} {Reviews of Modern Physics}\ }\textbf {\bibinfo {volume} {34}},\ \bibinfo {pages} {694} (\bibinfo {year} {1962})}\BibitemShut {NoStop}%
\bibitem [{\citenamefont {Pershan}(1988)}]{pershan1988structure}%
  \BibitemOpen
  \bibfield  {author} {\bibinfo {author} {\bibfnamefont {P.~S.}\ \bibnamefont {Pershan}},\ }\href@noop {} {\emph {\bibinfo {title} {Structure of liquid crystal phases}}},\ Vol.~\bibinfo {volume} {23}\ (\bibinfo  {publisher} {World Scientific},\ \bibinfo {year} {1988})\BibitemShut {NoStop}%
\bibitem [{\citenamefont {Izyumov}\ and\ \citenamefont {Syromyatnikov}(2012)}]{izyumov2012phase}%
  \BibitemOpen
  \bibfield  {author} {\bibinfo {author} {\bibfnamefont {Y.~A.}\ \bibnamefont {Izyumov}}\ and\ \bibinfo {author} {\bibfnamefont {V.~N.}\ \bibnamefont {Syromyatnikov}},\ }\href@noop {} {\emph {\bibinfo {title} {Phase transitions and crystal symmetry}}},\ Vol.~\bibinfo {volume} {38}\ (\bibinfo  {publisher} {Springer Science \& Business Media},\ \bibinfo {year} {2012})\BibitemShut {NoStop}%
\bibitem [{\citenamefont {Hirschfelder}\ \emph {et~al.}(1937)\citenamefont {Hirschfelder}, \citenamefont {Stevenson},\ and\ \citenamefont {Eyring}}]{hirschfelder1937theory}%
  \BibitemOpen
  \bibfield  {author} {\bibinfo {author} {\bibfnamefont {J.}~\bibnamefont {Hirschfelder}}, \bibinfo {author} {\bibfnamefont {D.}~\bibnamefont {Stevenson}},\ and\ \bibinfo {author} {\bibfnamefont {H.}~\bibnamefont {Eyring}},\ }\bibfield  {title} {\bibinfo {title} {A theory of liquid structure},\ }\href@noop {} {\bibfield  {journal} {\bibinfo  {journal} {The Journal of Chemical Physics}\ }\textbf {\bibinfo {volume} {5}},\ \bibinfo {pages} {896} (\bibinfo {year} {1937})}\BibitemShut {NoStop}%
\bibitem [{\citenamefont {Takayama}(1976)}]{takayama1976amorphous}%
  \BibitemOpen
  \bibfield  {author} {\bibinfo {author} {\bibfnamefont {S.}~\bibnamefont {Takayama}},\ }\bibfield  {title} {\bibinfo {title} {Amorphous structures and their formation and stability},\ }\href@noop {} {\bibfield  {journal} {\bibinfo  {journal} {Journal of Materials Science}\ }\textbf {\bibinfo {volume} {11}},\ \bibinfo {pages} {164} (\bibinfo {year} {1976})}\BibitemShut {NoStop}%
\bibitem [{\citenamefont {Jette}\ and\ \citenamefont {Foote}(1935)}]{jette1935precision}%
  \BibitemOpen
  \bibfield  {author} {\bibinfo {author} {\bibfnamefont {E.~R.}\ \bibnamefont {Jette}}\ and\ \bibinfo {author} {\bibfnamefont {F.}~\bibnamefont {Foote}},\ }\bibfield  {title} {\bibinfo {title} {Precision determination of lattice constants},\ }\href@noop {} {\bibfield  {journal} {\bibinfo  {journal} {The Journal of Chemical Physics}\ }\textbf {\bibinfo {volume} {3}},\ \bibinfo {pages} {605} (\bibinfo {year} {1935})}\BibitemShut {NoStop}%
\bibitem [{\citenamefont {Alexander}(1998)}]{alexander1998amorphous}%
  \BibitemOpen
  \bibfield  {author} {\bibinfo {author} {\bibfnamefont {S.}~\bibnamefont {Alexander}},\ }\bibfield  {title} {\bibinfo {title} {Amorphous solids: their structure, lattice dynamics and elasticity},\ }\href@noop {} {\bibfield  {journal} {\bibinfo  {journal} {Physics reports}\ }\textbf {\bibinfo {volume} {296}},\ \bibinfo {pages} {65} (\bibinfo {year} {1998})}\BibitemShut {NoStop}%
\bibitem [{\citenamefont {Drabold}(2009)}]{drabold2009topics}%
  \BibitemOpen
  \bibfield  {author} {\bibinfo {author} {\bibfnamefont {D.}~\bibnamefont {Drabold}},\ }\bibfield  {title} {\bibinfo {title} {Topics in the theory of amorphous materials},\ }\href@noop {} {\bibfield  {journal} {\bibinfo  {journal} {The European Physical Journal B}\ }\textbf {\bibinfo {volume} {68}},\ \bibinfo {pages} {1} (\bibinfo {year} {2009})}\BibitemShut {NoStop}%
\bibitem [{\citenamefont {Glazer}\ \emph {et~al.}(2012)\citenamefont {Glazer}, \citenamefont {Burns},\ and\ \citenamefont {Glazer}}]{glazer2012space}%
  \BibitemOpen
  \bibfield  {author} {\bibinfo {author} {\bibfnamefont {M.}~\bibnamefont {Glazer}}, \bibinfo {author} {\bibfnamefont {G.}~\bibnamefont {Burns}},\ and\ \bibinfo {author} {\bibfnamefont {A.~N.}\ \bibnamefont {Glazer}},\ }\href@noop {} {\emph {\bibinfo {title} {Space groups for solid state scientists}}}\ (\bibinfo  {publisher} {Elsevier},\ \bibinfo {year} {2012})\BibitemShut {NoStop}%
\bibitem [{\citenamefont {Aroyo}\ and\ \citenamefont {Wondratschek}(2013)}]{aroyo2013international}%
  \BibitemOpen
  \bibfield  {author} {\bibinfo {author} {\bibfnamefont {M.~I.}\ \bibnamefont {Aroyo}}\ and\ \bibinfo {author} {\bibfnamefont {H.}~\bibnamefont {Wondratschek}},\ }\href@noop {} {\emph {\bibinfo {title} {International tables for crystallography}}}\ (\bibinfo  {publisher} {Wiley Online Library},\ \bibinfo {year} {2013})\BibitemShut {NoStop}%
\bibitem [{\citenamefont {De~Wolff}(1974)}]{de1974pseudo}%
  \BibitemOpen
  \bibfield  {author} {\bibinfo {author} {\bibfnamefont {P.}~\bibnamefont {De~Wolff}},\ }\bibfield  {title} {\bibinfo {title} {The pseudo-symmetry of modulated crystal structures},\ }\href@noop {} {\bibfield  {journal} {\bibinfo  {journal} {Acta Crystallographica Section A: Crystal Physics, Diffraction, Theoretical and General Crystallography}\ }\textbf {\bibinfo {volume} {30}},\ \bibinfo {pages} {777} (\bibinfo {year} {1974})}\BibitemShut {NoStop}%
\bibitem [{\citenamefont {Dressel}\ \emph {et~al.}(2014)\citenamefont {Dressel}, \citenamefont {Reppe}, \citenamefont {Prehm}, \citenamefont {Brautzsch},\ and\ \citenamefont {Tschierske}}]{dressel2014chiral}%
  \BibitemOpen
  \bibfield  {author} {\bibinfo {author} {\bibfnamefont {C.}~\bibnamefont {Dressel}}, \bibinfo {author} {\bibfnamefont {T.}~\bibnamefont {Reppe}}, \bibinfo {author} {\bibfnamefont {M.}~\bibnamefont {Prehm}}, \bibinfo {author} {\bibfnamefont {M.}~\bibnamefont {Brautzsch}},\ and\ \bibinfo {author} {\bibfnamefont {C.}~\bibnamefont {Tschierske}},\ }\bibfield  {title} {\bibinfo {title} {Chiral self-sorting and amplification in isotropic liquids of achiral molecules},\ }\href@noop {} {\bibfield  {journal} {\bibinfo  {journal} {Nature chemistry}\ }\textbf {\bibinfo {volume} {6}},\ \bibinfo {pages} {971} (\bibinfo {year} {2014})}\BibitemShut {NoStop}%
\bibitem [{\citenamefont {Agarwala}\ \emph {et~al.}(2020)\citenamefont {Agarwala}, \citenamefont {Juri{\v{c}}i{\'c}},\ and\ \citenamefont {Roy}}]{agarwala2020higher}%
  \BibitemOpen
  \bibfield  {author} {\bibinfo {author} {\bibfnamefont {A.}~\bibnamefont {Agarwala}}, \bibinfo {author} {\bibfnamefont {V.}~\bibnamefont {Juri{\v{c}}i{\'c}}},\ and\ \bibinfo {author} {\bibfnamefont {B.}~\bibnamefont {Roy}},\ }\bibfield  {title} {\bibinfo {title} {Higher-order topological insulators in amorphous solids},\ }\href@noop {} {\bibfield  {journal} {\bibinfo  {journal} {Physical Review Research}\ }\textbf {\bibinfo {volume} {2}},\ \bibinfo {pages} {012067} (\bibinfo {year} {2020})}\BibitemShut {NoStop}%
\bibitem [{\citenamefont {Bernal}(1964)}]{bernal1964bakerian}%
  \BibitemOpen
  \bibfield  {author} {\bibinfo {author} {\bibfnamefont {J.~D.}\ \bibnamefont {Bernal}},\ }\bibfield  {title} {\bibinfo {title} {The bakerian lecture, 1962. the structure of liquids},\ }\href@noop {} {\bibfield  {journal} {\bibinfo  {journal} {Proceedings of the Royal Society of London. Series A, Mathematical and Physical Sciences}\ }\textbf {\bibinfo {volume} {280}},\ \bibinfo {pages} {299} (\bibinfo {year} {1964})}\BibitemShut {NoStop}%
\bibitem [{\citenamefont {Behler}\ and\ \citenamefont {Parrinello}(2007)}]{behler2007generalized}%
  \BibitemOpen
  \bibfield  {author} {\bibinfo {author} {\bibfnamefont {J.}~\bibnamefont {Behler}}\ and\ \bibinfo {author} {\bibfnamefont {M.}~\bibnamefont {Parrinello}},\ }\bibfield  {title} {\bibinfo {title} {Generalized neural-network representation of high-dimensional potential-energy surfaces},\ }\href@noop {} {\bibfield  {journal} {\bibinfo  {journal} {Physical review letters}\ }\textbf {\bibinfo {volume} {98}},\ \bibinfo {pages} {146401} (\bibinfo {year} {2007})}\BibitemShut {NoStop}%
\bibitem [{\citenamefont {Behler}(2011)}]{behler2011atom}%
  \BibitemOpen
  \bibfield  {author} {\bibinfo {author} {\bibfnamefont {J.}~\bibnamefont {Behler}},\ }\bibfield  {title} {\bibinfo {title} {Atom-centered symmetry functions for constructing high-dimensional neural network potentials},\ }\href@noop {} {\bibfield  {journal} {\bibinfo  {journal} {The Journal of chemical physics}\ }\textbf {\bibinfo {volume} {134}} (\bibinfo {year} {2011})}\BibitemShut {NoStop}%
\bibitem [{\citenamefont {Bart{\'o}k}\ \emph {et~al.}(2013)\citenamefont {Bart{\'o}k}, \citenamefont {Kondor},\ and\ \citenamefont {Cs{\'a}nyi}}]{bartok2013representing}%
  \BibitemOpen
  \bibfield  {author} {\bibinfo {author} {\bibfnamefont {A.~P.}\ \bibnamefont {Bart{\'o}k}}, \bibinfo {author} {\bibfnamefont {R.}~\bibnamefont {Kondor}},\ and\ \bibinfo {author} {\bibfnamefont {G.}~\bibnamefont {Cs{\'a}nyi}},\ }\bibfield  {title} {\bibinfo {title} {On representing chemical environments},\ }\href@noop {} {\bibfield  {journal} {\bibinfo  {journal} {Physical Review B—Condensed Matter and Materials Physics}\ }\textbf {\bibinfo {volume} {87}},\ \bibinfo {pages} {184115} (\bibinfo {year} {2013})}\BibitemShut {NoStop}%
\bibitem [{\citenamefont {Landau}(1936)}]{landau1936theory}%
  \BibitemOpen
  \bibfield  {author} {\bibinfo {author} {\bibfnamefont {L.}~\bibnamefont {Landau}},\ }\bibfield  {title} {\bibinfo {title} {The theory of phase transitions},\ }\href@noop {} {\bibfield  {journal} {\bibinfo  {journal} {Nature}\ }\textbf {\bibinfo {volume} {138}},\ \bibinfo {pages} {840} (\bibinfo {year} {1936})}\BibitemShut {NoStop}%
\bibitem [{\citenamefont {Landau}\ \emph {et~al.}(1937)\citenamefont {Landau} \emph {et~al.}}]{landau1937theory}%
  \BibitemOpen
  \bibfield  {author} {\bibinfo {author} {\bibfnamefont {L.~D.}\ \bibnamefont {Landau}} \emph {et~al.},\ }\bibfield  {title} {\bibinfo {title} {On the theory of phase transitions},\ }\href@noop {} {\bibfield  {journal} {\bibinfo  {journal} {Zh. eksp. teor. Fiz}\ }\textbf {\bibinfo {volume} {7}},\ \bibinfo {pages} {926} (\bibinfo {year} {1937})}\BibitemShut {NoStop}%
\bibitem [{\citenamefont {Steinhardt}\ \emph {et~al.}(1983)\citenamefont {Steinhardt}, \citenamefont {Nelson},\ and\ \citenamefont {Ronchetti}}]{steinhardt1983bond}%
  \BibitemOpen
  \bibfield  {author} {\bibinfo {author} {\bibfnamefont {P.~J.}\ \bibnamefont {Steinhardt}}, \bibinfo {author} {\bibfnamefont {D.~R.}\ \bibnamefont {Nelson}},\ and\ \bibinfo {author} {\bibfnamefont {M.}~\bibnamefont {Ronchetti}},\ }\bibfield  {title} {\bibinfo {title} {Bond-orientational order in liquids and glasses},\ }\href@noop {} {\bibfield  {journal} {\bibinfo  {journal} {Physical Review B}\ }\textbf {\bibinfo {volume} {28}},\ \bibinfo {pages} {784} (\bibinfo {year} {1983})}\BibitemShut {NoStop}%
\bibitem [{\citenamefont {Ten~Wolde}\ \emph {et~al.}(1995)\citenamefont {Ten~Wolde}, \citenamefont {Ruiz-Montero},\ and\ \citenamefont {Frenkel}}]{ten1995numerical}%
  \BibitemOpen
  \bibfield  {author} {\bibinfo {author} {\bibfnamefont {P.~R.}\ \bibnamefont {Ten~Wolde}}, \bibinfo {author} {\bibfnamefont {M.~J.}\ \bibnamefont {Ruiz-Montero}},\ and\ \bibinfo {author} {\bibfnamefont {D.}~\bibnamefont {Frenkel}},\ }\bibfield  {title} {\bibinfo {title} {Numerical evidence for bcc ordering at the surface of a critical fcc nucleus},\ }\href@noop {} {\bibfield  {journal} {\bibinfo  {journal} {Physical review letters}\ }\textbf {\bibinfo {volume} {75}},\ \bibinfo {pages} {2714} (\bibinfo {year} {1995})}\BibitemShut {NoStop}%
\bibitem [{\citenamefont {Zaccone}\ and\ \citenamefont {Scossa-Romano}(2011)}]{zaccone2011approximate}%
  \BibitemOpen
  \bibfield  {author} {\bibinfo {author} {\bibfnamefont {A.}~\bibnamefont {Zaccone}}\ and\ \bibinfo {author} {\bibfnamefont {E.}~\bibnamefont {Scossa-Romano}},\ }\bibfield  {title} {\bibinfo {title} {Approximate analytical description of the nonaffine response of amorphous solids},\ }\href@noop {} {\bibfield  {journal} {\bibinfo  {journal} {Physical Review B—Condensed Matter and Materials Physics}\ }\textbf {\bibinfo {volume} {83}},\ \bibinfo {pages} {184205} (\bibinfo {year} {2011})}\BibitemShut {NoStop}%
\bibitem [{\citenamefont {Milkus}\ and\ \citenamefont {Zaccone}(2016)}]{milkus2016local}%
  \BibitemOpen
  \bibfield  {author} {\bibinfo {author} {\bibfnamefont {R.}~\bibnamefont {Milkus}}\ and\ \bibinfo {author} {\bibfnamefont {A.}~\bibnamefont {Zaccone}},\ }\bibfield  {title} {\bibinfo {title} {Local inversion-symmetry breaking controls the boson peak in glasses and crystals},\ }\href@noop {} {\bibfield  {journal} {\bibinfo  {journal} {Physical Review B}\ }\textbf {\bibinfo {volume} {93}},\ \bibinfo {pages} {094204} (\bibinfo {year} {2016})}\BibitemShut {NoStop}%
\bibitem [{\citenamefont {Liu}\ \emph {et~al.}(2022)\citenamefont {Liu}, \citenamefont {B{\o}jesen}, \citenamefont {Tabor}, \citenamefont {Mudie}, \citenamefont {Zaccone}, \citenamefont {Harrowell},\ and\ \citenamefont {Petersen}}]{liu2022local}%
  \BibitemOpen
  \bibfield  {author} {\bibinfo {author} {\bibfnamefont {A.~C.}\ \bibnamefont {Liu}}, \bibinfo {author} {\bibfnamefont {E.~D.}\ \bibnamefont {B{\o}jesen}}, \bibinfo {author} {\bibfnamefont {R.~F.}\ \bibnamefont {Tabor}}, \bibinfo {author} {\bibfnamefont {S.~T.}\ \bibnamefont {Mudie}}, \bibinfo {author} {\bibfnamefont {A.}~\bibnamefont {Zaccone}}, \bibinfo {author} {\bibfnamefont {P.}~\bibnamefont {Harrowell}},\ and\ \bibinfo {author} {\bibfnamefont {T.~C.}\ \bibnamefont {Petersen}},\ }\bibfield  {title} {\bibinfo {title} {Local symmetry predictors of mechanical stability in glasses},\ }\href@noop {} {\bibfield  {journal} {\bibinfo  {journal} {Science Advances}\ }\textbf {\bibinfo {volume} {8}},\ \bibinfo {pages} {eabn0681} (\bibinfo {year} {2022})}\BibitemShut {NoStop}%
\bibitem [{\citenamefont {Zaccone}(2023)}]{zaccone2023theory}%
  \BibitemOpen
  \bibfield  {author} {\bibinfo {author} {\bibfnamefont {A.}~\bibnamefont {Zaccone}},\ }\bibfield  {title} {\bibinfo {title} {Theory of disordered solids},\ }\href@noop {} {\bibfield  {journal} {\bibinfo  {journal} {Cham: Springer.[Google Scholar]}\ } (\bibinfo {year} {2023})}\BibitemShut {NoStop}%
\bibitem [{\citenamefont {Patterson}(1934)}]{patterson1934fourier}%
  \BibitemOpen
  \bibfield  {author} {\bibinfo {author} {\bibfnamefont {A.~L.}\ \bibnamefont {Patterson}},\ }\bibfield  {title} {\bibinfo {title} {A fourier series method for the determination of the components of interatomic distances in crystals},\ }\href@noop {} {\bibfield  {journal} {\bibinfo  {journal} {Physical Review}\ }\textbf {\bibinfo {volume} {46}},\ \bibinfo {pages} {372} (\bibinfo {year} {1934})}\BibitemShut {NoStop}%
\bibitem [{\citenamefont {Sheng}\ \emph {et~al.}(2006)\citenamefont {Sheng}, \citenamefont {Luo}, \citenamefont {Alamgir}, \citenamefont {Bai},\ and\ \citenamefont {Ma}}]{sheng2006atomic}%
  \BibitemOpen
  \bibfield  {author} {\bibinfo {author} {\bibfnamefont {H.}~\bibnamefont {Sheng}}, \bibinfo {author} {\bibfnamefont {W.}~\bibnamefont {Luo}}, \bibinfo {author} {\bibfnamefont {F.}~\bibnamefont {Alamgir}}, \bibinfo {author} {\bibfnamefont {J.}~\bibnamefont {Bai}},\ and\ \bibinfo {author} {\bibfnamefont {E.}~\bibnamefont {Ma}},\ }\bibfield  {title} {\bibinfo {title} {Atomic packing and short-to-medium-range order in metallic glasses},\ }\href@noop {} {\bibfield  {journal} {\bibinfo  {journal} {Nature}\ }\textbf {\bibinfo {volume} {439}},\ \bibinfo {pages} {419} (\bibinfo {year} {2006})}\BibitemShut {NoStop}%
\bibitem [{\citenamefont {Bates}\ and\ \citenamefont {Fredrickson}(1999)}]{bates1999block}%
  \BibitemOpen
  \bibfield  {author} {\bibinfo {author} {\bibfnamefont {F.~S.}\ \bibnamefont {Bates}}\ and\ \bibinfo {author} {\bibfnamefont {G.~H.}\ \bibnamefont {Fredrickson}},\ }\bibfield  {title} {\bibinfo {title} {Block copolymers—designer soft materials},\ }\href@noop {} {\bibfield  {journal} {\bibinfo  {journal} {Physics today}\ }\textbf {\bibinfo {volume} {52}},\ \bibinfo {pages} {32} (\bibinfo {year} {1999})}\BibitemShut {NoStop}%
\bibitem [{\citenamefont {Thorpe}(1983)}]{thorpe1983continuous}%
  \BibitemOpen
  \bibfield  {author} {\bibinfo {author} {\bibfnamefont {M.~F.}\ \bibnamefont {Thorpe}},\ }\bibfield  {title} {\bibinfo {title} {Continuous deformations in random networks},\ }\href@noop {} {\bibfield  {journal} {\bibinfo  {journal} {Journal of Non-Crystalline Solids}\ }\textbf {\bibinfo {volume} {57}},\ \bibinfo {pages} {355} (\bibinfo {year} {1983})}\BibitemShut {NoStop}%
\bibitem [{\citenamefont {Frank}(1952)}]{frank1952supercooling}%
  \BibitemOpen
  \bibfield  {author} {\bibinfo {author} {\bibfnamefont {F.~C.}\ \bibnamefont {Frank}},\ }\bibfield  {title} {\bibinfo {title} {Supercooling of liquids},\ }\href@noop {} {\bibfield  {journal} {\bibinfo  {journal} {Proceedings of the Royal Society of London. Series A. Mathematical and Physical Sciences}\ }\textbf {\bibinfo {volume} {215}},\ \bibinfo {pages} {43} (\bibinfo {year} {1952})}\BibitemShut {NoStop}%
\bibitem [{\citenamefont {Finney}(1970)}]{finney1970random}%
  \BibitemOpen
  \bibfield  {author} {\bibinfo {author} {\bibfnamefont {J.}~\bibnamefont {Finney}},\ }\bibfield  {title} {\bibinfo {title} {Random packings and the structure of simple liquids. i. the geometry of random close packing},\ }\href@noop {} {\bibfield  {journal} {\bibinfo  {journal} {Proceedings of the Royal Society of London. A. Mathematical and Physical Sciences}\ }\textbf {\bibinfo {volume} {319}},\ \bibinfo {pages} {479} (\bibinfo {year} {1970})}\BibitemShut {NoStop}%
\bibitem [{\citenamefont {Bak}\ \emph {et~al.}(1987)\citenamefont {Bak}, \citenamefont {Tang},\ and\ \citenamefont {Wiesenfeld}}]{bak1987self}%
  \BibitemOpen
  \bibfield  {author} {\bibinfo {author} {\bibfnamefont {P.}~\bibnamefont {Bak}}, \bibinfo {author} {\bibfnamefont {C.}~\bibnamefont {Tang}},\ and\ \bibinfo {author} {\bibfnamefont {K.}~\bibnamefont {Wiesenfeld}},\ }\bibfield  {title} {\bibinfo {title} {Self-organized criticality: An explanation of the 1/f noise},\ }\href@noop {} {\bibfield  {journal} {\bibinfo  {journal} {Physical review letters}\ }\textbf {\bibinfo {volume} {59}},\ \bibinfo {pages} {381} (\bibinfo {year} {1987})}\BibitemShut {NoStop}%
\bibitem [{\citenamefont {Olami}\ \emph {et~al.}(1992)\citenamefont {Olami}, \citenamefont {Feder},\ and\ \citenamefont {Christensen}}]{olami1992self}%
  \BibitemOpen
  \bibfield  {author} {\bibinfo {author} {\bibfnamefont {Z.}~\bibnamefont {Olami}}, \bibinfo {author} {\bibfnamefont {H.~J.~S.}\ \bibnamefont {Feder}},\ and\ \bibinfo {author} {\bibfnamefont {K.}~\bibnamefont {Christensen}},\ }\bibfield  {title} {\bibinfo {title} {Self-organized criticality in a continuous, nonconservative cellular automaton modeling earthquakes},\ }\href@noop {} {\bibfield  {journal} {\bibinfo  {journal} {Physical review letters}\ }\textbf {\bibinfo {volume} {68}},\ \bibinfo {pages} {1244} (\bibinfo {year} {1992})}\BibitemShut {NoStop}%
\bibitem [{\citenamefont {Christensen}\ and\ \citenamefont {Moloney}(2005)}]{christensen2005complexity}%
  \BibitemOpen
  \bibfield  {author} {\bibinfo {author} {\bibfnamefont {K.}~\bibnamefont {Christensen}}\ and\ \bibinfo {author} {\bibfnamefont {N.~R.}\ \bibnamefont {Moloney}},\ }\href@noop {} {\emph {\bibinfo {title} {Complexity and criticality}}},\ Vol.~\bibinfo {volume} {1}\ (\bibinfo  {publisher} {World Scientific Publishing Company},\ \bibinfo {year} {2005})\BibitemShut {NoStop}%
\bibitem [{\citenamefont {Anderson}(1972)}]{anderson1972more}%
  \BibitemOpen
  \bibfield  {author} {\bibinfo {author} {\bibfnamefont {P.~W.}\ \bibnamefont {Anderson}},\ }\bibfield  {title} {\bibinfo {title} {More is different: Broken symmetry and the nature of the hierarchical structure of science.},\ }\href@noop {} {\bibfield  {journal} {\bibinfo  {journal} {Science}\ }\textbf {\bibinfo {volume} {177}},\ \bibinfo {pages} {393} (\bibinfo {year} {1972})}\BibitemShut {NoStop}%
\bibitem [{\citenamefont {Carleo}\ \emph {et~al.}(2019)\citenamefont {Carleo}, \citenamefont {Cirac}, \citenamefont {Cranmer}, \citenamefont {Daudet}, \citenamefont {Schuld}, \citenamefont {Tishby}, \citenamefont {Vogt-Maranto},\ and\ \citenamefont {Zdeborov{\'a}}}]{carleo2019machine}%
  \BibitemOpen
  \bibfield  {author} {\bibinfo {author} {\bibfnamefont {G.}~\bibnamefont {Carleo}}, \bibinfo {author} {\bibfnamefont {I.}~\bibnamefont {Cirac}}, \bibinfo {author} {\bibfnamefont {K.}~\bibnamefont {Cranmer}}, \bibinfo {author} {\bibfnamefont {L.}~\bibnamefont {Daudet}}, \bibinfo {author} {\bibfnamefont {M.}~\bibnamefont {Schuld}}, \bibinfo {author} {\bibfnamefont {N.}~\bibnamefont {Tishby}}, \bibinfo {author} {\bibfnamefont {L.}~\bibnamefont {Vogt-Maranto}},\ and\ \bibinfo {author} {\bibfnamefont {L.}~\bibnamefont {Zdeborov{\'a}}},\ }\bibfield  {title} {\bibinfo {title} {Machine learning and the physical sciences},\ }\href@noop {} {\bibfield  {journal} {\bibinfo  {journal} {Reviews of Modern Physics}\ }\textbf {\bibinfo {volume} {91}},\ \bibinfo {pages} {045002} (\bibinfo {year} {2019})}\BibitemShut {NoStop}%
\bibitem [{\citenamefont {Mehta}\ \emph {et~al.}(2019)\citenamefont {Mehta}, \citenamefont {Bukov}, \citenamefont {Wang}, \citenamefont {Day}, \citenamefont {Richardson}, \citenamefont {Fisher},\ and\ \citenamefont {Schwab}}]{mehta2019high}%
  \BibitemOpen
  \bibfield  {author} {\bibinfo {author} {\bibfnamefont {P.}~\bibnamefont {Mehta}}, \bibinfo {author} {\bibfnamefont {M.}~\bibnamefont {Bukov}}, \bibinfo {author} {\bibfnamefont {C.-H.}\ \bibnamefont {Wang}}, \bibinfo {author} {\bibfnamefont {A.~G.}\ \bibnamefont {Day}}, \bibinfo {author} {\bibfnamefont {C.}~\bibnamefont {Richardson}}, \bibinfo {author} {\bibfnamefont {C.~K.}\ \bibnamefont {Fisher}},\ and\ \bibinfo {author} {\bibfnamefont {D.~J.}\ \bibnamefont {Schwab}},\ }\bibfield  {title} {\bibinfo {title} {A high-bias, low-variance introduction to machine learning for physicists},\ }\href@noop {} {\bibfield  {journal} {\bibinfo  {journal} {Physics reports}\ }\textbf {\bibinfo {volume} {810}},\ \bibinfo {pages} {1} (\bibinfo {year} {2019})}\BibitemShut {NoStop}%
\bibitem [{\citenamefont {Radovic}\ \emph {et~al.}(2018)\citenamefont {Radovic}, \citenamefont {Williams}, \citenamefont {Rousseau}, \citenamefont {Kagan}, \citenamefont {Bonacorsi}, \citenamefont {Himmel}, \citenamefont {Aurisano}, \citenamefont {Terao},\ and\ \citenamefont {Wongjirad}}]{radovic2018machine}%
  \BibitemOpen
  \bibfield  {author} {\bibinfo {author} {\bibfnamefont {A.}~\bibnamefont {Radovic}}, \bibinfo {author} {\bibfnamefont {M.}~\bibnamefont {Williams}}, \bibinfo {author} {\bibfnamefont {D.}~\bibnamefont {Rousseau}}, \bibinfo {author} {\bibfnamefont {M.}~\bibnamefont {Kagan}}, \bibinfo {author} {\bibfnamefont {D.}~\bibnamefont {Bonacorsi}}, \bibinfo {author} {\bibfnamefont {A.}~\bibnamefont {Himmel}}, \bibinfo {author} {\bibfnamefont {A.}~\bibnamefont {Aurisano}}, \bibinfo {author} {\bibfnamefont {K.}~\bibnamefont {Terao}},\ and\ \bibinfo {author} {\bibfnamefont {T.}~\bibnamefont {Wongjirad}},\ }\bibfield  {title} {\bibinfo {title} {Machine learning at the energy and intensity frontiers of particle physics},\ }\href@noop {} {\bibfield  {journal} {\bibinfo  {journal} {Nature}\ }\textbf {\bibinfo {volume} {560}},\ \bibinfo {pages} {41} (\bibinfo {year} {2018})}\BibitemShut {NoStop}%
\bibitem [{\citenamefont {Carleo}\ and\ \citenamefont {Troyer}(2017)}]{carleo2017solving}%
  \BibitemOpen
  \bibfield  {author} {\bibinfo {author} {\bibfnamefont {G.}~\bibnamefont {Carleo}}\ and\ \bibinfo {author} {\bibfnamefont {M.}~\bibnamefont {Troyer}},\ }\bibfield  {title} {\bibinfo {title} {Solving the quantum many-body problem with artificial neural networks},\ }\href@noop {} {\bibfield  {journal} {\bibinfo  {journal} {Science}\ }\textbf {\bibinfo {volume} {355}},\ \bibinfo {pages} {602} (\bibinfo {year} {2017})}\BibitemShut {NoStop}%
\bibitem [{\citenamefont {Scarselli}\ \emph {et~al.}(2008)\citenamefont {Scarselli}, \citenamefont {Gori}, \citenamefont {Tsoi}, \citenamefont {Hagenbuchner},\ and\ \citenamefont {Monfardini}}]{scarselli2008graph}%
  \BibitemOpen
  \bibfield  {author} {\bibinfo {author} {\bibfnamefont {F.}~\bibnamefont {Scarselli}}, \bibinfo {author} {\bibfnamefont {M.}~\bibnamefont {Gori}}, \bibinfo {author} {\bibfnamefont {A.~C.}\ \bibnamefont {Tsoi}}, \bibinfo {author} {\bibfnamefont {M.}~\bibnamefont {Hagenbuchner}},\ and\ \bibinfo {author} {\bibfnamefont {G.}~\bibnamefont {Monfardini}},\ }\bibfield  {title} {\bibinfo {title} {The graph neural network model},\ }\href@noop {} {\bibfield  {journal} {\bibinfo  {journal} {IEEE transactions on neural networks}\ }\textbf {\bibinfo {volume} {20}},\ \bibinfo {pages} {61} (\bibinfo {year} {2008})}\BibitemShut {NoStop}%
\bibitem [{\citenamefont {Raissi}\ \emph {et~al.}(2019)\citenamefont {Raissi}, \citenamefont {Perdikaris},\ and\ \citenamefont {Karniadakis}}]{raissi2019physics}%
  \BibitemOpen
  \bibfield  {author} {\bibinfo {author} {\bibfnamefont {M.}~\bibnamefont {Raissi}}, \bibinfo {author} {\bibfnamefont {P.}~\bibnamefont {Perdikaris}},\ and\ \bibinfo {author} {\bibfnamefont {G.~E.}\ \bibnamefont {Karniadakis}},\ }\bibfield  {title} {\bibinfo {title} {Physics-informed neural networks: A deep learning framework for solving forward and inverse problems involving nonlinear partial differential equations},\ }\href@noop {} {\bibfield  {journal} {\bibinfo  {journal} {Journal of Computational physics}\ }\textbf {\bibinfo {volume} {378}},\ \bibinfo {pages} {686} (\bibinfo {year} {2019})}\BibitemShut {NoStop}%
\bibitem [{\citenamefont {Bapst}\ \emph {et~al.}(2020)\citenamefont {Bapst}, \citenamefont {Keck}, \citenamefont {Grabska-Barwi{\'n}ska}, \citenamefont {Donner}, \citenamefont {Cubuk}, \citenamefont {Schoenholz}, \citenamefont {Obika}, \citenamefont {Nelson}, \citenamefont {Back}, \citenamefont {Hassabis} \emph {et~al.}}]{bapst2020unveiling}%
  \BibitemOpen
  \bibfield  {author} {\bibinfo {author} {\bibfnamefont {V.}~\bibnamefont {Bapst}}, \bibinfo {author} {\bibfnamefont {T.}~\bibnamefont {Keck}}, \bibinfo {author} {\bibfnamefont {A.}~\bibnamefont {Grabska-Barwi{\'n}ska}}, \bibinfo {author} {\bibfnamefont {C.}~\bibnamefont {Donner}}, \bibinfo {author} {\bibfnamefont {E.~D.}\ \bibnamefont {Cubuk}}, \bibinfo {author} {\bibfnamefont {S.~S.}\ \bibnamefont {Schoenholz}}, \bibinfo {author} {\bibfnamefont {A.}~\bibnamefont {Obika}}, \bibinfo {author} {\bibfnamefont {A.~W.}\ \bibnamefont {Nelson}}, \bibinfo {author} {\bibfnamefont {T.}~\bibnamefont {Back}}, \bibinfo {author} {\bibfnamefont {D.}~\bibnamefont {Hassabis}}, \emph {et~al.},\ }\bibfield  {title} {\bibinfo {title} {Unveiling the predictive power of static structure in glassy systems},\ }\href@noop {} {\bibfield  {journal} {\bibinfo  {journal} {Nature physics}\ }\textbf {\bibinfo {volume} {16}},\ \bibinfo {pages} {448} (\bibinfo {year} {2020})}\BibitemShut {NoStop}%
\bibitem [{\citenamefont {Munkres}(2018)}]{munkres2018elements}%
  \BibitemOpen
  \bibfield  {author} {\bibinfo {author} {\bibfnamefont {J.~R.}\ \bibnamefont {Munkres}},\ }\href@noop {} {\emph {\bibinfo {title} {Elements of algebraic topology}}}\ (\bibinfo  {publisher} {CRC press},\ \bibinfo {year} {2018})\BibitemShut {NoStop}%
\bibitem [{\citenamefont {Edelsbrunner}\ \emph {et~al.}(2002)\citenamefont {Edelsbrunner}, \citenamefont {Letscher},\ and\ \citenamefont {Zomorodian}}]{edelsbrunner2002topological}%
  \BibitemOpen
  \bibfield  {author} {\bibinfo {author} {\bibnamefont {Edelsbrunner}}, \bibinfo {author} {\bibnamefont {Letscher}},\ and\ \bibinfo {author} {\bibnamefont {Zomorodian}},\ }\bibfield  {title} {\bibinfo {title} {Topological persistence and simplification},\ }\href@noop {} {\bibfield  {journal} {\bibinfo  {journal} {Discrete \& computational geometry}\ }\textbf {\bibinfo {volume} {28}},\ \bibinfo {pages} {511} (\bibinfo {year} {2002})}\BibitemShut {NoStop}%
\bibitem [{\citenamefont {Zomorodian}\ and\ \citenamefont {Carlsson}(2004)}]{zomorodian2004computing}%
  \BibitemOpen
  \bibfield  {author} {\bibinfo {author} {\bibfnamefont {A.}~\bibnamefont {Zomorodian}}\ and\ \bibinfo {author} {\bibfnamefont {G.}~\bibnamefont {Carlsson}},\ }\bibfield  {title} {\bibinfo {title} {Computing persistent homology},\ }in\ \href@noop {} {\emph {\bibinfo {booktitle} {Proceedings of the twentieth annual symposium on Computational geometry}}}\ (\bibinfo {year} {2004})\ pp.\ \bibinfo {pages} {347--356}\BibitemShut {NoStop}%
\bibitem [{\citenamefont {Ghrist}(2008)}]{ghrist2008barcodes}%
  \BibitemOpen
  \bibfield  {author} {\bibinfo {author} {\bibfnamefont {R.}~\bibnamefont {Ghrist}},\ }\bibfield  {title} {\bibinfo {title} {Barcodes: the persistent topology of data},\ }\href@noop {} {\bibfield  {journal} {\bibinfo  {journal} {Bulletin of the American Mathematical Society}\ }\textbf {\bibinfo {volume} {45}},\ \bibinfo {pages} {61} (\bibinfo {year} {2008})}\BibitemShut {NoStop}%
\bibitem [{\citenamefont {Xia}\ \emph {et~al.}(2019)\citenamefont {Xia}, \citenamefont {Anand}, \citenamefont {Shikhar},\ and\ \citenamefont {Mu}}]{xia2019persistent}%
  \BibitemOpen
  \bibfield  {author} {\bibinfo {author} {\bibfnamefont {K.}~\bibnamefont {Xia}}, \bibinfo {author} {\bibfnamefont {D.~V.}\ \bibnamefont {Anand}}, \bibinfo {author} {\bibfnamefont {S.}~\bibnamefont {Shikhar}},\ and\ \bibinfo {author} {\bibfnamefont {Y.}~\bibnamefont {Mu}},\ }\bibfield  {title} {\bibinfo {title} {Persistent homology analysis of osmolyte molecular aggregation and their hydrogen-bonding networks},\ }\href@noop {} {\bibfield  {journal} {\bibinfo  {journal} {Physical Chemistry Chemical Physics}\ }\textbf {\bibinfo {volume} {21}},\ \bibinfo {pages} {21038} (\bibinfo {year} {2019})}\BibitemShut {NoStop}%
\bibitem [{\citenamefont {Chen}\ \emph {et~al.}(2020)\citenamefont {Chen}, \citenamefont {Chen}, \citenamefont {Weng}, \citenamefont {Jiang}, \citenamefont {Wei},\ and\ \citenamefont {Pan}}]{chen2020topology}%
  \BibitemOpen
  \bibfield  {author} {\bibinfo {author} {\bibfnamefont {X.}~\bibnamefont {Chen}}, \bibinfo {author} {\bibfnamefont {D.}~\bibnamefont {Chen}}, \bibinfo {author} {\bibfnamefont {M.}~\bibnamefont {Weng}}, \bibinfo {author} {\bibfnamefont {Y.}~\bibnamefont {Jiang}}, \bibinfo {author} {\bibfnamefont {G.-W.}\ \bibnamefont {Wei}},\ and\ \bibinfo {author} {\bibfnamefont {F.}~\bibnamefont {Pan}},\ }\bibfield  {title} {\bibinfo {title} {Topology-based machine learning strategy for cluster structure prediction},\ }\href@noop {} {\bibfield  {journal} {\bibinfo  {journal} {The journal of physical chemistry letters}\ }\textbf {\bibinfo {volume} {11}},\ \bibinfo {pages} {4392} (\bibinfo {year} {2020})}\BibitemShut {NoStop}%
\bibitem [{\citenamefont {Anand}\ \emph {et~al.}(2020)\citenamefont {Anand}, \citenamefont {Meng}, \citenamefont {Xia},\ and\ \citenamefont {Mu}}]{anand2020weighted}%
  \BibitemOpen
  \bibfield  {author} {\bibinfo {author} {\bibfnamefont {D.~V.}\ \bibnamefont {Anand}}, \bibinfo {author} {\bibfnamefont {Z.}~\bibnamefont {Meng}}, \bibinfo {author} {\bibfnamefont {K.}~\bibnamefont {Xia}},\ and\ \bibinfo {author} {\bibfnamefont {Y.}~\bibnamefont {Mu}},\ }\bibfield  {title} {\bibinfo {title} {Weighted persistent homology for osmolyte molecular aggregation and hydrogen-bonding network analysis},\ }\href@noop {} {\bibfield  {journal} {\bibinfo  {journal} {Scientific reports}\ }\textbf {\bibinfo {volume} {10}},\ \bibinfo {pages} {9685} (\bibinfo {year} {2020})}\BibitemShut {NoStop}%
\bibitem [{\citenamefont {Hiraoka}\ \emph {et~al.}(2016)\citenamefont {Hiraoka}, \citenamefont {Nakamura}, \citenamefont {Hirata}, \citenamefont {Escolar}, \citenamefont {Matsue},\ and\ \citenamefont {Nishiura}}]{hiraoka2016hierarchical}%
  \BibitemOpen
  \bibfield  {author} {\bibinfo {author} {\bibfnamefont {Y.}~\bibnamefont {Hiraoka}}, \bibinfo {author} {\bibfnamefont {T.}~\bibnamefont {Nakamura}}, \bibinfo {author} {\bibfnamefont {A.}~\bibnamefont {Hirata}}, \bibinfo {author} {\bibfnamefont {E.~G.}\ \bibnamefont {Escolar}}, \bibinfo {author} {\bibfnamefont {K.}~\bibnamefont {Matsue}},\ and\ \bibinfo {author} {\bibfnamefont {Y.}~\bibnamefont {Nishiura}},\ }\bibfield  {title} {\bibinfo {title} {Hierarchical structures of amorphous solids characterized by persistent homology},\ }\href@noop {} {\bibfield  {journal} {\bibinfo  {journal} {Proceedings of the National Academy of Sciences}\ }\textbf {\bibinfo {volume} {113}},\ \bibinfo {pages} {7035} (\bibinfo {year} {2016})}\BibitemShut {NoStop}%
\bibitem [{\citenamefont {Kusano}\ \emph {et~al.}(2016)\citenamefont {Kusano}, \citenamefont {Hiraoka},\ and\ \citenamefont {Fukumizu}}]{kusano2016persistence}%
  \BibitemOpen
  \bibfield  {author} {\bibinfo {author} {\bibfnamefont {G.}~\bibnamefont {Kusano}}, \bibinfo {author} {\bibfnamefont {Y.}~\bibnamefont {Hiraoka}},\ and\ \bibinfo {author} {\bibfnamefont {K.}~\bibnamefont {Fukumizu}},\ }\bibfield  {title} {\bibinfo {title} {Persistence weighted gaussian kernel for topological data analysis},\ }in\ \href@noop {} {\emph {\bibinfo {booktitle} {International conference on machine learning}}}\ (\bibinfo {organization} {PMLR},\ \bibinfo {year} {2016})\ pp.\ \bibinfo {pages} {2004--2013}\BibitemShut {NoStop}%
\bibitem [{\citenamefont {Minamitani}\ \emph {et~al.}(2023)\citenamefont {Minamitani}, \citenamefont {Obayashi}, \citenamefont {Shimizu},\ and\ \citenamefont {Watanabe}}]{minamitani2023persistent}%
  \BibitemOpen
  \bibfield  {author} {\bibinfo {author} {\bibfnamefont {E.}~\bibnamefont {Minamitani}}, \bibinfo {author} {\bibfnamefont {I.}~\bibnamefont {Obayashi}}, \bibinfo {author} {\bibfnamefont {K.}~\bibnamefont {Shimizu}},\ and\ \bibinfo {author} {\bibfnamefont {S.}~\bibnamefont {Watanabe}},\ }\bibfield  {title} {\bibinfo {title} {Persistent homology-based descriptor for machine-learning potential of amorphous structures},\ }\href@noop {} {\bibfield  {journal} {\bibinfo  {journal} {The Journal of Chemical Physics}\ }\textbf {\bibinfo {volume} {159}} (\bibinfo {year} {2023})}\BibitemShut {NoStop}%
\bibitem [{\citenamefont {Poincar{\'e}}(1899)}]{poincare1899complementa}%
  \BibitemOpen
  \bibfield  {author} {\bibinfo {author} {\bibfnamefont {H.}~\bibnamefont {Poincar{\'e}}},\ }\bibfield  {title} {\bibinfo {title} {Compl{\'e}menta l’analysis situs},\ }\href@noop {} {\bibfield  {journal} {\bibinfo  {journal} {Rendiconti del Circolo matematico di Palermo}\ }\textbf {\bibinfo {volume} {13}},\ \bibinfo {pages} {10} (\bibinfo {year} {1899})}\BibitemShut {NoStop}%
\bibitem [{\citenamefont {Veblen}(1912)}]{veblen1912application}%
  \BibitemOpen
  \bibfield  {author} {\bibinfo {author} {\bibfnamefont {O.}~\bibnamefont {Veblen}},\ }\bibfield  {title} {\bibinfo {title} {An application of modular equations in analysis situs},\ }\href@noop {} {\bibfield  {journal} {\bibinfo  {journal} {Annals of Mathematics}\ }\textbf {\bibinfo {volume} {14}},\ \bibinfo {pages} {86} (\bibinfo {year} {1912})}\BibitemShut {NoStop}%
\bibitem [{\citenamefont {Hausdorff}(1914)}]{hausdorff1914grundzuge}%
  \BibitemOpen
  \bibfield  {author} {\bibinfo {author} {\bibfnamefont {F.}~\bibnamefont {Hausdorff}},\ }\href@noop {} {\emph {\bibinfo {title} {Grundz{\"u}ge der mengenlehre}}},\ Vol.~\bibinfo {volume} {7}\ (\bibinfo  {publisher} {von Veit},\ \bibinfo {year} {1914})\BibitemShut {NoStop}%
\bibitem [{\citenamefont {Kuratowski}(1920)}]{kuratowski1920notion}%
  \BibitemOpen
  \bibfield  {author} {\bibinfo {author} {\bibfnamefont {K.}~\bibnamefont {Kuratowski}},\ }\bibfield  {title} {\bibinfo {title} {Sur la notion d'ensemble fini},\ }\href@noop {} {\bibfield  {journal} {\bibinfo  {journal} {Fundamenta Mathematicae}\ }\textbf {\bibinfo {volume} {1}},\ \bibinfo {pages} {129} (\bibinfo {year} {1920})}\BibitemShut {NoStop}%
\bibitem [{\citenamefont {Alexandroff}(1929)}]{alexandroff1929memoire}%
  \BibitemOpen
  \bibfield  {author} {\bibinfo {author} {\bibfnamefont {P.~S.}\ \bibnamefont {Alexandroff}},\ }\bibfield  {title} {\bibinfo {title} {M{\'e}moire sur les espaces topologiques compacts},\ }\href@noop {} {\bibfield  {journal} {\bibinfo  {journal} {Verh. Kon. Akad. van Weten. Te Amsterdam}\ }\textbf {\bibinfo {volume} {14}},\ \bibinfo {pages} {1} (\bibinfo {year} {1929})}\BibitemShut {NoStop}%
\bibitem [{\citenamefont {Adams}\ \emph {et~al.}(2017)\citenamefont {Adams}, \citenamefont {Emerson}, \citenamefont {Kirby}, \citenamefont {Neville}, \citenamefont {Peterson}, \citenamefont {Shipman}, \citenamefont {Chepushtanova}, \citenamefont {Hanson}, \citenamefont {Motta},\ and\ \citenamefont {Ziegelmeier}}]{adams2017persistence}%
  \BibitemOpen
  \bibfield  {author} {\bibinfo {author} {\bibfnamefont {H.}~\bibnamefont {Adams}}, \bibinfo {author} {\bibfnamefont {T.}~\bibnamefont {Emerson}}, \bibinfo {author} {\bibfnamefont {M.}~\bibnamefont {Kirby}}, \bibinfo {author} {\bibfnamefont {R.}~\bibnamefont {Neville}}, \bibinfo {author} {\bibfnamefont {C.}~\bibnamefont {Peterson}}, \bibinfo {author} {\bibfnamefont {P.}~\bibnamefont {Shipman}}, \bibinfo {author} {\bibfnamefont {S.}~\bibnamefont {Chepushtanova}}, \bibinfo {author} {\bibfnamefont {E.}~\bibnamefont {Hanson}}, \bibinfo {author} {\bibfnamefont {F.}~\bibnamefont {Motta}},\ and\ \bibinfo {author} {\bibfnamefont {L.}~\bibnamefont {Ziegelmeier}},\ }\bibfield  {title} {\bibinfo {title} {Persistence images: A stable vector representation of persistent homology},\ }\href@noop {} {\bibfield  {journal} {\bibinfo  {journal} {Journal of Machine Learning Research}\ }\textbf {\bibinfo {volume} {18}},\ \bibinfo {pages} {1} (\bibinfo {year} {2017})}\BibitemShut {NoStop}%
\bibitem [{\citenamefont {Betti}(1870)}]{betti1870sopra}%
  \BibitemOpen
  \bibfield  {author} {\bibinfo {author} {\bibfnamefont {E.}~\bibnamefont {Betti}},\ }\bibfield  {title} {\bibinfo {title} {Sopra gli spazi di un numero qualunque di dimensioni},\ }\href@noop {} {\bibfield  {journal} {\bibinfo  {journal} {Annali di Matematica Pura ed Applicata (1867-1897)}\ }\textbf {\bibinfo {volume} {4}},\ \bibinfo {pages} {140} (\bibinfo {year} {1870})}\BibitemShut {NoStop}%
\bibitem [{\citenamefont {Poincar{\'e}}(1895)}]{poincare1895analysis}%
  \BibitemOpen
  \bibfield  {author} {\bibinfo {author} {\bibfnamefont {H.}~\bibnamefont {Poincar{\'e}}},\ }\href@noop {} {\emph {\bibinfo {title} {Analysis situs}}}\ (\bibinfo  {publisher} {Gauthier-Villars Paris, France},\ \bibinfo {year} {1895})\BibitemShut {NoStop}%
\bibitem [{\citenamefont {Parzen}(1962)}]{parzen1962estimation}%
  \BibitemOpen
  \bibfield  {author} {\bibinfo {author} {\bibfnamefont {E.}~\bibnamefont {Parzen}},\ }\bibfield  {title} {\bibinfo {title} {On estimation of a probability density function and mode},\ }\href@noop {} {\bibfield  {journal} {\bibinfo  {journal} {The annals of mathematical statistics}\ }\textbf {\bibinfo {volume} {33}},\ \bibinfo {pages} {1065} (\bibinfo {year} {1962})}\BibitemShut {NoStop}%
\bibitem [{\citenamefont {Davis}\ \emph {et~al.}(2011)\citenamefont {Davis}, \citenamefont {Lii},\ and\ \citenamefont {Politis}}]{davis2011remarks}%
  \BibitemOpen
  \bibfield  {author} {\bibinfo {author} {\bibfnamefont {R.~A.}\ \bibnamefont {Davis}}, \bibinfo {author} {\bibfnamefont {K.-S.}\ \bibnamefont {Lii}},\ and\ \bibinfo {author} {\bibfnamefont {D.~N.}\ \bibnamefont {Politis}},\ }\bibfield  {title} {\bibinfo {title} {Remarks on some nonparametric estimates of a density function},\ }\href@noop {} {\bibfield  {journal} {\bibinfo  {journal} {Selected Works of Murray Rosenblatt}\ ,\ \bibinfo {pages} {95}} (\bibinfo {year} {2011})}\BibitemShut {NoStop}%
\bibitem [{\citenamefont {Gauss}(1877)}]{gauss1877theoria}%
  \BibitemOpen
  \bibfield  {author} {\bibinfo {author} {\bibfnamefont {C.~F.}\ \bibnamefont {Gauss}},\ }\href@noop {} {\emph {\bibinfo {title} {Theoria motus corporum coelestium in sectionibus conicis solem ambientium}}},\ Vol.~\bibinfo {volume} {7}\ (\bibinfo  {publisher} {FA Perthes},\ \bibinfo {year} {1877})\BibitemShut {NoStop}%
\bibitem [{\citenamefont {Bauer}(2021)}]{bauer2021ripser}%
  \BibitemOpen
  \bibfield  {author} {\bibinfo {author} {\bibfnamefont {U.}~\bibnamefont {Bauer}},\ }\bibfield  {title} {\bibinfo {title} {Ripser: efficient computation of vietoris--rips persistence barcodes},\ }\href@noop {} {\bibfield  {journal} {\bibinfo  {journal} {Journal of Applied and Computational Topology}\ }\textbf {\bibinfo {volume} {5}},\ \bibinfo {pages} {391} (\bibinfo {year} {2021})}\BibitemShut {NoStop}%
\bibitem [{\citenamefont {Einstein}(1905)}]{einstein1905molekularkinetischen}%
  \BibitemOpen
  \bibfield  {author} {\bibinfo {author} {\bibfnamefont {A.}~\bibnamefont {Einstein}},\ }\bibfield  {title} {\bibinfo {title} {{\"U}ber die von der molekularkinetischen theorie der w{\"a}rme geforderte bewegung von in ruhenden fl{\"u}ssigkeiten suspendierten teilchen},\ }\href@noop {} {\bibfield  {journal} {\bibinfo  {journal} {Annalen der physik}\ }\textbf {\bibinfo {volume} {4}} (\bibinfo {year} {1905})}\BibitemShut {NoStop}%
\bibitem [{\citenamefont {Jungblut}\ and\ \citenamefont {Dellago}(2016)}]{jungblut2016pathways}%
  \BibitemOpen
  \bibfield  {author} {\bibinfo {author} {\bibfnamefont {S.}~\bibnamefont {Jungblut}}\ and\ \bibinfo {author} {\bibfnamefont {C.}~\bibnamefont {Dellago}},\ }\bibfield  {title} {\bibinfo {title} {Pathways to self-organization: Crystallization via nucleation and growth},\ }\href@noop {} {\bibfield  {journal} {\bibinfo  {journal} {The European Physical Journal E}\ }\textbf {\bibinfo {volume} {39}},\ \bibinfo {pages} {1} (\bibinfo {year} {2016})}\BibitemShut {NoStop}%
\bibitem [{\citenamefont {Heaviside}(2003)}]{heaviside2003electromagnetic}%
  \BibitemOpen
  \bibfield  {author} {\bibinfo {author} {\bibfnamefont {O.}~\bibnamefont {Heaviside}},\ }\href@noop {} {\emph {\bibinfo {title} {Electromagnetic theory}}},\ Vol.\ \bibinfo {volume} {237}\ (\bibinfo  {publisher} {American Mathematical Soc.},\ \bibinfo {year} {2003})\BibitemShut {NoStop}%
\bibitem [{\citenamefont {Candelier}\ \emph {et~al.}(2010)\citenamefont {Candelier}, \citenamefont {Widmer-Cooper}, \citenamefont {Kummerfeld}, \citenamefont {Dauchot}, \citenamefont {Biroli}, \citenamefont {Harrowell},\ and\ \citenamefont {Reichman}}]{candelier2010spatiotemporal}%
  \BibitemOpen
  \bibfield  {author} {\bibinfo {author} {\bibfnamefont {R.}~\bibnamefont {Candelier}}, \bibinfo {author} {\bibfnamefont {A.}~\bibnamefont {Widmer-Cooper}}, \bibinfo {author} {\bibfnamefont {J.~K.}\ \bibnamefont {Kummerfeld}}, \bibinfo {author} {\bibfnamefont {O.}~\bibnamefont {Dauchot}}, \bibinfo {author} {\bibfnamefont {G.}~\bibnamefont {Biroli}}, \bibinfo {author} {\bibfnamefont {P.}~\bibnamefont {Harrowell}},\ and\ \bibinfo {author} {\bibfnamefont {D.~R.}\ \bibnamefont {Reichman}},\ }\bibfield  {title} {\bibinfo {title} {Spatiotemporal hierarchy of relaxation events, dynamical heterogeneities, and structural reorganization in a supercooled liquid},\ }\href@noop {} {\bibfield  {journal} {\bibinfo  {journal} {Physical review letters}\ }\textbf {\bibinfo {volume} {105}},\ \bibinfo {pages} {135702} (\bibinfo {year} {2010})}\BibitemShut {NoStop}%
\bibitem [{\citenamefont {Smessaert}\ and\ \citenamefont {Rottler}(2013)}]{smessaert2013distribution}%
  \BibitemOpen
  \bibfield  {author} {\bibinfo {author} {\bibfnamefont {A.}~\bibnamefont {Smessaert}}\ and\ \bibinfo {author} {\bibfnamefont {J.}~\bibnamefont {Rottler}},\ }\bibfield  {title} {\bibinfo {title} {Distribution of local relaxation events in an aging three-dimensional glass: Spatiotemporal correlation and dynamical heterogeneity},\ }\href@noop {} {\bibfield  {journal} {\bibinfo  {journal} {Physical Review E}\ }\textbf {\bibinfo {volume} {88}},\ \bibinfo {pages} {022314} (\bibinfo {year} {2013})}\BibitemShut {NoStop}%
\bibitem [{\citenamefont {Schoenholz}\ \emph {et~al.}(2016)\citenamefont {Schoenholz}, \citenamefont {Cubuk}, \citenamefont {Sussman}, \citenamefont {Kaxiras},\ and\ \citenamefont {Liu}}]{schoenholz2016structural}%
  \BibitemOpen
  \bibfield  {author} {\bibinfo {author} {\bibfnamefont {S.~S.}\ \bibnamefont {Schoenholz}}, \bibinfo {author} {\bibfnamefont {E.~D.}\ \bibnamefont {Cubuk}}, \bibinfo {author} {\bibfnamefont {D.~M.}\ \bibnamefont {Sussman}}, \bibinfo {author} {\bibfnamefont {E.}~\bibnamefont {Kaxiras}},\ and\ \bibinfo {author} {\bibfnamefont {A.~J.}\ \bibnamefont {Liu}},\ }\bibfield  {title} {\bibinfo {title} {A structural approach to relaxation in glassy liquids},\ }\href@noop {} {\bibfield  {journal} {\bibinfo  {journal} {Nature Physics}\ }\textbf {\bibinfo {volume} {12}},\ \bibinfo {pages} {469} (\bibinfo {year} {2016})}\BibitemShut {NoStop}%
\bibitem [{\citenamefont {Cortes}\ and\ \citenamefont {Vapnik}(1995)}]{cortes1995support}%
  \BibitemOpen
  \bibfield  {author} {\bibinfo {author} {\bibfnamefont {C.}~\bibnamefont {Cortes}}\ and\ \bibinfo {author} {\bibfnamefont {V.}~\bibnamefont {Vapnik}},\ }\bibfield  {title} {\bibinfo {title} {Support-vector networks},\ }\href@noop {} {\bibfield  {journal} {\bibinfo  {journal} {Machine learning}\ }\textbf {\bibinfo {volume} {20}},\ \bibinfo {pages} {273} (\bibinfo {year} {1995})}\BibitemShut {NoStop}%
\bibitem [{\citenamefont {Pearson}(1901)}]{pearson1901liii}%
  \BibitemOpen
  \bibfield  {author} {\bibinfo {author} {\bibfnamefont {K.}~\bibnamefont {Pearson}},\ }\bibfield  {title} {\bibinfo {title} {Liii. on lines and planes of closest fit to systems of points in space},\ }\href@noop {} {\bibfield  {journal} {\bibinfo  {journal} {The London, Edinburgh, and Dublin philosophical magazine and journal of science}\ }\textbf {\bibinfo {volume} {2}},\ \bibinfo {pages} {559} (\bibinfo {year} {1901})}\BibitemShut {NoStop}%
\bibitem [{\citenamefont {Matthews}(1975)}]{matthews1975comparison}%
  \BibitemOpen
  \bibfield  {author} {\bibinfo {author} {\bibfnamefont {B.~W.}\ \bibnamefont {Matthews}},\ }\bibfield  {title} {\bibinfo {title} {Comparison of the predicted and observed secondary structure of t4 phage lysozyme},\ }\href@noop {} {\bibfield  {journal} {\bibinfo  {journal} {Biochimica et Biophysica Acta (BBA)-Protein Structure}\ }\textbf {\bibinfo {volume} {405}},\ \bibinfo {pages} {442} (\bibinfo {year} {1975})}\BibitemShut {NoStop}%
\bibitem [{\citenamefont {Snoek}\ \emph {et~al.}(2012)\citenamefont {Snoek}, \citenamefont {Larochelle},\ and\ \citenamefont {Adams}}]{snoek2012practical}%
  \BibitemOpen
  \bibfield  {author} {\bibinfo {author} {\bibfnamefont {J.}~\bibnamefont {Snoek}}, \bibinfo {author} {\bibfnamefont {H.}~\bibnamefont {Larochelle}},\ and\ \bibinfo {author} {\bibfnamefont {R.~P.}\ \bibnamefont {Adams}},\ }\bibfield  {title} {\bibinfo {title} {Practical bayesian optimization of machine learning algorithms},\ }\href@noop {} {\bibfield  {journal} {\bibinfo  {journal} {Advances in neural information processing systems}\ }\textbf {\bibinfo {volume} {25}} (\bibinfo {year} {2012})}\BibitemShut {NoStop}%
\bibitem [{\citenamefont {Cubuk}\ \emph {et~al.}(2015)\citenamefont {Cubuk}, \citenamefont {Schoenholz}, \citenamefont {Rieser}, \citenamefont {Malone}, \citenamefont {Rottler}, \citenamefont {Durian}, \citenamefont {Kaxiras},\ and\ \citenamefont {Liu}}]{cubuk2015identifying}%
  \BibitemOpen
  \bibfield  {author} {\bibinfo {author} {\bibfnamefont {E.~D.}\ \bibnamefont {Cubuk}}, \bibinfo {author} {\bibfnamefont {S.~S.}\ \bibnamefont {Schoenholz}}, \bibinfo {author} {\bibfnamefont {J.~M.}\ \bibnamefont {Rieser}}, \bibinfo {author} {\bibfnamefont {B.~D.}\ \bibnamefont {Malone}}, \bibinfo {author} {\bibfnamefont {J.}~\bibnamefont {Rottler}}, \bibinfo {author} {\bibfnamefont {D.~J.}\ \bibnamefont {Durian}}, \bibinfo {author} {\bibfnamefont {E.}~\bibnamefont {Kaxiras}},\ and\ \bibinfo {author} {\bibfnamefont {A.~J.}\ \bibnamefont {Liu}},\ }\bibfield  {title} {\bibinfo {title} {Identifying structural flow defects in disordered solids using machine-learning methods},\ }\href@noop {} {\bibfield  {journal} {\bibinfo  {journal} {Physical review letters}\ }\textbf {\bibinfo {volume} {114}},\ \bibinfo {pages} {108001} (\bibinfo {year} {2015})}\BibitemShut {NoStop}%
\bibitem [{\citenamefont {Shapley}\ \emph {et~al.}(1953)\citenamefont {Shapley} \emph {et~al.}}]{shapley1953value}%
  \BibitemOpen
  \bibfield  {author} {\bibinfo {author} {\bibfnamefont {L.~S.}\ \bibnamefont {Shapley}} \emph {et~al.},\ }\bibfield  {title} {\bibinfo {title} {A value for n-person games},\ }\href@noop {} {\bibfield  {journal} {\bibinfo  {journal} {Contributions to the Theory of Games}\ }\textbf {\bibinfo {volume} {2}},\ \bibinfo {pages} {307} (\bibinfo {year} {1953})}\BibitemShut {NoStop}%
\bibitem [{\citenamefont {Winter}(2002)}]{winter2002shapley}%
  \BibitemOpen
  \bibfield  {author} {\bibinfo {author} {\bibfnamefont {E.}~\bibnamefont {Winter}},\ }\bibfield  {title} {\bibinfo {title} {The shapley value},\ }\href@noop {} {\bibfield  {journal} {\bibinfo  {journal} {Handbook of game theory with economic applications}\ }\textbf {\bibinfo {volume} {3}},\ \bibinfo {pages} {2025} (\bibinfo {year} {2002})}\BibitemShut {NoStop}%
\bibitem [{\citenamefont {Alder}\ and\ \citenamefont {Wainwright}(1957)}]{alder1957phase}%
  \BibitemOpen
  \bibfield  {author} {\bibinfo {author} {\bibfnamefont {B.~J.}\ \bibnamefont {Alder}}\ and\ \bibinfo {author} {\bibfnamefont {T.~E.}\ \bibnamefont {Wainwright}},\ }\bibfield  {title} {\bibinfo {title} {Phase transition for a hard sphere system},\ }\href@noop {} {\bibfield  {journal} {\bibinfo  {journal} {The Journal of chemical physics}\ }\textbf {\bibinfo {volume} {27}},\ \bibinfo {pages} {1208} (\bibinfo {year} {1957})}\BibitemShut {NoStop}%
\bibitem [{\citenamefont {Rahman}(1964)}]{rahman1964correlations}%
  \BibitemOpen
  \bibfield  {author} {\bibinfo {author} {\bibfnamefont {A.}~\bibnamefont {Rahman}},\ }\bibfield  {title} {\bibinfo {title} {Correlations in the motion of atoms in liquid argon},\ }\href@noop {} {\bibfield  {journal} {\bibinfo  {journal} {Physical review}\ }\textbf {\bibinfo {volume} {136}},\ \bibinfo {pages} {A405} (\bibinfo {year} {1964})}\BibitemShut {NoStop}%
\bibitem [{\citenamefont {Hirschfelder}\ \emph {et~al.}(1964)\citenamefont {Hirschfelder}, \citenamefont {Curtiss},\ and\ \citenamefont {Bird}}]{hirschfelder1964molecular}%
  \BibitemOpen
  \bibfield  {author} {\bibinfo {author} {\bibfnamefont {J.~O.}\ \bibnamefont {Hirschfelder}}, \bibinfo {author} {\bibfnamefont {C.~F.}\ \bibnamefont {Curtiss}},\ and\ \bibinfo {author} {\bibfnamefont {R.~B.}\ \bibnamefont {Bird}},\ }\href@noop {} {\emph {\bibinfo {title} {The molecular theory of gases and liquids}}}\ (\bibinfo  {publisher} {John Wiley \& Sons},\ \bibinfo {year} {1964})\BibitemShut {NoStop}%
\bibitem [{\citenamefont {Hoover}\ and\ \citenamefont {Holian}(1996)}]{hoover1996kinetic}%
  \BibitemOpen
  \bibfield  {author} {\bibinfo {author} {\bibfnamefont {W.~G.}\ \bibnamefont {Hoover}}\ and\ \bibinfo {author} {\bibfnamefont {B.~L.}\ \bibnamefont {Holian}},\ }\bibfield  {title} {\bibinfo {title} {Kinetic moments method for the canonical ensemble distribution},\ }\href@noop {} {\bibfield  {journal} {\bibinfo  {journal} {Physics Letters A}\ }\textbf {\bibinfo {volume} {211}},\ \bibinfo {pages} {253} (\bibinfo {year} {1996})}\BibitemShut {NoStop}%
\bibitem [{\citenamefont {Martyna}\ \emph {et~al.}(1994)\citenamefont {Martyna}, \citenamefont {Tobias},\ and\ \citenamefont {Klein}}]{martyna1994constant}%
  \BibitemOpen
  \bibfield  {author} {\bibinfo {author} {\bibfnamefont {G.~J.}\ \bibnamefont {Martyna}}, \bibinfo {author} {\bibfnamefont {D.~J.}\ \bibnamefont {Tobias}},\ and\ \bibinfo {author} {\bibfnamefont {M.~L.}\ \bibnamefont {Klein}},\ }\bibfield  {title} {\bibinfo {title} {Constant pressure molecular dynamics algorithms},\ }\href@noop {} {\bibfield  {journal} {\bibinfo  {journal} {The Journal of chemical physics}\ }\textbf {\bibinfo {volume} {101}},\ \bibinfo {pages} {4177} (\bibinfo {year} {1994})}\BibitemShut {NoStop}%
\bibitem [{\citenamefont {Wang}\ and\ \citenamefont {Sosso}(2024)}]{wang2024graph}%
  \BibitemOpen
  \bibfield  {author} {\bibinfo {author} {\bibfnamefont {A.}~\bibnamefont {Wang}}\ and\ \bibinfo {author} {\bibfnamefont {G.~C.}\ \bibnamefont {Sosso}},\ }\bibfield  {title} {\bibinfo {title} {Graph-based descriptors for condensed matter},\ }\href@noop {} {\bibfield  {journal} {\bibinfo  {journal} {arXiv preprint arXiv:2408.06156}\ } (\bibinfo {year} {2024})}\BibitemShut {NoStop}%
\bibitem [{\citenamefont {Blow}\ \emph {et~al.}(2021)\citenamefont {Blow}, \citenamefont {Quigley},\ and\ \citenamefont {Sosso}}]{blow2021seven}%
  \BibitemOpen
  \bibfield  {author} {\bibinfo {author} {\bibfnamefont {K.~E.}\ \bibnamefont {Blow}}, \bibinfo {author} {\bibfnamefont {D.}~\bibnamefont {Quigley}},\ and\ \bibinfo {author} {\bibfnamefont {G.~C.}\ \bibnamefont {Sosso}},\ }\bibfield  {title} {\bibinfo {title} {The seven deadly sins: When computing crystal nucleation rates, the devil is in the details},\ }\href@noop {} {\bibfield  {journal} {\bibinfo  {journal} {The Journal of Chemical Physics}\ }\textbf {\bibinfo {volume} {155}} (\bibinfo {year} {2021})}\BibitemShut {NoStop}%
\bibitem [{\citenamefont {Kob}\ and\ \citenamefont {Andersen}(1995)}]{kob1995testing}%
  \BibitemOpen
  \bibfield  {author} {\bibinfo {author} {\bibfnamefont {W.}~\bibnamefont {Kob}}\ and\ \bibinfo {author} {\bibfnamefont {H.~C.}\ \bibnamefont {Andersen}},\ }\bibfield  {title} {\bibinfo {title} {Testing mode-coupling theory for a supercooled binary lennard-jones mixture i: The van hove correlation function},\ }\href@noop {} {\bibfield  {journal} {\bibinfo  {journal} {Physical Review E}\ }\textbf {\bibinfo {volume} {51}},\ \bibinfo {pages} {4626} (\bibinfo {year} {1995})}\BibitemShut {NoStop}%
\bibitem [{\citenamefont {Pedersen}\ \emph {et~al.}(2018)\citenamefont {Pedersen}, \citenamefont {Schr{\o}der},\ and\ \citenamefont {Dyre}}]{pedersen2018phase}%
  \BibitemOpen
  \bibfield  {author} {\bibinfo {author} {\bibfnamefont {U.~R.}\ \bibnamefont {Pedersen}}, \bibinfo {author} {\bibfnamefont {T.~B.}\ \bibnamefont {Schr{\o}der}},\ and\ \bibinfo {author} {\bibfnamefont {J.~C.}\ \bibnamefont {Dyre}},\ }\bibfield  {title} {\bibinfo {title} {Phase diagram of kob-andersen-type binary lennard-jones mixtures},\ }\href@noop {} {\bibfield  {journal} {\bibinfo  {journal} {Physical review letters}\ }\textbf {\bibinfo {volume} {120}},\ \bibinfo {pages} {165501} (\bibinfo {year} {2018})}\BibitemShut {NoStop}%
\bibitem [{\citenamefont {Ganapathi}\ \emph {et~al.}(2021)\citenamefont {Ganapathi}, \citenamefont {Chakrabarti}, \citenamefont {Sood},\ and\ \citenamefont {Ganapathy}}]{ganapathi2021structure}%
  \BibitemOpen
  \bibfield  {author} {\bibinfo {author} {\bibfnamefont {D.}~\bibnamefont {Ganapathi}}, \bibinfo {author} {\bibfnamefont {D.}~\bibnamefont {Chakrabarti}}, \bibinfo {author} {\bibfnamefont {A.}~\bibnamefont {Sood}},\ and\ \bibinfo {author} {\bibfnamefont {R.}~\bibnamefont {Ganapathy}},\ }\bibfield  {title} {\bibinfo {title} {Structure determines where crystallization occurs in a soft colloidal glass},\ }\href@noop {} {\bibfield  {journal} {\bibinfo  {journal} {Nature Physics}\ }\textbf {\bibinfo {volume} {17}},\ \bibinfo {pages} {114} (\bibinfo {year} {2021})}\BibitemShut {NoStop}%
\bibitem [{\citenamefont {Boser}\ \emph {et~al.}(1992)\citenamefont {Boser}, \citenamefont {Guyon},\ and\ \citenamefont {Vapnik}}]{boser1992training}%
  \BibitemOpen
  \bibfield  {author} {\bibinfo {author} {\bibfnamefont {B.~E.}\ \bibnamefont {Boser}}, \bibinfo {author} {\bibfnamefont {I.~M.}\ \bibnamefont {Guyon}},\ and\ \bibinfo {author} {\bibfnamefont {V.~N.}\ \bibnamefont {Vapnik}},\ }\bibfield  {title} {\bibinfo {title} {A training algorithm for optimal margin classifiers},\ }in\ \href@noop {} {\emph {\bibinfo {booktitle} {Proceedings of the fifth annual workshop on Computational learning theory}}}\ (\bibinfo {year} {1992})\ pp.\ \bibinfo {pages} {144--152}\BibitemShut {NoStop}%
\bibitem [{\citenamefont {Gu}\ and\ \citenamefont {Yau}(2002)}]{gu2002computing}%
  \BibitemOpen
  \bibfield  {author} {\bibinfo {author} {\bibfnamefont {X.}~\bibnamefont {Gu}}\ and\ \bibinfo {author} {\bibfnamefont {S.-T.}\ \bibnamefont {Yau}},\ }\bibfield  {title} {\bibinfo {title} {Computing conformal structure of surfaces},\ }\href@noop {} {\bibfield  {journal} {\bibinfo  {journal} {arXiv preprint cs/0212043}\ } (\bibinfo {year} {2002})}\BibitemShut {NoStop}%
\bibitem [{\citenamefont {Gu}\ and\ \citenamefont {Yau}(2003)}]{gu2003global}%
  \BibitemOpen
  \bibfield  {author} {\bibinfo {author} {\bibfnamefont {X.}~\bibnamefont {Gu}}\ and\ \bibinfo {author} {\bibfnamefont {S.-T.}\ \bibnamefont {Yau}},\ }\bibfield  {title} {\bibinfo {title} {Global conformal surface parameterization},\ }in\ \href@noop {} {\emph {\bibinfo {booktitle} {Proceedings of the 2003 Eurographics/ACM SIGGRAPH symposium on Geometry processing}}}\ (\bibinfo {year} {2003})\ pp.\ \bibinfo {pages} {127--137}\BibitemShut {NoStop}%
\bibitem [{\citenamefont {Gu}\ \emph {et~al.}(2010)\citenamefont {Gu}, \citenamefont {Luo},\ and\ \citenamefont {Yau}}]{gu2010fundamentals}%
  \BibitemOpen
  \bibfield  {author} {\bibinfo {author} {\bibfnamefont {D.~X.}\ \bibnamefont {Gu}}, \bibinfo {author} {\bibfnamefont {F.}~\bibnamefont {Luo}},\ and\ \bibinfo {author} {\bibfnamefont {S.-T.}\ \bibnamefont {Yau}},\ }\bibfield  {title} {\bibinfo {title} {Fundamentals of computational conformal geometry},\ }\href@noop {} {\bibfield  {journal} {\bibinfo  {journal} {Mathematics in Computer Science}\ }\textbf {\bibinfo {volume} {4}},\ \bibinfo {pages} {389} (\bibinfo {year} {2010})}\BibitemShut {NoStop}%
\end{thebibliography}
%

\end{document}